\documentclass[twocolumn,tighten,times]{aastex62}   %

\usepackage{amsmath,amssymb,listings}
\usepackage{mathptmx}   %
\usepackage{ulem}

\usepackage{etoolbox}

\makeatletter

\patchcmd{\NAT@citex}
  {\@citea\NAT@hyper@{%
     \NAT@nmfmt{\NAT@nm}%
     \hyper@natlinkbreak{\NAT@aysep\NAT@spacechar}{\@citeb\@extra@b@citeb}%
     \NAT@date}}
  {\@citea\NAT@nmfmt{\NAT@nm}%
   \NAT@aysep\NAT@spacechar\NAT@hyper@{\NAT@date}}{}{}

\patchcmd{\NAT@citex}
  {\@citea\NAT@hyper@{%
     \NAT@nmfmt{\NAT@nm}%
     \hyper@natlinkbreak{\NAT@spacechar\NAT@@open\if*#1*\else#1\NAT@spacechar\fi}%
       {\@citeb\@extra@b@citeb}%
     \NAT@date}}
  {\@citea\NAT@nmfmt{\NAT@nm}%
   \NAT@spacechar\NAT@@open\if*#1*\else#1\NAT@spacechar\fi\NAT@hyper@{\NAT@date}}
  {}{}

\makeatother
\newcommand{\PLUTO}{{\texttt{PLUTO}}}
\newcommand{\mkmk}{{\texttt{Makemake}}}

\newcommand{\mesa}{{\texttt{mesa}}}

\newcommand{\nnBegut}[1]{#1}
\newcommand{\wBegut}[1]{}
\def\tauZeitskala{t}

\makeatletter
\let\jnl@style=\rm
\def\ref@jnl#1{{\jnl@style#1}}

\def\aj{\ref@jnl{AJ}}                   %
\def\actaa{\ref@jnl{Acta Astron.}}      %
\def\araa{\ref@jnl{ARA\&A}}             %
\def\apj{\ref@jnl{ApJ}}                 %
\def\apjl{\ref@jnl{ApJ}}                %
\def\apjs{\ref@jnl{ApJS}}               %
\def\ao{\ref@jnl{Appl.~Opt.}}           %
\def\apss{\ref@jnl{Ap\&SS}}             %
\def\aap{\ref@jnl{A\&A}}                %
\def\aapr{\ref@jnl{A\&A~Rev.}}          %
\def\aaps{\ref@jnl{A\&AS}}              %
\def\azh{\ref@jnl{AZh}}                 %
\def\baas{\ref@jnl{BAAS}}               %
\def\bac{\ref@jnl{Bull. astr. Inst. Czechosl.}}
\def\caa{\ref@jnl{Chinese Astron. Astrophys.}}
\def\cjaa{\ref@jnl{Chinese J. Astron. Astrophys.}}
\def\icarus{\ref@jnl{Icarus}}           %
\def\jcap{\ref@jnl{J. Cosmology Astropart. Phys.}}
\def\jrasc{\ref@jnl{JRASC}}             %
\def\memras{\ref@jnl{MmRAS}}            %
\def\mnras{\ref@jnl{MNRAS}}             %
\def\na{\ref@jnl{New A}}                %
\def\nar{\ref@jnl{New A Rev.}}          %
\def\pra{\ref@jnl{Phys.~Rev.~A}}        %
\def\prb{\ref@jnl{Phys.~Rev.~B}}        %
\def\prc{\ref@jnl{Phys.~Rev.~C}}        %
\def\prd{\ref@jnl{Phys.~Rev.~D}}        %
\def\pre{\ref@jnl{Phys.~Rev.~E}}        %
\def\prl{\ref@jnl{Phys.~Rev.~Lett.}}    %
\def\pasa{\ref@jnl{PASA}}               %
\def\pasp{\ref@jnl{PASP}}               %
\def\pasj{\ref@jnl{PASJ}}               %
\def\rmxaa{\ref@jnl{Rev. Mexicana Astron. Astrofis.}}%
\def\qjras{\ref@jnl{QJRAS}}             %
\def\skytel{\ref@jnl{S\&T}}             %
\def\solphys{\ref@jnl{Sol.~Phys.}}      %
\def\sovast{\ref@jnl{Soviet~Ast.}}      %
\def\ssr{\ref@jnl{Space~Sci.~Rev.}}     %
\def\zap{\ref@jnl{ZAp}}                 %
\def\nat{\ref@jnl{Nature}}              %
\def\iaucirc{\ref@jnl{IAU~Circ.}}       %
\def\aplett{\ref@jnl{Astrophys.~Lett.}} %
\def\apspr{\ref@jnl{Astrophys.~Space~Phys.~Res.}}
\def\bain{\ref@jnl{Bull.~Astron.~Inst.~Netherlands}} 
\def\fcp{\ref@jnl{Fund.~Cosmic~Phys.}}  %
\def\gca{\ref@jnl{Geochim.~Cosmochim.~Acta}}   %
\def\grl{\ref@jnl{Geophys.~Res.~Lett.}} %
\def\jcp{\ref@jnl{J.~Chem.~Phys.}}      %
\def\jgr{\ref@jnl{J.~Geophys.~Res.}}    %
\def\jqsrt{\ref@jnl{J.~Quant.~Spec.~Radiat.~Transf.}}
\def\memsai{\ref@jnl{Mem.~Soc.~Astron.~Italiana}}
\def\nphysa{\ref@jnl{Nucl.~Phys.~A}}   %
\def\physrep{\ref@jnl{Phys.~Rep.}}   %
\def\physscr{\ref@jnl{Phys.~Scr}}   %
\def\planss{\ref@jnl{Planet.~Space~Sci.}}   %
\def\procspie{\ref@jnl{Proc.~SPIE}}   %

\def\ptp{\ref@jnl{Prog.~Th.~Phys.}}   %

\makeatother

\defcitealias{goli04}{G04}
\defcitealias{marl07}{M07}
\defcitealias{scvh}{SCvH}
\defcitealias{bl94}{BL94}
\defcitealias{mc14}{MC14}
\defcitealias{ensman94}{E94}
\defcitealias{lever81}{LP81}
\defcitealias{semenov03}{S03}

\newcommand{\Vekt}[1]{\mathbf{#1}}  %

\newcommand{\eqsep}{\;\;\;}
\newcommand{\giltwenn}[1]{\eqsep\eqsep(#1)}  %

\newcommand{\dd}{{\rm d}}

\def\upartial{\partial}

\newcommand{\overbar}[1]{\mkern 1.5mu\overline{\mkern-1.5mu#1\mkern-1.5mu}\mkern 1.5mu}

\usepackage{grffile}

\newcommand{\K}[1]{}

\definecolor{Hellgrau}{gray}{0.7}

\newcommand{\MJ}{{M_{\textnormal{J}}}}
\newcommand{\RJ}{{R_{\textnormal{J}}}}

\newcommand{\ME}{{M_{\oplus}}}

\newcommand{\LSonne}{{L_{\odot}}}

\newcommand{\sigSB}{{\sigma}}

\newcommand{\mH}{{m_{\textnormal{H}}}}

\newcommand{\kB}{{k_{\textnormal{B}}}}

\newcommand{\MP}{{M_{\textnormal{p}}}}
\newcommand{\RP}{{R_{\textnormal{p}}}}

\newcommand{\LP}{{L_{\textnormal{p}}}}
\newcommand{\Lint}{{L_{\textnormal{int}}}}

\newcommand{\RAkk}{{R_{\textnormal{acc}}}}

\newcommand{\RHill}{{R_{\textnormal{Hill}}}}
\newcommand{\RBondi}{{R_{\textnormal{Bondi}}}}
\newcommand{\kLiss}{{k_{\textnormal{Lissauer}}}}
\newcommand{\MPunkt}{{\dot{M}}}

\newcommand{\TNeb}{{T_{\textnormal{neb}}}}

\newcommand{\rhoVorSch}{{\rho_{\textnormal{pre}}}}
\newcommand{\rhoNachSch}{{\rho_{\textnormal{post}}}}
\newcommand{\Mach}{{\mathcal{M}}}

\newcommand{\Ppost}{{P_{\textnormal{post}}}}

\newcommand{\Snach}{{s_{\rm post}}}

\newcommand{\kapR}{{\kappa_{\textnormal{R}}}}
\newcommand{\kapP}{{\kappa_{\textnormal{P}}}}
\newcommand{\kapE}{{\kappa_{\textnormal{E}}}}
\newcommand{\TZerst}{{T_{\textnormal{dest}}}}

\newcommand{\tauR}{{\tau_{\mathrm{R}}}}

\newcommand{\vFf}{{v_{\textnormal{ff}}}}
\newcommand{\rhoFf}{{\rho_{\textnormal{ff}}}}

\newcommand{\tKH}{{t_{\rm KH}}}

\newcommand{\tFliess}{{\tauZeitskala_{\textnormal{flow}}}}

\newcommand{\tAkk}{{\tauZeitskala_{\rm acc}}}

\newcommand{\tnitti}{{\tauZeitskala_{90}}}
\newcommand{\texp}{{\tauZeitskala_{\textnormal{exp}}}}

\newcommand{\Teff}{{T_{\textnormal{eff}}}}
\newcommand{\fred}{{f_{\textnormal{red}}}}

\newcommand{\ceff}{{c_{\textnormal{eff}}}}

\newcommand{\Eint}{{E_{\textnormal{int}}}}
\newcommand{\Ekin}{{E_{\textnormal{kin}}}}

\newcommand{\Erad}{{E_{\textnormal{rad}}}}
\newcommand{\Trad}{{T_{\textnormal{rad}}}}

\newcommand{\TGas}{{T_{\textnormal{gas}}}}

\newcommand{\eint}{{e_{\textnormal{int}}}}
\newcommand{\ekin}{{e_{\textnormal{kin}}}}

\newcommand{\DF}[1][]{{D_{\textnormal{F}#1}}}  %

\newcommand{\rmin}{{r_{\rm min}}}

\newcommand{\rmax}{{r_{\rm max}}}

\newcommand{\QSchock}{{Q^+_{\rm shock}}}
\newcommand{\qSchock}{{q^+_{\rm shock,\,rel}}}

\newcommand{\rSchock}{{r_{\rm shock}}}
\newcommand{\LAkk}{{L_{\rm acc}}}

\newcommand{\LAkkmax}{{L_{\rm acc,\,max}}}
\newcommand{\TSchock}{{T_{\rm shock}}}
\newcommand{\TSchockff}{{T_{\rm sh,\,fs}}}  %
\newcommand{\TSchockdiff}{{T_{\rm sh,\,diff}}}  %
\newcommand{\LkonstLfred}{{L_{(\ref{Gl:T(r) konst L/fred})}}}  %
\def\TAkk{T_{\mathrm{acc}}}   %

\newcommand{\Llinks}{{L_{\rm dnstr}}}
\newcommand{\LKomp}{{L_{\rm compr}}}

\newcommand{\geff}{{\gamma_{\rm eff}}}  %

\newcommand{\tEnde}{{t_{\textnormal{stop}}}}

\newcommand{\Frad}{{F_{\rm rad}}}

\def\vSch{{ v_{\rm shock} }}
\def\etaklassisch{ \eta^{\rm kin} }
\def\etaphys{      \eta^{\rm phys} }

\def\fmin{      f_{\rm min} }

\def\Pram{P_{\rm ram}}

\newcommand{\cs}{{c_{\rm s}}}

\newcommand{\EPkt}{{\dot{E}}}

\received{15 November 2018}
\revised{01 May 2019}
\accepted{22 May 2019}

\defcitealias{m16Schock}{Paper~I}

\shorttitle{%
Planet accretion shock. II.\ Grid of simulations%
}
\shortauthors{Marleau, Mordasini, \&\ Kuiper}

\begin{document}

\title{The Planetary Accretion Shock.~II.~Grid of Post-Shock Entropies and Radiative Shock Efficiencies\\
for Non-Equilibrium Radiation Transport} %

\email{gabriel.marleau@uni-tuebingen.de}  %

\author{Gabriel-Dominique Marleau}
\affil{%
Physikalisches Institut,
Universit\"{a}t Bern,
Gesellschaftsstr.~6,
3012 Bern, Switzerland}
\affiliation{%
Institut f\"ur Astronomie und Astrophysik,
Eberhard Karls Universit\"at T\"ubingen,
Auf der Morgenstelle 10,
72076 T\"ubingen, Germany
}
\affiliation{%
Max-Planck-Institut f\"ur Astronomie,
K\"onigstuhl 17,
69117 Heidelberg, Germany
}

\author{Christoph Mordasini}
\affiliation{%
Physikalisches Institut,
Universit\"{a}t Bern,
Gesellschaftsstr.~6,
3012 Bern, Switzerland}

\author{Rolf Kuiper}
\affiliation{%
Institut f\"ur Astronomie und Astrophysik,
Eberhard Karls Universit\"at T\"ubingen,
Auf der Morgenstelle 10,
72076 T\"ubingen, Germany
}
\affiliation{%
Max-Planck-Institut f\"ur Astronomie,
K\"onigstuhl 17,
69117 Heidelberg, Germany
}

\begin{abstract}
In the core-accretion formation scenario of gas giants, most of the gas accreting onto a planet
is processed through an accretion shock.
In this series of papers we study
this shock since it is key in setting the forming planet's structure
and thus its post-formation luminosity, with dramatic observational consequences.
We perform
one-dimensional grey radiation-hydrodynamical simulations
with non-equilibrium (two-temperature) radiation transport and up-to-date opacities.
We survey the parameter space of accretion rate, planet mass, and planet radius
and obtain post-shock temperatures, pressures, and entropies, as well as global radiation efficiencies.
We find that usually, the shock temperature $\TSchock$ is given by the ``free-streaming'' limit.
At low temperatures the dust opacity can make the shock hotter
but not significantly. %
We corroborate this with an original semi-analytical derivation
of $\TSchock$.
We also estimate the change in luminosity between the shock and the nebula. %
Neither $\TSchock$ nor the luminosity profile
depend directly on the optical depth between the shock and the nebula.
Rather, $\TSchock$ depends on the immediate pre-shock opacity,
and the luminosity change on the equation of state (EOS).
We find quite high immediate post-shock entropies ($S\approx13$--20~$\kB\,\mH^{-1}$),
which makes it seem unlikely that the shock can cool the planet.
The global radiation efficiencies are high ($\etaphys\gtrsim97\,\%$) but
the remainder of the total incoming energy, which is brought into the planet,
exceeds the internal luminosity of classical cold starts by orders of magnitude.
Overall, these findings suggest that warm or hot starts are more plausible.
\end{abstract}

\section{Introduction}
With its first direct detections already some ten to fifteen years ago \citep{chauvin04,marois08},
the technique of direct imaging has started to reveal a scarce but interesting population
of planets or very-low-mass substellar objects at large separations from their host stars
\citep{bowler16,bowler18,wagner19}.
The formation mechanism of individual detections is often not obvious
but gravitational instability as well as core accretion
(with the inclusion of $N$-body interactions during the formation phase and in the first few million years afterwards)
are likely candidates to explain the origin
of at least some of these systems (e.g., \citealp{m18hip}).  %
These different formation pathways may imprint into the observed brightness
of the planets \citep{bbmm16,mordasini17}.

To interpret the brightness measurements however requires knowing
the post-formation luminosity of planets of different masses.
Formation models%
, principally the ones of the California group \citep{pollack96,boden00,marl07,lissauer09,boden13}  %
and of the Bern group \citep{alibert05,morda12_I,morda12_II,mordasini17}
seek to predict this luminosity within the approximation of spherical accretion.
They need to assume something
about the efficiency of the gas accretion shock at the surface of the planet
during runaway gas accretion.
This efficiency is defined as the fraction of the total energy influx
which is re-radiated into the local disc and thus does not end up being added to the planet.
The extremes are known as ``cold starts'' and ``hot starts'' \citep{marl07}
and their post-formation luminosities can differ by orders of magnitude.

In a recent series of papers, \citet{berardo17}, \citet{berardocumming17}, and \citet{cumming18}
have begun calculating the structure of accreting planets following \citet{stahler88}.
Crucially, they take into account that in the settling zone below the shock,
the continuing compression of the post-shock layers leads to a non-constant luminosity.
They find that the thermal influence of the shock on the evolution of the planet during accretion depends
on the contrast between the entropy of the (outer convective zone of the)
planet and that of the post-shock gas.
This approach promises eventually to lead to more realistic predictions of the post-formation luminosity\footnote{
 This statement holds given an accretion history, which, admittedly,
 is however uncertain since it depends in part on the migration behaviour of the planet,
 itself fraught with uncertainty,}
but does require, as a boundary condition, knowledge of the temperature of the shock.

While global three-dimensional (radiation-)hydrodynamical simulations of the protoplanetary
and of the circumplanetary discs \citep{klahrkley06,machida08,tanigawa12,dangeloboden13,szul16,szul17}
have the potential of providing a realistic answer as to the post-shock conditions and its efficiency,
they
still have a limited spatial resolution of $\Delta x\sim1~\RJ$ at best at the position of the planet,
despite their high dynamical range of spatial scales.
Also due to the computational cost, they are (currently) unable to survey the large input
parameter space, which covers a factor of a few in planet radius,
an order of magnitude in mass, and several orders or magnitude in accretion rate
and internal luminosity (e.g., \citealp{morda12_I,mordasini17}).

In the present series of papers, we use one-dimensional models of the gas accretion
to take a careful look at shock microphysics. In \citet[][hereafter \citetalias{m16Schock}]{m16Schock},
we introduced our approach, presented a detailed analysis of results for one combination
of formation parameters, and discussed the shock efficiency for a certain range of parameters.
However, we restricted ourselves to equilibrium radiation transport,
in which the gas and radiation temperatures are assumed to be equal everywhere,
and assumed a perfect\footnote{
  This is also sometimes termed a ``constant EOS'' but should not
  be refered to only as an ``ideal gas'', as is unfortunately often done.
  Indeed, the latter only needs fulfill $P=\rho/(\mu\mH)\kB T$
  with $\mu$ not necessarily constant.
  A non-ideal EOS also includes for example quantum degeneracy effects.
  }
equation of state (EOS), i.e., a constant mean molecular weight $\mu$
and heat capacity ratio\footnote{
  For a perfect EOS, the various adiabatic indices $\Gamma_{1,2,3}$,
  the heat capacity ratio $\gamma$, as well as $\geff\equiv P/\eint+1$,
  where $\eint$ is the internal energy, are all equal.
  Therefore we always write $\gamma$ in this paper.}
$\gamma$.  %
We found that the shock is isothermal and supercritical
and that the efficiencies can be as low as 40\,\%.

Here, we relax the assumption of equilibrium radiation transport,
look more carefully at the role of opacity,
and consider a wider parameter space than in \citetalias{m16Schock},
exploring systematically the dependence on accretion rate, mass, and planet radius.
We also extend considerably our analytical derivations.
We again assume a perfect EOS, varying $\mu$ and $\gamma$, and focus on the cases in which
the internal luminosity $\Lint$ is much smaller than the shock luminosity $\LAkk$.

This paper is structured as follows:
In Section~\ref{Theil:LintLAkk} we estimate in what regions of $(\MPunkt,\MP,\RP,\Lint)$,
where $\MPunkt$ is the accretion rate while $\MP$ and $\RP$ are the planet mass and radius,
the assumption
$\Lint\ll\LAkk$ holds, which guides our choice of parameters.
Section~\ref{Theil:Modell} reviews briefly our set-up and details the relevant microphysics,
including the updated opacities.
The main thrust of this paper is in Section~\ref{Theil:Gitter+Analyse},
which presents and analyses results, including shock temperatures,
global radiative shock efficiencies, and post-shock entropies,
for a grid of simulations.
In Section~\ref{Theil:AnalytischT} we present semi-analytical derivations of the shock temperature
and of the temperature profile in the accretion flow and compare these to our results.
In Section~\ref{Theil:einzelne} we
investigate carefully the effect of different perfect EOS in non-equilibrium radiation transport
and of dust destruction in the Zel'dovich spike.
This motivates us to derive analytically the drop or increase
of the luminosity across the Hill sphere.
While we do not calculate the structure of forming planets using our shock results yet,
Section~\ref{Theil:Disk} explores whether hot or cold starts are expected,
and presents a further discussion.  %
Finally, Section~\ref{Theil:Zus} summarizes this work and presents our conclusions.

\section{Estimate of negligible internal luminosity}
 \label{Theil:LintLAkk}

The main formation parameters are
the accretion rate onto the planet, the planet mass, the planet radius,
and the internal luminosity of the planet, denoted respectively by
$\MPunkt$, $\MP$, $\RP=\rSchock$
(the position of the shock $\rSchock$ defining the radius of the very nearly hydrostatic protoplanet),
and $\Lint$.
In \citet{m16Schock} and this work, we focus on the case of negligible $\Lint$.
This luminosity comes from the contraction and cooling of the planet interior
and is generated almost entirely within a small fraction of the planet's volume,
where most of the mass resides.
Specifically, this means $\Lint\ll \Delta L$, where the luminosity jump at the shock $\Delta L \sim \LAkk$.
This reduces the number of free parameters in our study  %
but, mainly, is also expected to be the limit in which the shock simulations we perform are the most relevant.

\begin{figure}
 \epsscale{1.2}
 \plotone{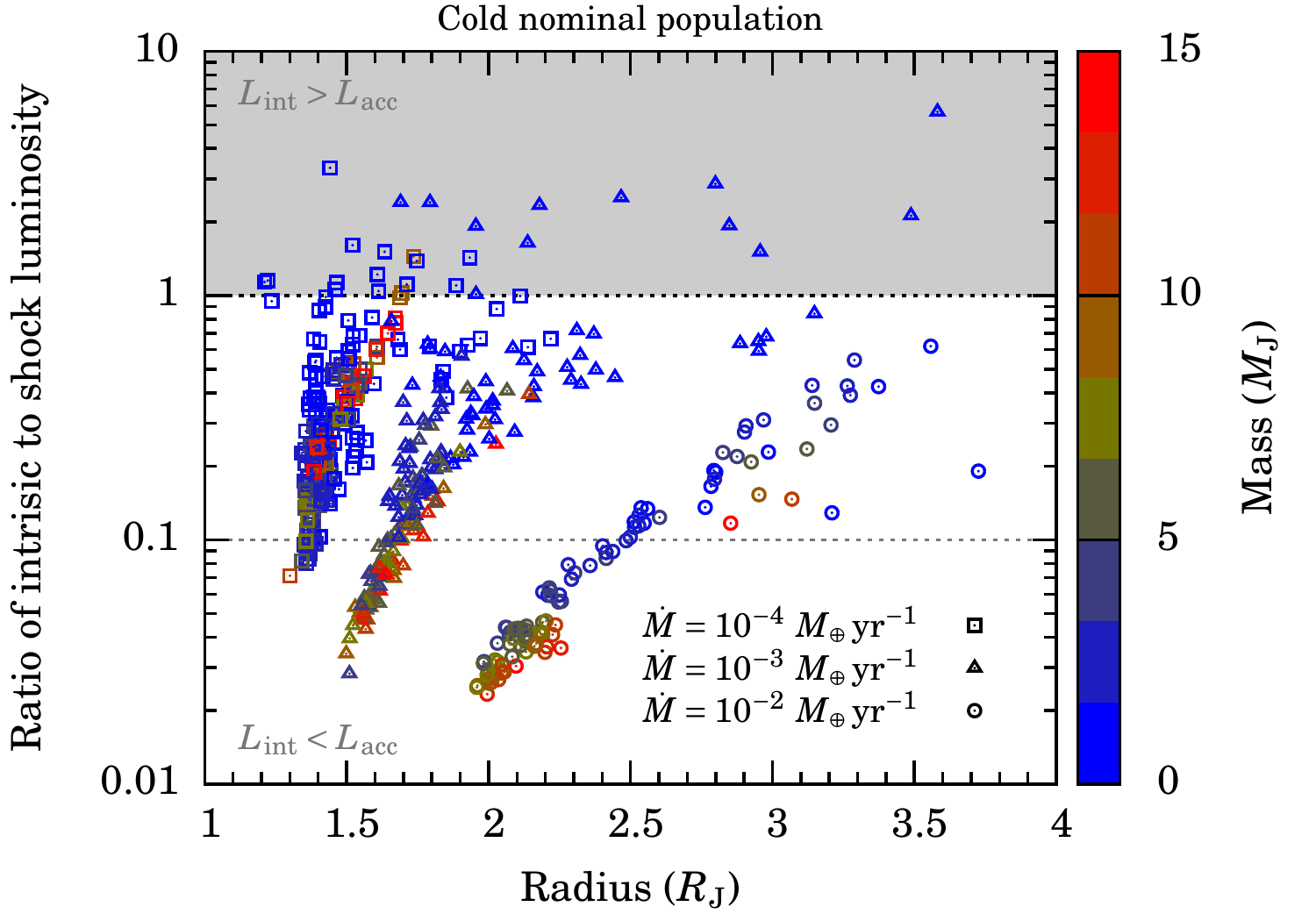}
\caption{
Ratio of the intrinsic to the shock luminosity in the cold nominal %
population syntheses of \citet{mordasini17}.
The colour of the points encodes the mass, with planets between $\MP=0.3$ and 15~$\MJ$ shown.
The three groups of points
are for accretion rates %
of $\MPunkt=10^{-4}$--$10^{-2}~\ME\,\textrm{yr}^{-1}$ within 0.05~dex (left to right; see legend).
}
\label{Abb:Lint/LSchock}
\end{figure}

We can
estimate in what cases neglecting the interior luminosity should be best justified.
In the \citet{marl07} formation calculations, which represent the extreme case
of cold accretion \citep{mordasini17}, it is blatantly obvious that $\Lint$ is negligible
compared to $\LAkk$ (see their figure~3 and the discussion in Section~\ref{Theil:H oder K?}).
For more ``moderate'' (i.e., warmer) versions of cold starts,
Figure~\ref{Abb:Lint/LSchock} shows the ratio $\Lint/\LAkk$ for cold-start
population synthesis planets as in figure~7d of \citep{mordasini17}.
Most points are below $\Lint/\LAkk\sim1$.
Overall, the higher the accretion rate or the mass, the less important the interior luminosity.
Using the hot-start population should yield somewhat weaker shock since the radii are larger
(leading to the ``core mass effect''; \citealp{morda13})
but overall the results should be similar (see figure~13 of \citealp{mordasini17}).

If we focus on $\MPunkt\gtrsim10^{-3}~\ME\,\textrm{yr}^{-1}$ and masses above a few $\MJ$,
simulations with $\RP\approx1.5$--3~$\RJ$
should be the ones in which the accretion shock is the most relevant.
At $\MPunkt\lesssim10^{-4}~\ME\,\textrm{yr}^{-1}$, the interior luminosity
frequently represents an appreciable fraction of the accretion luminosity,
even up to high planets masses.
Therefore, we do not consider these lower rates in this paper.
Of course, this estimate is not self-consistent since these formation and evolution calculations
assume a fixed shock efficiency $\eta=100\,\%$,   %
whereas we find clearly smaller values (in the sense that the heating of the planet
by the post-shock material is important relative to the planet's interior luminosity;
see Section~\ref{Theil:H oder K?} and \citetalias{m16Schock}).
Nevertheless, the estimate should provide reasonable guidance.

\section{Simulation approach}
 \label{Theil:Modell}

As in \citetalias{m16Schock}, we use the \PLUTO\ code \citep{mignone07,mignone12}
to solve in a time-explicit fashion for the hydrodynamics.
The one-dimensional, spherically-symmetric mass and momentum conservation equations are respectively
\begin{subequations}
\label{Gl:GG} %
\begin{align}
  \frac{\partial\rho}{\partial t} + \frac{1}{r^2}\frac{\partial}{\partial r}\left(r^2\rho v\right) &=0\label{Gl:GGa}\\
  \frac{\partial\left(\rho v\right)}{\partial t} + \frac{1}{r^2}\frac{\partial}{\partial r}\left(r^2 \varrho v^2 \right) + \frac{\partial P}{\partial r} &= -\varrho g, \label{Gl:GGb}
\end{align}
\end{subequations}
where $P$ is the pressure and the local gravitational acceleration is $g=|-G\MP/r^2|$, the self-gravity of the gas being negligible.
Note that, using mass conservation, the momentum equation can also be written in the form
\begin{equation}
  \rho\frac{\partial v}{\partial t} + \rho v\frac{\partial v}{\partial r} + \frac{\partial P}{\partial r} = -\varrho g,
\end{equation}
i.e., without the geometry factors $r^{\pm2}$ in the flux term.

The energy equation is similar to in \citetalias{m16Schock} but
we make use of the newest version of the flux-limited diffusion (FLD) solver \mkmk.
We return to this in Section~\ref{Theil:StrT}.

\subsection{Set-up}
 \label{Theil:Modell: Numerik}

The simulation domain extends from close to the planet's surface
almost out to the ``accretion radius'' $\RAkk$ (\citealp{boden00}; \citetalias{m16Schock}),
at one third \citep[$\kLiss=1/3$][]{lissauer09} of the Hill radius.
A grid with 1000~uniformly-spaced cells between the inner edge of the domain, $\rmin$, and $\rmin+0.5~\RJ$
and 1000~geometrically stretched cells from $\rmin+0.5~\RJ$ to the outer edge of the domain, $\rmax$, was found
to yield good results for several parameter combinations.
For the other cases, an increase to 2000+2000 cells sufficed to obtain smooth profiles
(e.g., in the local accretion rate $\MPunkt(r) = 4\pi r^2 \rho(v) v(r)$ below the shock).
We verified that increasing the resolution does not produce different results
except for sharpening the Zel'dovich spike (see below) and changing its peak value.
Thus there are no qualitative consequences on the pre- or post-shock structure.  %

To avoid numerical issues at early times, we have revised the initial set-up.
We do not use a hydrostatic atmosphere as was done in \citet{m16Schock}
but rather begin directly with an accretion profile in the density $\rho$ and velocity $v$.
The initial radiation and gas temperatures are set to the nebula temperature $\TNeb=150$~K \citep{poll94} at the outer edge $\rmax$
and increase linearly up to $1.1\TNeb$ at $\rmin$. The radiation and gas temperatures are set equal
and the initial pressure is obtained from $\rho$ and $\TGas$.

We use a CFL number of 0.8 (we verified that $\textrm{CFL}=0.4$ gives identical profiles),
and use the \texttt{tvdlf} Riemann solver along with the \texttt{MinMod} slope limiter.
This somewhat diffusive combination ensures numerical stability
while not significantly influencing the outcome.

The inner wall is reflective and the radiation flux there is set to zero.
We let gas fall in free-fall from the outer edge under the action of the central potential,
enforcing an accretion rate at the outer edge only.
The gas piles up at the bottom of the domain, forming a shock which defines the surface
of the planet at $\rSchock=\RP$ (both variables are used interchangeably throughout,
with $\rSchock^-$ ($\rSchock^+$) denoting the downstream (upstream) location).
This shock surface often, but not always, moves outwards over time, although at a rate
which is slow even compared to the strongly sub-sonic post-shock velocity.
The deepest pressure in the atmosphere we simulate is greater than the post-shock (ram) pressure
by one to several orders of magnitude, depending on the amount of accumulated mass
and the local gravitational acceleration $g=G\MP/r^2$, where $r$ is the radial coordinate in spherical geometry
and $G$ the gravitational constant.

While we use a fully time-dependent code %
our simulations represent steady-state snapshots
in the formation-parameter space, as discussed in \citetalias{m16Schock}.
For reference, the profiles for the runs presented here needed on the order of $5\times10^6$~s %
to come into equilibrium.
This is entirely negligible compared to the timescale on which the protoplanet grows,
which is on the order of $10^4$--$10^6$~yr.
(Even after $\tEnde=2\times10^7$~s, our usual stopping time,
some starting structures were still visible at a position $r$ deep below the shock
for which $\tEnde\lesssim-\int_r^{\rSchock}(1/v)\dd x$, in at least some simulations;
however, these regions are not of interest here.)  %

\subsection{Radiation transport}
 \label{Theil:StrT}
 
A significant improvement since \citetalias{m16Schock} is
the \nnBegut{change from equilibrium to}
non-equilibrium (or two-temperature, 2-$T$)
\nnBegut{flux-limited diffusion (FLD) in the radiation transport routine
\mkmk\ (\citealp{kuiper10}; Kuiper, Yorke, \&\ Mignone, in prep.)}
\nnBegut{In this approach,}
the gas and radiation temperatures are not enforced to be equal,
and the full energy equation reads
(see also \citealp{mihalas84,kley89a,turnerstone01,kuiper10,klassen14,commer11ramses})
\begin{subequations}
\label{Gl:GG2T}
\begin{align}
  \frac{\upartial\left(\Ekin+\Eint\right)}{\upartial t} + \nabla\left(\left[\Ekin+\Eint+P\right]v \right)
  & = - \Lambda - \rho g v,\label{Gl:GG2Ta}\\
  \frac{\upartial\Erad}{\upartial t} + \nabla\Frad & = + \Lambda,\label{Gl:GG2Tb}
\end{align}
\end{subequations}
omitting the radiation pressure since it is negligible in this problem,
and with $\Frad$ the radiative flux.
In FLD,
one writes \citepalias{m16Schock}
\begin{subequations}
 \label{Gl:Frad in FLD}
\begin{align}
 \Frad &= -\DF \Vekt{\nabla}\Erad,\label{Gl:FStrahl}\\
 \DF &\equiv \frac{\lambda(R) c}{\kapR\rho},\label{Gl:DF-Def}\\
 R &\equiv \frac{\| \Vekt{\nabla}\ln \Erad \|}{\kapR\,\rho},\label{Gl:RDef}  %
\end{align}
\end{subequations}
where $\DF$ is the diffusion coefficient, $\lambda$ the flux limiter (see Section~\ref{Gl:T-Profile im Akkfl} below),
$\kapR$ the Rosseland mean,
and $R$ the ``local radiation quantity'' (see \citetalias{m16Schock}).
Thus the radiative flux is assumed to be colinear with and opposite in direction to the gradient
of radiation energy density\footnote{In angularly-resolved radiation transport methods,
  it is in fact possible for the radiation to flow \textit{up} the $\Erad$ gradient
  \citep{mcclarrendrake10,jiang14}. However, this can occur only for subcritical shocks
  and does not represent a large effect, thus not affecting our work.}.
In Equation~(\ref{Gl:GG2T}), the combined cooling and heating or exchange term is
\begin{subequations}
 \label{Gl:Lambda}
\begin{align}
  \Lambda &\equiv c\kapP \rho\frac{4\pi S}{c} - c\kapE \rho\Erad\\
          &= c\kapP\rho \left(a \TGas^4 - \Erad\right).
\end{align}
\end{subequations}
We use the second line, since we take
the source function $S$ to be the Planck function $B = \sigSB\TGas^4/\pi$,  %
with $4\sigSB=ac$ ($\sigSB$ being the Stefan--Boltzmann constant
and $a$ and $c$ the radiation constant and speed of light).
This assumes local thermodynamic equilibrium (LTE).  %
Our grey approximation suffices to make $\kapP$ and $\kapE$,
the Planck (blackbody-weighted) and $\Erad$-weighted mean opacities respectively, be the same.  %
Equations~(\ref{Gl:GG2Ta}) and~(\ref{Gl:GG2Tb}) are followed separately to prevent
the $\Lambda$ terms from cancelling.
This system states that the material (gas or dust, assumed here to have the same temperature)
is losing energy at the rate $c\kapP \rho a \TGas^4$ but absorbing energy (photons)  %
at the rate $c\kapE \rho \Erad$, and conversely for the radiation.

\nnBegut{To solve the energy equation~(\ref{Gl:GG2Tb}) in the radiation transport substep,
the radiative flux $\Frad$ is replaced by its expression from Equation~(\ref{Gl:Frad in FLD}),
so that the FLD approach in fact does not naturally yield $\Frad$ (nor thus the luminosity).
While it is possible, at a given time, to calculate it from the output quantities
$\rho(r)$, $\Erad(r)$, $\kapR(r)$ by Equation~(\ref{Gl:Frad in FLD}),
this is approximate since in our implicit approach
the diffusion coefficient $\DF$ (Equation~\ref{Gl:DF-Def}) is computed from the quantities
at the current time while the $\nabla\Erad$ factor is written with $\Erad$ at the following timestep.
To avoid any numerical noise and inaccuracy, we therefore store $\DF$ before the FLD step
and combine it with $\Erad$ after, thus reconstructing exactly the flux
as it is effectively computed (albeit otherwise not explicitly).
This was also done in \citetalias{m16Schock} but not reported there.}

We use the same outer boundary condition for $\Erad$ as in \citetalias{m16Schock} but verified
that changing the outer temperature boundary condition (in particular to a Dirichlet boundary condition)
does not, except at the largest radii, affect the temperature or luminosity structure.
\subsection{Microphysics: opacities}

With respect to \citetalias{m16Schock}, we now always use the dust opacities of \citet{semenov03},
taking by default their simplest model for the dust grains, the ``normal homogeneous spheres'' (\texttt{nrm\_h\_s}).
It features more evaporation transitions than in \citet{bl94}, which we used originally.
We modified slightly the opacity routine\footnote{Original version available under \url{http://www2.mpia-hd.mpg.de/home/henning/Dust_opacities/Opacities/opacities.html}.}
to leave the evaporation of the most refractory component to be handled in the main code
and to not add the contribution from the \citet{helling00} gas opacities.

Instead, gas opacity is provided by \citet{malygin14}, and we take their one-temperature Planck mean,  %
as opposed to the 2-$T$ Planck mean (see their equation~(4)).
We found that it is crucial for numerical stability, especially towards higher constant $\mu$ and $\gamma$,
to evaluate the single-temperature $\kapP$
at the radiation temperature $\Trad \equiv (\Erad/a)^{1/4}$ and not at the gas temperature $\TGas$.
Fortunately, it is also justified.
Indeed, looking at $\kapP(\Trad,\TGas)$ for fixed densities, one generally incurs a smaller mistake
when using $\kapP\approx\kapP(\Trad,\Trad)$ than with $\kapP\approx\kapP(\TGas,\TGas)$.
We also evaluate the dust opacities at $\Trad$ but note that this makes barely a difference
since we will find that the radiation and gas are always well coupled ($\TGas=\Trad$)
in the region where the dust opacity matters ($\TGas\lesssim1500$~K).

We remind that the Planck opacity is relevant for the coupling between dust/gas and radiation,
whereas the Rosseland mean determines whether the radiation can stream freely
or has to diffuse. This is discussed in more detail in Section~\ref{Disk:kappa}.

\begin{figure*}
 \epsscale{1.1}
 \plottwo{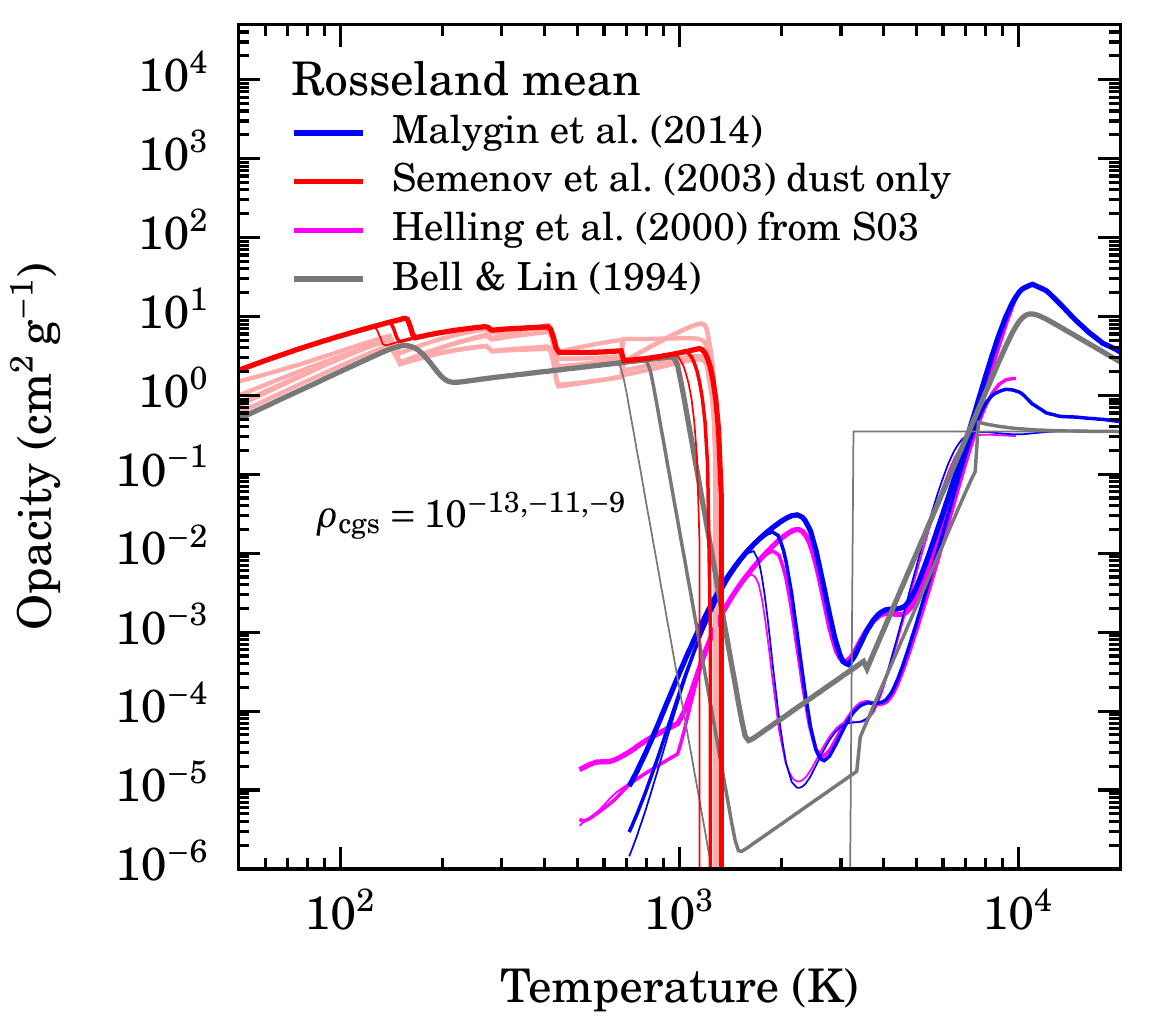}{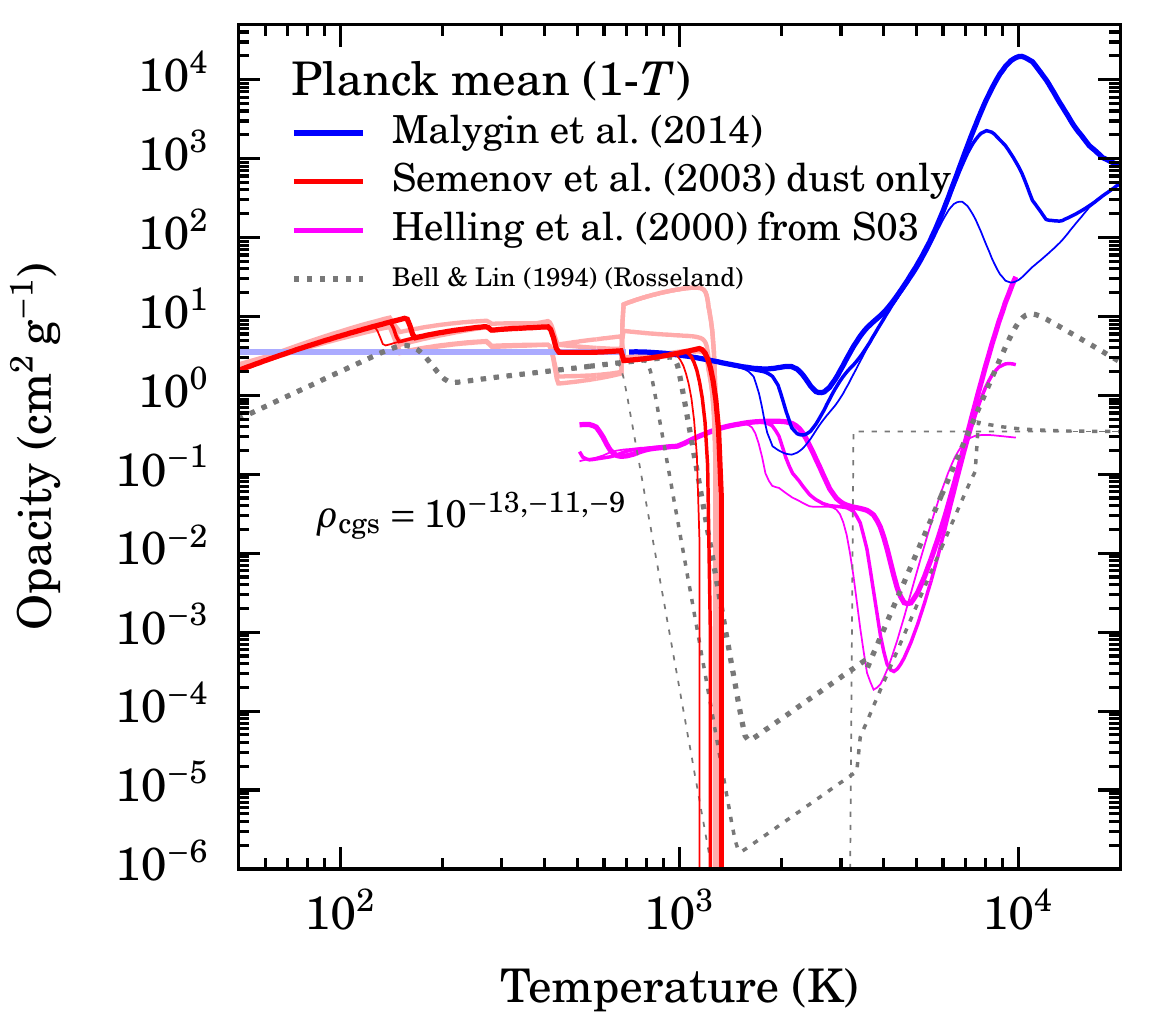}
\caption{
Rosseland ($\kapR$) and Planck (one-temperature; $\kapP$) mean opacities (left and right panels respectively),
from
\protect\citet{malygin14} for the gas
and
\protect\citet{semenov03} %
for the dust.
The total opacity is taken as the sum of the gas and dust contributions.
Three densities are shown: $\rho=10^{-13,-11,-9}$~g\,cm$^{-3}$ (thin to thick lines).
The \protect\citet{malygin14} opacities are kept constant below the table limit of $T=700$~K
(pale blue $\rho$-independent lines in the left panel).
\nnBegut{Since} \protect\citet{bl94} do not provide $\kapP$, their $\kapR$ are displayed for comparison.
Their curves reach down to $\kappa=10^{-7}$--10$^{-4}$~cm$^2$\,g$^{-1}$,
\protect\citepalias{m16Schock},
roughly four orders of magnitude smaller than the \citet{malygin14} Planck values.
We also display the \protect\citet{helling00} $\kapP$ opacities,
which are also too low \protect\citep{malygin14}.
For the \protect\citet{semenov03} opacities, we show their ``nrm.h.s'' model (thick red lines).
The other available ``homogeneous'' and ``composite'' models (thin pale red lines) are also shown
at $\rho=10^{-11}$~g\,cm$^{-3}$,
with the curve sticking out mostly in the Planck mean being the five-layer composite model ``nrm.c.5''.
}
\label{Abb:kappaSchnitt}
\end{figure*}

The Planck opacity $\kapP$ is shown in Figure~\ref{Abb:kappaSchnitt}
and compared to the Rosseland mean $\kapR$.
In the low-temperature, dust-dominated regime $\kapP$ is similar to $\kapR$
with $\kapP\sim\kapR\sim1$--10~cm$^2$\,g$^{-1}$.
For comparison, the different models of \citet{semenov03} are shown,
with the exception of the porous models.
Indeed, fluffy dust aggregates are not expected in protoplanetary discs
due to compactification from collisions, as seems to be borne out by observations
\citep[][and references therein]{cuzzi14,kataoka14}.

While the Rosseland mean opacity past dust evaporation (near $T\approx1500$~K)
drops to $\kapR\approx10^{-2}$~cm$^2$\,g$^{-1}$ (see figure~1 of \citetalias{m16Schock}),
the Planck mean does not drop much below $\kapP\approx1$~cm$^2$\,g$^{-1}$
and even increases (between a few thousand to $10^4$~K) to $\kapP=10^2$--10$^4$~cm$^2$\,g$^{-1}$,
depending on the density.

Notice that the \citet{bl94} Rosseland mean opacities are three to six orders of magnitude~(!) lower
than the Plank average above the dust destruction temperature.
This implies that studies using non-equilibrium radiation transport with
the \citet{bl94} Rosseland mean
as their Plank opacity are effectively assuming much less coupling between the opacity carrier
(dust or gas) and the radiation.
As \citet{malygin14} found out, a similar word of caution applies to the gas opacities of \citet{helling00},
which are included in \citet{semenov03}.
Whether using these low opacities would actually lead to a strong disequilibrium between the matter and radiation
will however depend also on the local density and velocity, as briefly discussed in Section~\ref{Disk:kappa}.
\section{Grid of simulations: results and analysis}
 \label{Theil:Gitter+Analyse}

We now present and discuss results for a grid of simulations
for the macrophysical parameters
\begin{align}
 10^{-3} \leqslant &~\MPunkt  \leqslant 10^{-2}~\ME\,\mathrm{yr}^{-1}\notag\\
     1.3 \leqslant &~\MP      \leqslant 10~\MJ\notag\\
       1.6\lesssim &~\RP   \lesssim 3~\RJ.\notag
\end{align}
We consider 
a mixture of molecular hydrogen and helium
with helium mass fraction $Y=0.25$, yielding $\mu=2.29$ and $\gamma=1.44$.
At high temperatures, the hydrogen should dissociate \citep{szulmorda16}
but we defer simulations taking this into account to
another paper in this series (Marleau et al., in prep.).

\begin{figure*}
 \epsscale{1.1}
 \plotone{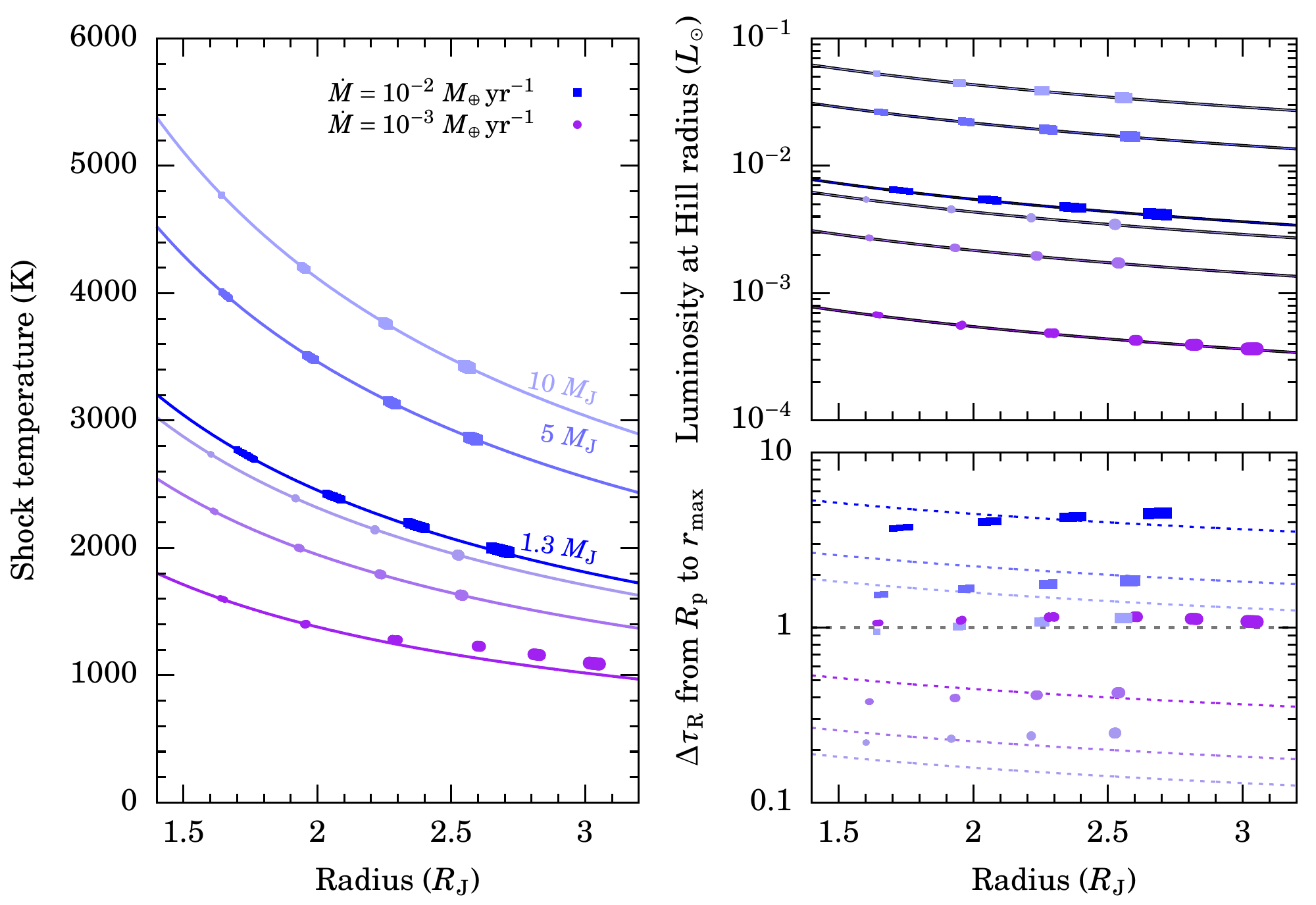}
\caption{
Shock properties in the perfect-gas case with $(\mu=2.29,\gamma=1.44)$ and with tabulated opacities \citep{malygin14,semenov03}
for a grid of
accretion rates $\MPunkt=10^{-3}$--$10^{-2}~\ME$\,yr$^{-1}$ (colour and symbol shape);  %
masses $\MP=1.3$, 5, 10~$\MJ$ (line hue),
and planet radii, i.e., shock positions $\RP\approx1.5$--3~$\RJ$.
The symbol size scales with the position of the inner edge of the respective simulation box $\rmin=1.6$--2.9~$\RJ$. %
Shown are %
\textit{(left column)} the shock temperature $\TSchock$; %
\textit{(top right)} the luminosity at $\rmax$, which is roughly the luminosity at the accretion radius $L(\rmax)\approx L(\frac{1}{3}\RHill)$; and
\textit{(bottom right)} the Rosseland optical depth $\Delta\tauR$ from $\RP$ to $\rmax$.
Note that $L(\RAkk)$ depends on $(\mu,\gamma)$ (see Equations~(\ref{Gl:dLdr mit gamma}) and~(\ref{Gl:Delta L bis RAkk approx})).
We compare in the temperature panel to Equation~(\ref{Gl:TSchock_rauh b}), \nnBegut{i.e., with $\etaklassisch=\Delta\fred=1$;}
in the luminosity panel to Equation~(\ref{Gl:LAkk ab unendlich});
and in the $\Delta\tauR$ panel to the rough estimate Equation~(\ref{Gl:DeltaUngef}) with $\kapR=1$~cm$^2$\,g$^{-1}$,
the grey long-dashed line highlighting $\Delta\tauR\sim1$.
}
\label{Abb:Gitter}
\end{figure*}

The results are summarised in Figures~\ref{Abb:Gitter}, \ref{Abb:Gitter-etaphys}, and~\ref{Abb:Gitter Joseph}
and discussed in the following:
Figure~\ref{Abb:Gitter} shows, as a function of the macrophysical parameters,
the resulting shock temperature (Section~\ref{Theil:T}),
the luminosity at roughly the Hill radius (Section~\ref{Theil:L bei Hill-Sphaere}),
and the optical depth to the Hill radius (Section~\ref{Theil:Delta tau bis rmax}),
whereas Figure~\ref{Abb:Gitter-etaphys} shows the global physical efficiency
as a function of the pre-shock Mach number (Section~\ref{Theil:eta(Mach)}).
Finally, in Figure~\ref{Abb:Gitter Joseph} we display, as a function of the macrophysical parameters,
the efficiency (Section~\ref{Theil:eta(MPkt,MP,RP)})
and the post-shock entropy (Section~\ref{Theil:Spost}).

Note that the grid is irregular in shock position because we considered the same $\rmin$ values
for all masses and accretion rates but the shock moves at different rates $\dd\rSchock/\dd t$; %
over the course of $2\times10^7$~s, which we use as the maximal simulation time
since it is more than enough for the profiles to reach a quasi-steady state,
the ranges of radial positions which the shock covers often do not overlap
between different $(\MPunkt,\MP,\rmin)$.
We added several simulations at higher $\rmin\leqslant2.9~\RJ$
for $(\MPunkt=10^{-3}~\ME\,\mathrm{yr}^{-1},\MP=1.3~\MJ)$.

\subsection{Shock temperature}
 \label{Theil:T}

For our choice of macrophysical parameters, the shock temperature is always above 1000~K.
In Figure~\ref{Abb:Gitter}, simulations with a given $(\MPunkt,\MP)$
but different $\rmin$ (symbol size) %
are seen to lead to the same shock temperature at a given radius $\rSchock=\RP$.
This confirms that, even though the post-shock density structures differ,
the shock properties do not depend on our placement of the inner boundary,
thus supporting the robustness of our results. We have also verified that varying
other numerical parameters such as the resolution or the tolerance in the FLD solver step (Kuiper et al., in prep.)
does not modify the simulation outcomes.

An interesting result is that in all cases, the post-shock gas and radiation (not shown) are in equilibrium,
with $\TGas=\Trad$ to better than 1\,\%, so that one can speak of a single temperature.
The same applies to the pre-shock temperature.
(For other choices of $\gamma$ the equality
is less strict but still within a few percent.)
In Figure~\ref{Abb:Gitter}a, the values obtained are compared to the analytical shock estimate
from equation~(27) of \citetalias{m16Schock},
\begin{subequations}
\label{Gl:TSchock_rauh}
\begin{align}
  \sigSB \TSchock^4 &= \frac{\etaklassisch}{4\Delta\fred} \frac{\rhoFf \vSch^3}{2} \label{Gl:TSchock_rauh a}\\
                    &\approx \frac{G\MP\MPunkt}{16\pi\RP^3}, \label{Gl:TSchock_rauh b} %
\end{align}
\end{subequations}
where $\etaklassisch\equiv\Delta\Frad/(0.5\rho v^3)\approx \LAkk/(G\MP\MPunkt/\RP)$ is the normalised jump in luminosity
and $\Delta\fred\equiv\fred(\rSchock^+)-\fred(\rSchock^-)$ the jump in $\fred$ at the shock.
This equation will be revisited in Section~\ref{Theil:TSchock analytisch}.  %
The free-fall density $\rhoFf=\MPunkt/(4\pi r^2\vFf)$ is evaluated ahead of the shock,
with $\vFf=\sqrt{2G\MP/r}$ the approximate free-fall velocity \citepalias{m16Schock}.
The reduced flux
\begin{equation}
 \label{Gl:fred-Def}
 \fred \equiv \Frad/c\Erad
\end{equation}
is also termed the ``streaming factor'' \citep{kley89a}
because it indicates to what extent the radiation is freely streaming ($\fred\rightarrow1$)
or diffusing ($\fred\rightarrow0$).
In usual shock terminology, free-streaming regions are called ``transmissive'' (see figure~8 of \citealt{vaytetgonz13}; \citealp{drake06}).
(Recall in passing that Equation~(\ref{Gl:TSchock_rauh a}) is valid for $\Lint\gtrless\LAkk$.)
The simulations have pre-shock Mach numbers $\Mach\approx7$--35 (Figure~\ref{Abb:Gitter-etaphys}) and therefore $\etaklassisch=1$
(used in going from Equation~(\ref{Gl:TSchock_rauh a}) to~(\ref{Gl:TSchock_rauh b}))
since this is above $\Mach\approx2.5$ (see the thin line in Figure~\ref{Abb:Gitter-etaphys}; \citetalias{m16Schock}).
Also, we find that $\Delta\fred\approx1$ for a large part of the parameter space,
i.e., the downstream regions
are in the diffusion limit with $\fred(\rSchock^-)\approx0.03$--0.05 (equivalently, a flux limiter $\lambda\approx1/3$),
while the pre-shock region
is in the free-streaming regime, with $\fred(\rSchock^+)\approx1$.
Thus the shock is a ``thick--thin'' shock in the classification of \citet{drake06}.

As a consequence of this, Equation~(\ref{Gl:TSchock_rauh b}) holds very accurately for almost
all simulations; we are almost always in the limit of equation~(28a) of \citetalias{m16Schock},
discussed as Equation~(\ref{Gl:TSchockff}) below.
There is an exception to this,
namely for the lowest mass (1.3~$\MJ$) at a low accretion rate ($10^{-3}~\ME\,\mathrm{yr}^{-1}$)
and towards larger radii ($\RP\gtrsim2.5~\RJ$).
In this case, the post-shock temperature is higher than predicted from Equation~(\ref{Gl:TSchock_rauh b})
by $\Delta T\approx50$~K.  %
This is because $\fred$ is lower ahead of the shock, with e.g.\ $\fred^+\equiv\fred(\rSchock^+)=0.65$
for the $\TSchock\approx1100$~K case.
Indeed, since $\Frad\propto\fred\Trad^4$ and we find $\TGas=\Trad$,
a lower $\fred$ at a location requires a higher temperature for the same radiation flux
(equal to the kinetic energy flux) to flow through that location.
As discussed in \citetalias{m16Schock},  %
this smaller $\fred$ can also be pictured as a smaller effective speed of light,
so that $\Erad$ must increase so as to have the same $\Frad = \ceff \Erad$.
We return to these points in Section~\ref{Theil:TSchock analytisch}
and show that also another effect is at play.
Note however that this is the gas temperature but not the \textit{effective} temperature,
with only the latter setting the spectral shape.

\subsection{Luminosity at the Hill sphere}
 \label{Theil:L bei Hill-Sphaere}
Figure~\ref{Abb:Gitter}b shows the luminosity at the outer edge of the grid,
which is very nearly equal to the luminosity at the Hill sphere,
as will be shown in Section~\ref{Theil:L(r; ZG)}.
The Mach numbers we find here are all $\Mach>2.5$, so that essentially the entire kinetic energy
is converted to radiation (see pale grey curve in Figure~\ref{Abb:Gitter-etaphys}).
Since for the specific choice of $(\mu=2.29, \gamma=1.44)$, the decrease in $L$
from the shock to $\RHill$ is insignificant (see Section~\ref{Theil:L,eta von ZG}),
the entire kinetic energy is \nnBegut{transformed into radiation},
according to the usual expression for the shock luminosity
\begin{equation}
 \label{Gl:LAkk ab unendlich}
 \LAkk=\frac{G\MP\MPunkt}{\RP}.
\end{equation}
This equation is shown as solid lines and seen to match very well
(note that the symbols are almost smaller than the line), even though we neglected
the finite ``accretion radius'' $\RAkk$.
This radius is however much larger
than the planet radius since we are considering the detached phase during giant planet formation \citep{morda12_I}.

We also produced a grid with $(\mu=1.1,\gamma=1.1)$ (not shown),
which increases the contrast with the present situation.  %
One example will be discussed in Section~\ref{Theil:L,eta von ZG} below.
Over the grid, though,  %
the shock temperatures were the same, as one would expect from Equation~(\ref{Gl:TSchock_rauh})
since they do not depend on the EOS explicitly.
The situation could be different
for an ideal but non-perfect EOS, however, due to the potential sink of energy (dissociation and ionization).
As for the luminosity at the Hill radius, it was different from the $\mu=1.23$ case
and was lower by at most $\approx10\,\%$ compared to the immediate shock upstream luminosity.
The luminosity profile is discussed in Section~\ref{Theil:L,eta von ZG}.

\subsection{Optical depth of the infalling gas}
 \label{Theil:Delta tau bis rmax}

Figure~\ref{Abb:Gitter}c displays the Rosseland mean optical depth from the shock
to the outer edge of the computational domain, $\Delta\tauR=-\int_{\rmax}^{\rSchock}\rho\kapR\,\dd r$.
Especially for the lower masses, the contribution from the outer layers
is actually significant despite their low density,
and $\Delta\tauR$ does depend on the choice of the domain size (here, $\rmax=0.7\RAkk$ as in \citetalias{m16Schock}).
We find that \nnBegut{$\Delta\tauR\approx0.2$--5}, increasing with accretion rate and decreasing with mass.
This is qualitatively as expected from equation~(24) of \citetalias{m16Schock},
\begin{align}
 \Delta\tauR \sim 3 \left(\frac{\kapR}{1~\mbox{cm$^2$\,g$^{-1}$}}\right) & \left(\frac{\MPunkt}{10^{-2}~\ME\,{\rm yr}^{-1}}\right)\notag\\
                      &\times\sqrt{ \left(\frac{1~\MJ}{\MP}\right) \left(\frac{2~\RJ}{\rSchock}\right) },\label{Gl:DeltaUngef}
\end{align}
shown as dotted lines in Figure~\ref{Abb:Gitter}c.
The agreement with the nominal models is not too rough (to at worse 1~dex)
considering that we took a constant $\kapR=1$~cm$^2$\,g$^{-1}$ in Equation~(\ref{Gl:DeltaUngef})
in this estimate for all simulations.
Looking at the radial profiles, the maximum value of the actual $\kapR(r)$
is near~7 (3)~cm$^2$\,g$^{-1}$ for $\MPunkt=10^{-3}$ ($10^{-2}$)~$\ME$\,yr$^{-1}$.
However, contrary to what was explained in \citetalias{m16Schock},  %
what is relevant is the \textit{immediate} pre-shock opacity, as we shall see in Section~\ref{Theil:AnalytischT}.

The significance of Figure~\ref{Abb:Gitter}c can be appreciated only in conjunction with panels~(a) and~(b).
It shows that the total optical depth, at least up to moderate depths of $\Delta\tauR\sim10$,
does not set the shock temperature nor the shock luminosity
(which is nearly identical to the Hill-radius luminosity in the present case).
In fact, the deviations of $\TSchock$ from Equation~(\ref{Gl:TSchock_rauh}) do not occur at the highest optical depths.
(What causes these deviations is explained in Section~\ref{Theil:AnalytischT}.)
Also, when, as is often the case, the layers between the $\Delta\tauR\sim1$ surface
and the shock are not sufficiently diffusive ($\fred\lesssim0.1$),
it does not hold that $\TSchock^4 \approx \frac{3}{4}(\Delta\tauR(r) +\frac{2}{3}) \Teff^4$,  %
where $\Teff$ is the temperature at the radius where $\Delta\tauR=2/3$, the photosphere,
and where $\Delta\tauR(r)=-\int_{\rmax}^{r}\rho\kapR\,\dd r'$ is the optical depth
measured from the outer radius inwards.
The temperature rise from the photosphere to the shock is much higher because the optical depth increases more slowly
when the radiation is less diffusive, i.e., when $\dd\tau/\dd r=\kapR\rho$ is small.

Finally,
we note that there is nothing special about the $\Delta\tauR\sim1$ surface.
As Figure~\ref{Abb:Gitter}c shows, most simulations at $\MPunkt=10^{-3}~\ME$\,yr$^{-1}$ remain optically thin or barely optically thick
above the shock, yet their temperature structure (not shown) is qualitatively identical to cases with an optically thick
accretion flow.
In any case, $\Delta\tauR$ is not well defined due to its dependence on the outer integration radius.
We come back to considerations of shock temperature in Section~\ref{Theil:TSchock analytisch}.
\subsection{Shock efficiency against Mach number}
  \label{Theil:eta(Mach)}
When calculating a planet structure for given $(\MPunkt,\MP,\RP)$, only the post-shock $(\Ppost,\TSchock)$ point
is used in an approach equivalent to the one used in \mesa\ by \citet{berardo17} and \citet{berardocumming17}.
 This yields the density $\rho$; from this and $\MPunkt$, the boundary condition on the velocity $v$
 and thus the luminosity profile $L(r)$ in the post-shock region follow \citep{berardo17}.
When however following only the global energetics (see the review in sect.~2.1 of \citealp{berardo17}),
we argued in \citetalias{m16Schock} that one should use $\etaphys$
and not $\etaklassisch$.

\begin{figure*}
 \epsscale{.9}
 \plotone{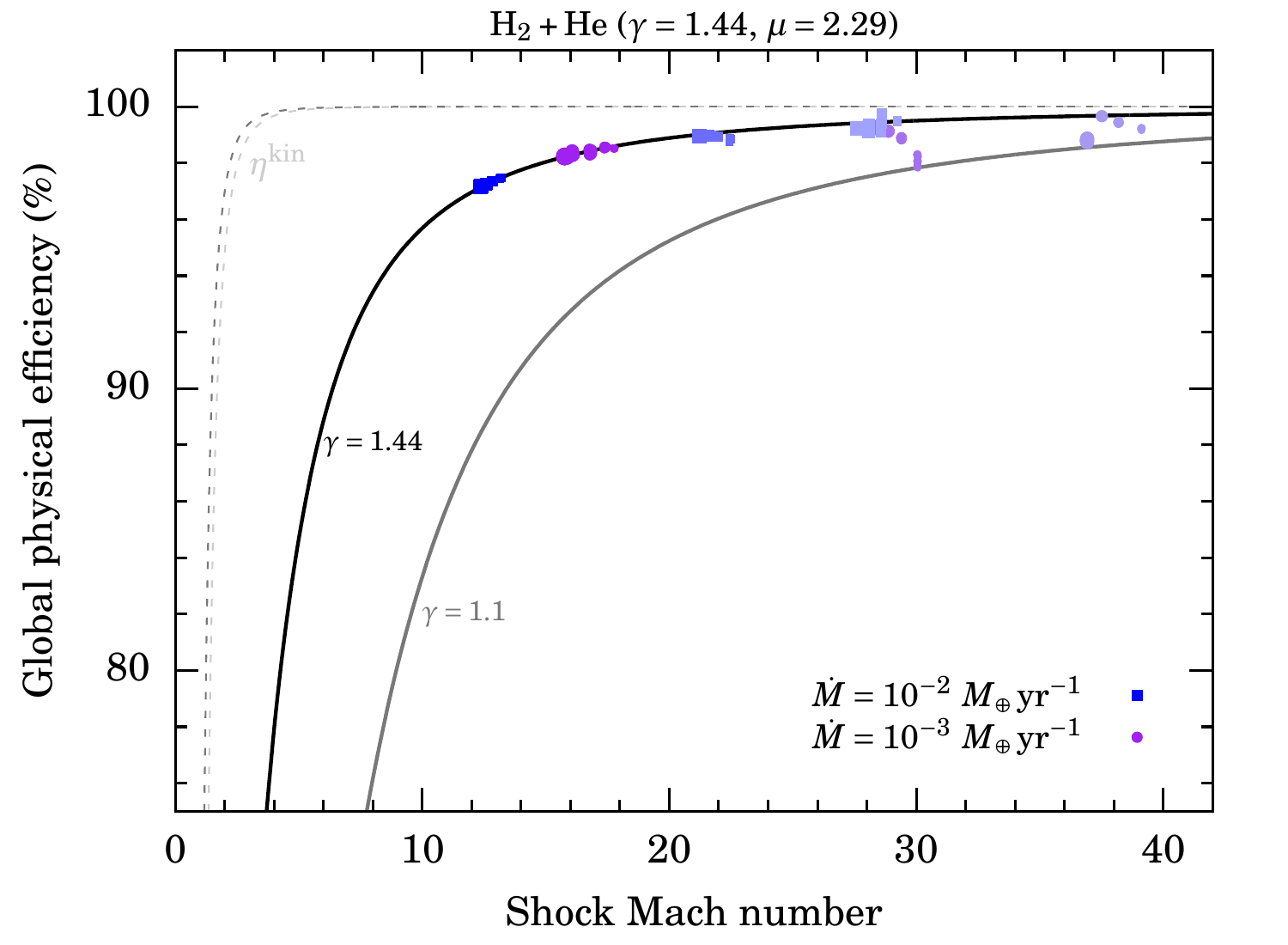}
\caption{Global physical shock efficiency $\etaphys$ (Equation~(\ref{Gl:etaphys_DeltaLE})) against the pre-shock Mach number $\Mach$
in the perfect-gas case with $(\mu=2.29,\gamma=1.44)$ and with tabulated opacities \citep{malygin14,semenov03}
(grid of accretion rates, masses, and radii of Figure~\ref{Abb:Gitter}; see colour, hue, and shape meaning there).
We compare to Equation~(\ref{Gl:etaisoth_phys}) with $\gamma=1.44$ (black) and $\gamma=1.1$ for reference (grey),
and also show $\etaklassisch$ (Equation~(\ref{Gl:etaisoth_klass}) for these two $\gamma$ values (thin dashed grey lines).
\nnBegut{There} is some noise in some simulations but this is only cosmetic.
}
\label{Abb:Gitter-etaphys}
\end{figure*}

We show in Figure~\ref{Abb:Gitter-etaphys} the ``global physical efficiency'' $\etaphys$ of the shock
against the pre-shock Mach number.
The quantity $\etaphys$ is measured as \citepalias[][their equation~(18)]{m16Schock}
\begin{equation}
\label{Gl:etaphys_DeltaLE}
 \etaphys \equiv \frac{ \EPkt(\rmax) - \EPkt(\rSchock^-) }{ \EPkt(\rmax) },
\end{equation}
where $\rSchock^-$ is immediately downstream of the shock.
The
material-energy flow rate is defined as
\begin{equation}
\label{Gl:EPkt(r)}
 \EPkt(r) \equiv -|\MPunkt|\left[\ekin(r) + h(r) + \Delta\Phi(r,\rSchock)\right],  %
\end{equation}
where $\ekin=\frac{1}{2}v^2$, $\eint$, and $h=\eint+P/\rho$ are respectively the
kinetic energy,
internal energy density, and the enthalpy per unit mass, and $\Phi$ is the external potential.
The $\Delta\Phi$ term in Equation~(\ref{Gl:EPkt(r)}) accounts for the work done by the potential
on the gas down to the shock, with the potential difference from $r_0$ to $r$ given by
\begin{equation}
 \label{Gl:DeltaPhi}
 \Delta\Phi(r,r_0) = -G\MP\left(\frac{1}{r} - \frac{1}{r_0}\right).
\end{equation}
We found that for all simulations here, using the tenth cell below the shock
as the post-shock location was a robust prescription.
The shock itself was identified by the $\dd v/\dd r<0$ and $\Delta P/\min(P)>5$ criterion
in appendix~B of \citet{mignone12}.

The efficiency $\etaphys$ measured from Equation~(\ref{Gl:etaphys_DeltaLE}) is compared
in Figure~\ref{Abb:Gitter-etaphys} as a function of the Mach number to the analytical result
in the isothermal limit \citepalias[][their equation~(36)]{m16Schock},
\begin{subequations}
\label{Gl:etaisoth_phys}  %
\begin{align}
 \etaphys_{\textrm{isoth}} &= \etaklassisch_{\textrm{isoth}} \times \left(1 + \frac{2}{\gamma-1}\frac{1}{\Mach^2}\right)^{-1}\\
                           &= \frac{\left(\gamma^2\Mach^4-1\right)(\gamma-1)}{\gamma^2\Mach^2\left[(\gamma-1)\Mach^2+2\right]},
\end{align}
\end{subequations}
with the the isothermal ``kinetic efficiency'' given by \citep{commer11}
\begin{equation}
 \label{Gl:etaisoth_klass}
 \etaklassisch_{\textrm{isoth}} =1 - \frac{1}{\gamma^2\Mach^4}
\end{equation}
This definition was derived from energy conservation for the 1-$T$ case but it is also meaningful here
since shocks are isothermal in the radiation temperature and we find that the gas and radiation
are tightly coupled.

As expected, we find essentially perfect agreement between the measured (Equation~(\ref{Gl:etaphys_DeltaLE}))
and the theoretical (Equation~(\ref{Gl:etaisoth_phys})) efficiencies.
This reflects both energy conservation by our radiation--hydrodynamical code
and the isothermality---i.e., supercriticality---of the shock.

We also computed the kinetic efficiency
\begin{equation}
 \etaklassisch \equiv \frac{\Delta \Frad}{\frac{1}{2}\rho_+{v_+}^3},
\end{equation}
where $\Delta \Frad$ is the jump in radiative flux at the shock and $\rho_+$ and $v_+$ the density and velocity upstream of the shock.
This measured $\etaklassisch$ was found
to match Equation~(\ref{Gl:etaisoth_klass}).
As mentioned above, since $\Mach\gtrsim2.5$ and the shock is isothermal, we have $\etaklassisch\approx100$\,\%.
Thus, locally at the shock, the whole incoming kinetic energy is converted to radiation.

\subsection{Shock efficiency against formation parameters}
  \label{Theil:eta(MPkt,MP,RP)}

\begin{figure*}
 \epsscale{1.1}
 \plotone{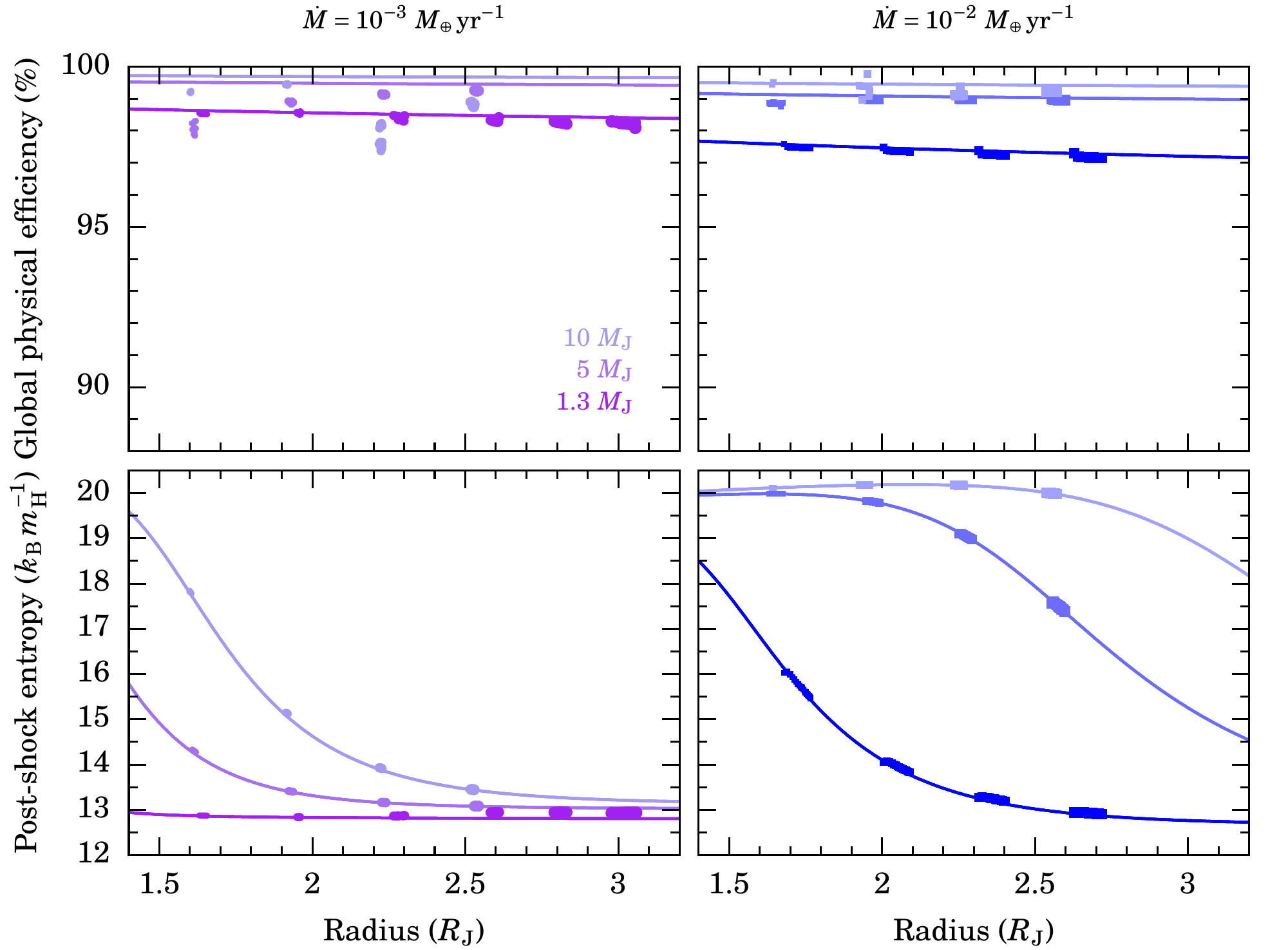}
\caption{
\textit{Top row}:
Global physical efficiency $\etaphys$
for the simulations shown
in Figure~\ref{Abb:Gitter} using the same colour and symbol coding.
The theoretical curves (solid lines) use Equation~(\ref{Gl:etaisoth_phys}) with $\Mach=\vFf/\cs$,
with the sound speed $\cs$ set by the temperature from Equation~(\ref{Gl:TSchock_rauh b}).
Note the small vertical range.
The deviations of the simulation data from the analytical expression at $\MPunkt=10^{-3}~\ME\,\textrm{yr}^{-1}$
are discussed in the text.
\textit{Bottom row}:
Post-shock entropy $s(\MPunkt,\MP,\RP)$ for the same simulations.
The entropy is calculated with
a full, non-perfect EOS \citep{berardo17},
which formally is not self-consistent
with the assumption of a perfect gas for the post-shock temperature $\TSchock$ and pressure $\Pram$
(dots: simulation results; lines: Equations~(\ref{Gl:TSchock_rauh b}) and~(\ref{Gl:Pram})).
However, since $\TSchock$ and $\Pram$ should be independent of the EOS, the entropy values are probably realistic.
The zero-point of the entropy (relevant only when comparing to other work) is discussed in the text.
}
\label{Abb:Gitter Joseph}
\end{figure*}

Figure~\ref{Abb:Gitter Joseph} shows the global physical efficiency $\etaphys$
but now as an explicit function of the (possibly observable, macrophysical) formation parameters $(\MPunkt,\MP,\RP)$.
The efficiencies range from 97~\%\ at high accretion rate to almost 100~\%\ at low $\MPunkt$,
increasing both with decreasing radius and increasing mass.
The precise range depends on the assumed EOS but qualitatively our results should be robust.  %
Typically, in core accretion formation, the radius decreases as the mass grows in the detached phase\footnote{
  This can easily be verified for instance with the results of the Bern planet formation code
  in the ``Evolution'' section
  of the Data Analysis Centre for Exoplanets (DACE) platform
  under \url{https://dace.unige.ch} by plotting $\RP(t)$ against $\MP(t)$.
}.
Since the gas accretion onto a planet usually slows down with time,
at least in the single-embryo-per-disc simulations of \citet{morda12_I},
our simulations clearly suggest that the efficiency $\etaphys$ increases
over time.
Thus, the accretion of the outer layers is associated with less energy recycling \citepalias{m16Schock}
in the accretion flow, and a greater net fraction of the kinetic energy
escapes the system.

However, as discussed in \citet{morda13}, there could be a self-amplifying memory effect:
an $\etaphys$ that is small early after detachment should lead to the accretion
of ``hot'' (high-entropy) material into the planet. If the planet's accretion timescale
is much shorter than its global cooling timescale---i.e., $\tAkk\equiv\MP/\MPunkt \ll \tKH\equiv G\MP^2/(\RP \LP)$,
where $\tAkk$ and $\tKH$ are respectively the accretion and Kelvin--Helmholtz timescales,
with $L$ the luminosity at the radiative--convective boundary \citep{berardo17}---,  %
a large planet radius should ensue.
This large radius would in turn keep $\etaphys$ small,
so that the planet would always remain in a regime where it is accreting
rather high-entropy material. Ultimately, this would lead to a hot (or at least warm) start.
This scenario should be tested with dedicated evolutionary simulations coupling
the shock efficiency self-consistently with the interior structure.
It might also be instructive to compare this with accretion in the context of
star formation, where a similar phenomenon is seen \citep{hosokawa09,hosokawa13,kuiper13b}.
Care should be taken to distinguish, in the analysis, the \textit{local} accretion and cooling times
of the material immediately below the accretion shock from the global ones $\tAkk$ and $\tKH$ (i.e., for the planet as a whole).

\subsection{Post-shock entropy}
 \label{Theil:Spost}

Our detailed calculations of the accretion shock are meant to serve as outer boundary conditions
for calculating the structure of accreting planets as in \citet{morda12_I,berardo17,berardocumming17,cumming18}.
Since these planet structure calculations always use a full EOS including chemical reactions (dissociation and ionisation),
we use this to compute the entropy corresponding to the shock temperatures and pressures we obtained above.
This is formally
\textit{not} self-consistent given that we assumed a perfect EOS
for the shock simulations.
However, since $\TSchock$ should be independent of the EOS,
as we argue in Section~\ref{Theil:L,eta von ZG},
the entropy values are possibly realistic.
In any event they
will serve as a comparison point for simulations
with a realistic EOS (Marleau et al., in prep.).

To calculate the entropy,
we can use the ideal-gas form $P=\rho/(\mu\mH)\kB T$ (but with variable $\mu$)
since degeneracy \nnBegut{starts}  %
being relevant only at a conservative limit of $\rho\sim10^{-2}$~g\,cm$^{-3}$
(cf.\ figure~1 of \citetalias{scvh}), several orders of magnitude above
even the highest post-shock densities $\rhoNachSch\propto\rhoVorSch\Mach^2$,
with $\rhoVorSch\sim10^{-13}$--$10^{-10}$~g\,cm$^{-3}$ (see figure~4 of \citetalias{m16Schock})
and $\Mach\ll100$, so that $\rhoNachSch\ll10^{-6}$~g\,cm$^{-3}$.
We use the Saha equation and Sackur--Tetrode formula as
implemented\footnote{See %
  \url{https://github.com/andrewcumming/gasgiant}.
   }
in
\citet{berardo17}. The entropy zero-point (see appendix~B of \citealt{mc14}
and footnote~2 of \citealt{mordasini17}) is the same as in the published
version of \citetalias{scvh} and \mesa\ \citep{paxton11,paxton13,paxton15,paxton18},
and thus the entropy values reported here are higher by $(1-Y)\ln 2 = 0.52~\kB\,\mH^{-1}$,
where $Y=0.25$ is the helium mass fraction,  %
than the ones in, e.g., \citet{mordasini17}. The zero-point of the entropy
is not physically meaningful
but does have to be taken into account when comparing entropies from different works.

These effective post-shock entropies $\Snach$ are shown in Figure~\ref{Abb:Gitter Joseph}.
For the range of shock positions $\rSchock\approx1.5$--3~$\RJ$
and masses $\MP=1.3$--10~$\MJ$ shown, the entropies are,
for $\MPunkt=10^{-3}~\ME\,\textrm{yr}^{-1}$,
mostly around
$\Snach\approx12$--14 but go up to $\Snach\approx19$
(dropping, also in the following, the usual units of~$\kB\,\mH^{-1}$ for clarity).
For $\MPunkt=10^{-2}~\ME\,\textrm{yr}^{-1}$,
the whole range $\Snach\approx13$--20 is covered.
These values are high compared to the post-formation entropy of planets,
which is at most around 10--14~$\kB\,\mH^{-1}$ according to current, though not definitive,
predictions \citep{berardocumming17,berardo17,morda13,mordasini17}.
However, we caution and emphasize that this post-shock entropy is not the same
as the entropy below the post-shock settling layer; this latter quantity is most likely 
the one most relevant in setting the entropy of the planet as it accretes.

Before calculating the (non-self-consistent) post-shock entropy analytically,
we briefly discuss the ram pressure
\begin{equation}
 \label{Gl:Pram}
 \Pram = \rho \vFf^2 = \frac{\MPunkt}{4\pi\RP^2}\vFf=\frac{\sqrt{2G}}{4\pi} \frac{\MP^{1/2}\MPunkt}{\RP^{5/2}}.
\end{equation}
We inserted the expressions for free-fall from infinity
and find that
Equation~(\ref{Gl:Pram}) holds very well
for all simulations, even for those for which $\Delta\fred\neq1$.
At high shock temperatures, the small pressure build-up ahead
of the shock slows down the gas slightly,
making Equation~(\ref{Gl:Pram}) less accurate by at most 3.5~\%\   %
for the range of parameters shown.
Given the relatively weak (logarithmic, with a small pre-factor)
dependence of the entropy $s(P,T)$ on $P$, outside of dissociation or ionisation regions,
this will not be an important source of inaccuracy.
The ram pressure varies from $\Pram\approx10^{-4}$~bar %
to $0.2$~bar %
for the range of parameters discussed here.

We compare the post-shock entropies $\Snach$ in Figure~\ref{Abb:Gitter Joseph}
based on the actual $T$ to $\Snach$ %
using as input %
$T=\TSchock$ from Equation~(\ref{Gl:TSchock_rauh b}), i.e., taking $\etaklassisch=\Delta\fred=1$
for all simulations.
The match is very good, which reflects the overall good match %
of temperature, on which the entropy depends only logarithmically.  %
\section[Analytics of the temperature ahead of and at the shock]%
{Analytics of the temperature\\ ahead of and at the shock}
 \label{Theil:AnalytischT}

As will be seen in Figure~\ref{Abb:Abbgammamu}, the temperature profile upstream
of the shock shows variations which appear related to variations in opacity.
This modifies the temperature at the outer edge of the accretion flow
(i.e., the local nebula temperature)
compared to what a naive extrapolation $T(r)=\TSchock \left(\rSchock/r\right)^{-1/2}$
would predict.
Also, and even more importantly,
the temperature at the shock is obviously a key outcome of our simulations.
The usual expression for the shock temperature,
with our modification of a factor $(1/4)^{1/4}$ (Equation~(\ref{Gl:TSchock_rauh})),
provides in general a very good estimate.
However, we have seen in Figure~\ref{Abb:Gitter} that there can be small deviations.
We now turn to the task of understanding both the temperature profile
in the accretion flow and the shock temperature by analytical means.
We also discuss the link between the reduced flux and the Rosseland opacity.

\subsection{Temperature profile in the accretion flow}
 \label{Gl:T-Profile im Akkfl}
To derive the slope of the temperature throughout the infalling gas,
let us begin with the general relationships
\begin{align}
 ac\Trad^4 &= \frac{L}{4\pi r^2}\frac{1}{\fred} \label{Gl:aT^4=L/r^2fred}\\
 \fred &= R \lambda(R) \label{Gl:fredR},
\end{align}
with $a\Trad^4\equiv\Erad$.
The first equation is nothing but a rewriting of the definition of $\fred$,
and the second follows from Equation~(\ref{Gl:aT^4=L/r^2fred})
and the definition of the radiation quantity $R$, given by
\begin{equation}
 \label{Gl:R-Def}
 R \equiv \frac{1}{\kapR\rho}\left|\frac{\dd \ln\Erad}{\dd r}\right|
\end{equation}
in spherical symmetry. (In this section we use the symbol ``$\dd$'' instead of ``$\partial$''
because of time independence.)
It is equal to the ratio of the photon mean free path
to the ``$\Erad$ scale height'' \citepalias{m16Schock}.
Taking the derivative of Equation~(\ref{Gl:aT^4=L/r^2fred}) yields
\begin{equation}
 \label{Gl:dlnT^4dlnr}
 \frac{\dd\ln \Trad^4}{\dd\ln r} = -2 + \frac{\dd\ln L}{\dd\ln r} - \frac{\dd\ln \fred}{\dd\ln r}.
\end{equation}
The last term, by Equation~(\ref{Gl:fredR}), is
\begin{align}
 \label{Gl:dlnfreddlnr}
 \frac{\dd\ln \fred}{\dd\ln r} &= \frac{\dd\ln R}{\dd\ln r}\left(1+\frac{\dd\ln\lambda}{\dd\ln R}\right).
\end{align}
These expressions are exact and general.
We now proceed to expand the last equation.

By the definition of $R$, the first factor in Equation~(\ref{Gl:dlnfreddlnr})  %
is
\begin{equation}
 \label{Gl:dlnRdlnr}
 \frac{\dd\ln R}{\dd\ln r} = \frac{\dd}{\dd\ln r}\left( \ln\left[\frac{\dd\ln \Trad^4}{\dd\ln r}\right] -\ln\left[\kapR\rho r\right] \right)
\end{equation}
since $\Erad=a\Trad^4$.
Inserting recursively Equation~(\ref{Gl:dlnT^4dlnr}) into Equation~(\ref{Gl:dlnRdlnr})
is likely not fruitful as it generates derivatives of $\fred$ of ever-higher order.
However, it will prove instructive to perform this once.
In full generality, this yields
\begin{align}
 \label{Gl:dlnRdlnr mehr}
 \frac{\dd\ln R}{\dd\ln r} = &\frac{ \dd^2_{(\ln r)^2} \ln L - \dd^2_{(\ln r)^2} \ln\fred
                                 }{ -2 + \dd_{\ln r}\ln L - \dd_{\ln r}\ln\fred }
                             - \frac{\dd\ln\kapR}{\dd\ln r} + \frac{1}{2},
\end{align}
where $\dd^n_{x^n}f(x)\equiv \dd^nf(x)/\dd x^n$.
For this derivation we have made use of the fact that $\rho\propto r^{-3/2}$ in the accretion flow.

The second factor in Equation~(\ref{Gl:dlnfreddlnr}) depends only on the choice
of the flux limiter
and can be computed easily independently of a simulation.
If one takes the flux limiter used in \citet{ensman94},
\begin{equation}
 \label{Gl:Flubeg E94}
 \lambda(R) = \frac{1}{3+R},
\end{equation}
one obtains for the derivative term the simple expression
\begin{equation}
 \label{Gl:1+dlnlambdadlnR E94}
 1+\frac{\dd\ln\lambda}{\dd\ln R} = 3\lambda.
\end{equation}
Note that this flux limiter is actually not physical \citep{lever84}
but that it recovers the correct limits
of free-streaming and diffusion (see \citetalias{m16Schock}).
Using instead the rational approximation of the \citet{lever81} flux limiter,
\begin{equation}
 \lambda(R) = \frac{2+R}{6+3R+R^2},
\end{equation}
which we take by default in this work, the result is %
\begin{equation}
 \label{Gl:1+dlnlambdadlnR LP81}
 1+\frac{\dd\ln\lambda}{\dd\ln R} = 1 - \frac{R^2\left(R+4\right)}{(2+R)(6+3R+R^2)}.
\end{equation}
For $R\ll2$ (the diffusion limit), this function (the right-hand side) is equal to $3\lambda$.
In the other limit of $R\gg2$, it converges to simply $\lambda$; in any case it remains
roughly proportional to the flux limiter $\lambda$, a fact we shall use in the discussion below.

Combining Equations~(\ref{Gl:dlnT^4dlnr}), (\ref{Gl:dlnfreddlnr}), (\ref{Gl:dlnRdlnr mehr}), and~(\ref{Gl:1+dlnlambdadlnR E94}), we obtain
\begin{equation}
\label{Gl:dlnT/dlnr}
 \frac{\dd\ln T}{\dd\ln r} \approx -\frac{1}{2} + \frac{ 3\lambda(R)}{4}  %
                         \left[\frac{ \dd^2_{(\ln r)^2}\ln\fred }{2+\dd_{\ln r}\ln\fred} - \frac{\dd\ln\kapR}{\dd\ln r} + \frac{1}{2} \right]
\end{equation}
in the limit of a radially constant luminosity
(in the sense that $\left|\dd\ln L/\dd\ln r\right|\ll 2$, which is the case here as shown below).
The temperature here was written as $T$ since we find that $\TGas=\Trad$;
if the gas and radiation are not coupled, $T$ should be taken to refer to $\Trad$ only.
We used Equation~(\ref{Gl:1+dlnlambdadlnR E94}) since it makes it explicit
that the second term on the righthand side of Equation~(\ref{Gl:dlnT/dlnr})
is (in general roughly) proportional to the flux limiter $\lambda(R)$.
While, at least in this form, Equation~(\ref{Gl:dlnT/dlnr}) is not predictive
since $\fred(r)$ is needed for its derivatives, it does exhibit the link
between the temperature and the opacity slopes, as we now discuss.

Looking in detail at Equation~(\ref{Gl:dlnT/dlnr}), we see that
the first term on the right-hand side corresponds to the result
that $T\propto r^{-1/2}$ for a radially-constant luminosity.
It seems likely that this term will dominate in the free-streaming
(often termed ``optically-thin'') regime since then $\lambda$ (and thus the second term) goes to zero.

For the temperature slope to differ significantly from $-1/2$, %
it is sufficient for either the opacity slope or the term with derivatives of $\fred$
to be large, and necessary for the radiation to be sufficiently diffusive ($\lambda$ not too small).

The second term of Equation~(\ref{Gl:dlnT/dlnr}) is interesting: it relates the local slope of the opacity
to that of the temperature.
In the free-streaming regime, the flux limiter $\lambda$ goes to zero
and thus also the second term in Equation~(\ref{Gl:dlnT/dlnr} since it is multiplied by $\lambda$).
Therefore, even strong variations of $\kapR$ with radius will only minorly affect
the temperature structure; this is as expected for the limit of infinite
optical mean free path, in which the radiation does not interact
with the opacity carrier. In the other limiting case of diffusion,
$\lambda\approx1/3$ and the opacity variations are important.
The evaporation of dust as the material moves in
leads to a large $\dd\ln\kapR/\dd\ln r$,
and if $\lambda$ is not too small, this will be able
to slow the decrease of the temperature outwards.

Also, it does not seem necessary
that the radiation be 
free streaming ($\fred\rightarrow1$)
in order to have $T\propto r^{-1/2}$, which is usually associated with free streaming.
If $\kapR$ is constant with radius and (as perhaps a consequence) $\fred$ is sufficiently constant,
and $\lambda$ has an intermediate value (e.g., $\lambda\sim0.1$),
the $-1/2$ term in Equation~(\ref{Gl:dlnT/dlnr}) can dominate.

\begin{figure}
 \epsscale{1.2}
 \plotone{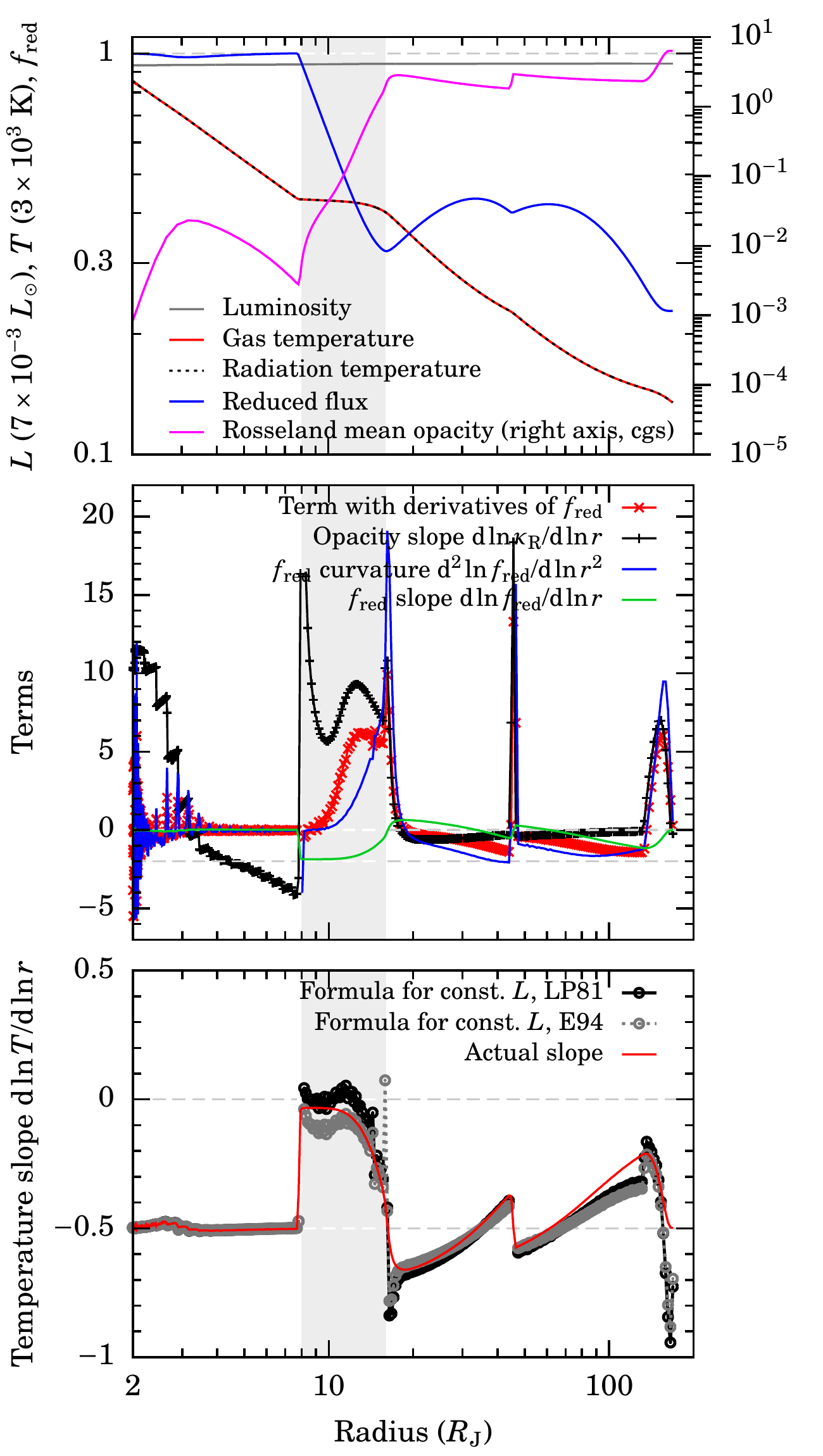}
\caption{
\textit{Top panel:} Normalised luminosity, temperatures (gas and radiation), reduced flux (against the left axis),
and opacity $\kapR$ (against the outer right axis) for the nominal case presented in Figure~\ref{Abb:Abbgammamu}.
Note the logarithmic scale. The region with a temperature flattening ($r=8$--$16~\RJ$) due to dust destruction
is highlighted.
\textit{Middle panel:} First two lines: different terms in the brackets in Equation~(\ref{Gl:dlnT/dlnr})
(see legend). Some of the factors are plotted individually (blue and green lines).
\textit{Bottom panel:} Temperature slope obtained from Equation~(\ref{Gl:dlnT/dlnr}) for two flux limiters
(black and grey lines; in both cases the same simulations are used, with $\lambda$ changing only in the formula).
This is compared to the exact result (red line).
}
\label{Abb:Jakobus}
\end{figure}

We show in the top panel of Figure~\ref{Abb:Jakobus} the luminosity, temperature, reduced flux,
and opacity on logarithmic scales for the nominal case shown in Figure~\ref{Abb:Abbgammamu}.
Indeed, the luminosity is effectively constant radially.

The middle panel of Figure~\ref{Abb:Jakobus} shows the terms on the righthand side of Equation~(\ref{Gl:dlnT/dlnr})
and parts thereof. In the free-streaming region at $r=2$--8~$\RJ$, the first and second derivatives of $\fred$
are nearly zero, and the strong values of the opacity slope do not bring $T$ away from an $r^{-1/2}$ scaling
because the radiation is nearly free-streaming: $\fred\approx1$ (see top panel),
so that $\lambda\approx0$.
In the flat temperature part at $r=8$--16~$\RJ$ (shaded region), the opacity slope term dominates
but the term with the derivatives of $\fred$ is similar in magnitude, with their signed sum dominating
over the $-1/2$ term. The result, along with a lower $\fred$ (i.e., higher $\lambda$),
is a different (namely, almost zero) temperature slope.
At $r>16~\RJ$, the radiation remains somewhat diffusive with $\fred\sim0.3$
but the opacity is nearly constant.
The temperature slope is therefore set by the leading $-1/2$ term as well as the two other terms
in the brackets and the factor of $\approx3\lambda$ (depending on the flux limiter model).
The net result is that the $-1/2$ dominates.

The slope from Equation~(\ref{Gl:dlnT/dlnr}) is shown in the bottom panel of Figure~\ref{Abb:Jakobus}
and compared to the actual value. The agreement is excellent, which mainly confirms
the approximation $\dd L/\dd r\approx0$ since the equation is otherwise exact
for a free-fall density profile.
The choice of the flux limiter barely makes a difference.

Note finally that Equations~(\ref{Gl:R-Def}) and~(\ref{Gl:dlnT^4dlnr}) imply that, when the ratio $L/\fred$ is radially constant,
the radiation quantity $R$ is given by
\begin{equation}
 \label{Gl:R konst L}
 R = \frac{2}{\kapR\rho r}\giltwenn{\mbox{constant $L/\fred$}}.
\end{equation}
This had been derived in \citetalias[][above their equation~(25)]{m16Schock}.

This analysis provides an approximate analytical understanding of the link between
the opacity and the temperature.
What is less clear from this derivation is how much the geometry accurately reflects
the realistic transport of radiation, even within the grey approximation,
and to what extent it depends on the approximations (for instance concerning
the angular distribution of the specific intensity) inherent to moments-based methods
such as FLD.
Nevertheless, it might prove insightful to attempt a similar derivation using M1 radiation transport
(e.g., \citealp{gonz07,hanawa14}).

\subsection{Link between reduced flux and opacity}

Figure~\ref{Abb:Jakobus} suggests that there is a relationship between the reduced flux $\fred$
and the Rosseland mean opacity $\kapR$.
Defining the logarithmic slope
\begin{equation}
 \label{Gl:beta-Def}
 \beta\equiv \frac{\dd\ln (L/\fred)}{\dd\ln r},
\end{equation}
this relationship can be obtained by combining Equations~(\ref{Gl:aT^4=L/r^2fred}--\ref{Gl:R-Def})
to yield
\begin{subequations}
 \label{Gl:fred og kapparhor}
\begin{align}
 \frac{1}{\fred} &= 1 + \frac{3}{\left|2- \beta\right|}\kapR\rho r \label{Gl:fred og kapparhor a}\\
                 &= 1 + \frac{3}{2}\kapR\rho r \giltwenn{\mbox{if constant $L/\fred$}}, \label{Gl:fred og kapparhor b}
\end{align}
\end{subequations}
which holds anywhere in the flow.
This generalises equation~(25) in \citetalias{m16Schock}.
We used  %
$\lambda$ from Equation~(\ref{Gl:Flubeg E94}) but
it is trivial to repeat the derivation for another flux limiter.

Equation~(\ref{Gl:fred og kapparhor})
explains why $\fred$ drops in the dust destruction region
in Figure~\ref{Abb:Jakobus} (grey band there):
the opacity increases such that $\kapR\rho r \gtrsim 1$,
thereby leading to a drop in $\fred$.
Physically, this has the intuitive explanation that the radiation becomes less freely streaming
and starts to diffuse.
In the dust destruction region, $\beta\approx1.87$ (and relatively constant)
due to the outwards decreasing $\fred$.
From Equation~(\ref{Gl:fred og kapparhor a}) this implies that $1/\fred$ will be larger,
i.e., $\fred$ smaller, than predicted by Equation~(\ref{Gl:fred og kapparhor b}),
but the trend is the same.
Quantitiatively, this works well in this example but to obtain the exact value of $\fred$
one would have to use the \citet{lever81} flux limiter since this was chosen for the simulations.
Thus, Equation~(\ref{Gl:fred og kapparhor}) explains the sudden drop of $\fred$ where the opacity suddenly increases
to become important in the sense of $\kapR\rho r\gtrsim1$.

\subsection{Schock temperature} %
 \label{Theil:TSchock analytisch}
\subsubsection{Calculation set-up}
 \label{Theil:Aufsetzen}
 
We now turn to explaining the shock temperatures
found in Figure~\ref{Abb:Gitter}, in particular the points deviating
from Equation~(\ref{Gl:TSchock_rauh b}). We recall that this temperature
more precisely refers to the radiation temperature $\Trad$,
should it ever be found to differ from $\TGas$, which is here however not the case.

Looking at the results of Section~\ref{Theil:T} again, %
there is a tight anticorrelation (not shown) between $\fred$ %
and $\kapR\rho r$ both evaluated immediately upstream of the shock.
Empirically across our grid of models, $\beta(\rSchock^+) \sim-0.1$ typically,
which in absolute value is~$\ll2$.
Therefore, the tight relationship
comes from Equation~(\ref{Gl:fred og kapparhor b}).
Thus while in general, $\beta(\rSchock^+)$
is not known but could perhaps be estimated,
in the following analysis we will restrict ourselves to $\beta(\rSchock^+)\ll2$
and consequently use $R(\rSchock^+)=2/\kapR\rhoFf\RP$ at the shock (Eq.~(\ref{Gl:R konst L})).

With this, Equation~(\ref{Gl:TSchock_rauh a}) can be rewritten as
\begin{subequations}
\label{Gl:TSchock implizit beide}
\begin{align}
 \sigSB \TSchock^4 &= \frac{\Llinks + \etaklassisch\!(\TSchock)\frac{G\MP\MPunkt}{\RP}}{\fred^+\!(\TSchock)16\pi\RP^2} \label{Gl:TSchock implizit}\\
                   &\approx \left(1+\frac{3}{2}\kapR\!(\TSchock)\rhoFf\RP\right) \frac{\Llinks + \frac{G\MP\MPunkt}{\RP}}{16\pi\RP^2}, \label{Gl:TSchock implizit approx}
\end{align}
\end{subequations}
which is an implicit equation for $\TSchock$.
The second line is valid when the resulting $\Mach\gtrsim2.5$ (so that $\etaklassisch\approx1$),
which should be verified \textit{a posteriori}, and was written for the \citet{ensman94} flux limiter
through Equation~(\ref{Gl:fred og kapparhor}).
The downstream luminosity $\Llinks$ is the sum of the luminosity coming from the deep interior
and of the compression luminosity: $\Llinks=\Lint+\LKomp$.
While $\LKomp$ might depend on $\TSchock$ we effectively absorb this dependency
into $\Lint$, treated as a free parameter.
Still, the generalisation $\Llinks\rightarrow\Llinks(\TSchock)$ could be readily made.
With these assumptions, in Equation~(\ref{Gl:TSchock implizit approx}) the shock temperature
enters on the righthand side only through the opacity.

In the 100-\%\ efficiency limit,
the shock temperature that solves Equation~(\ref{Gl:TSchock implizit approx})
thus has the limiting cases
\begin{subequations}
 \label{Gl:TSchockff}
\begin{align}
 \TSchockff \equiv&~\TSchock(\kapR\rho\RP\ll1)\\  %
 =& \ell^{1/4} \left( \frac{G}{16\pi\sigSB} \right)^{1/4} \frac{\MP^{1/4}\MPunkt^{1/4}}{\RP^{3/4}}\\
                   = &~2315~{\rm K}~\ell^{1/4}\left(\frac{\RP}{2~\RJ}\right)^{-3/4} \notag \\
                          &\times\left(\frac{\MPunkt}{10^{-2}~\ME\,{\rm yr}^{-1}}\right)^{1/4} \left(\frac{\MP}{1~\MJ}\right)^{1/4},
\end{align}
\end{subequations}
\begin{subequations}
 \label{Gl:TSchock fuer kapparhor>>1}
\begin{align}
 \TSchockdiff \equiv &~\TSchock(\kapR\rho\RP\gg1)\\
 =&~\ell^{1/4} \left(\frac{3\sqrt{G}}{128\sqrt{2}\pi^2\sigSB}\right)^{1/4}
      \frac{\MPunkt^{1/2}\kapR^{1/4}\MP^{1/8}}{\RP^{7/8}} \label{Gl:TSchock fuer kapparhor>>1 b}\\
                    = &~3220~{\rm K}~\ell^{1/4}\left(\frac{\RP}{2~\RJ}\right)^{-7/8}\left(\frac{\kapR}{1~{\rm cm}^2\,{\rm g}^{-1}}\right)^{1/4} \notag \\
                                   &\times\left(\frac{\MPunkt}{10^{-2}~\ME\,{\rm yr}^{-1}}\right)^{1/2} 
                                     \left(\frac{\MP}{1~\MJ}\right)^{1/8} \label{Gl:TSchock fuer kapparhor>>1 c},
\end{align}
\end{subequations}
with $\ell=1+\Llinks/\LAkkmax$, without needing to assume that $\Llinks/\LAkkmax$ is small. %
We defined $\TSchockff$ as the ``free-streaming shock temperature'' given by Equation~(\ref{Gl:TSchock_rauh b})
but now generalised to include a downstream luminosity $\Llinks$ (we consider $\Llinks=0$ in this work).
Similarly, $\TSchockdiff$ is the ``diffusive shock temperature'', which obtains when the pre-shock material is diffusive.
The maximum accretion luminosity is $\LAkkmax = G\MP\MPunkt/\RP$,
with the actual luminosity at the shock $\LAkk=\etaklassisch\LAkkmax$ \citepalias{m16Schock}.
The opacity $\kapR$ appearing in the expressions for $\TSchockdiff$ is either the constant opacity
or less trivially the value satisfying Equation~(\ref{Gl:TSchock implizit approx}), as detailed below.
Equation~(\ref{Gl:TSchock fuer kapparhor>>1 c}) should replace equation~(28b) of \citetalias{m16Schock}
since the prefactor there was inexact
and the exponent of the $\RP$ factor was missing a minus sign.
Equations~(\ref{Gl:TSchock fuer kapparhor>>1 b}) and~(\ref{Gl:TSchock fuer kapparhor>>1 c}) were
written to leading order in the quantity $\kapR\rhoFf\rSchock$,
in which case they are also independent of the flux limiter.
Note that should $\Llinks$ dominate over $\LAkkmax$ ($\ell\gg1$),
the functional dependence of the shock temperature on the various parameters
would be different from what Equations~(\ref{Gl:TSchockff}) and~(\ref{Gl:TSchock fuer kapparhor>>1}) naively suggest.

As an example,
for the $(\gamma=1.1, \kappa=1$~cm$^2$\,g$^{-1})$ simulation in figure~2 of \citetalias{m16Schock},
Equation~(\ref{Gl:TSchock fuer kapparhor>>1}) predicts $\TSchockff\approx3600$~K,
which is close to the actual shock temperature 3500~K.
This is satisfying, especially given that there $\kapR\rho\RP=2.3$,
which is not entirely $\gg1$,
and that the \citet{lever81} flux limiter had been used and not the \citet{ensman94} one
as for Equations~(\ref{Gl:TSchock fuer kapparhor>>1 b}) and~(\ref{Gl:TSchock fuer kapparhor>>1 c}),
which makes a difference in this transition regime between free streaming and diffusion.

In general, we solve Equation~(\ref{Gl:TSchock implizit approx}) by writing it as
\begin{equation}
 \label{Gl:TSchock implizit approx, fuer Abb}
 \left(\frac{\TSchock}{\TSchockff}\right)^4
 = \begin{cases}
   1 + \frac{3}{2}x & \mbox{\citepalias{ensman94}}\\
   \left(1+\frac{3}{2}x+\frac{3}{2}x^2\right)/\left(1+x\right) & \mbox{\citepalias{lever81}}
\end{cases},
\end{equation}
with
\begin{equation}
 x(\TSchock)\equiv \kapR(\TSchock)\rhoFf\RP.
\end{equation}
The first and second versions of the righthand side of Equation~(\ref{Gl:TSchock implizit approx, fuer Abb}) hold
for the flux limiters of \citet[][\citetalias{ensman94}]{ensman94}
and \citet[][\citetalias{lever81}]{lever81}, respectively.
The equation is solved by root finding,
\nnBegut{simply starting slightly below} $T=\TSchockff$ and stepping up in temperature
\nnBegut{to find an}
intersection of the two sides of Equation~(\ref{Gl:TSchock implizit approx, fuer Abb}).

\subsubsection{Semi-analytical shock temperature solutions}

\begin{figure*}
 \epsscale{1.1}
 \plottwo{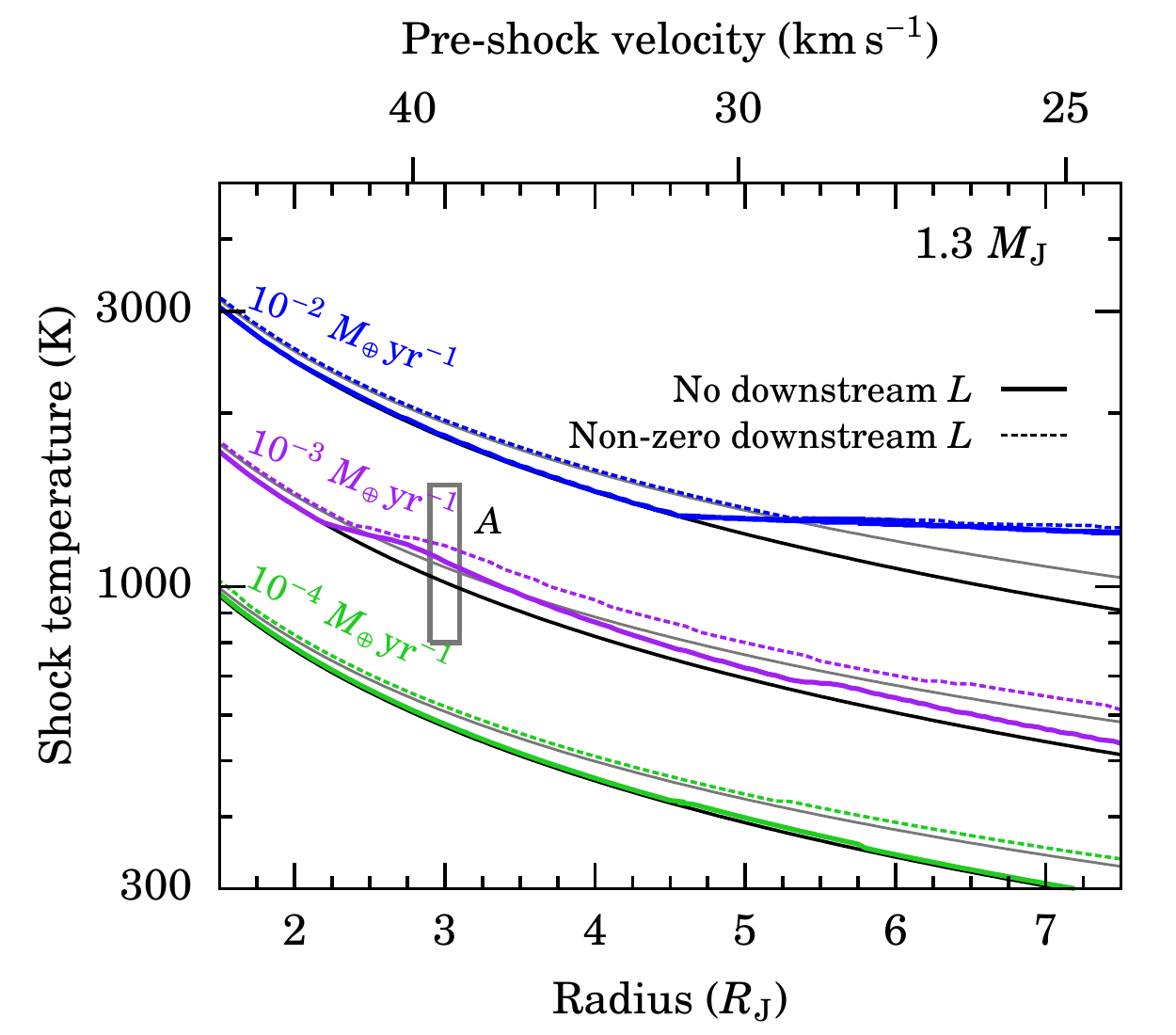}%
{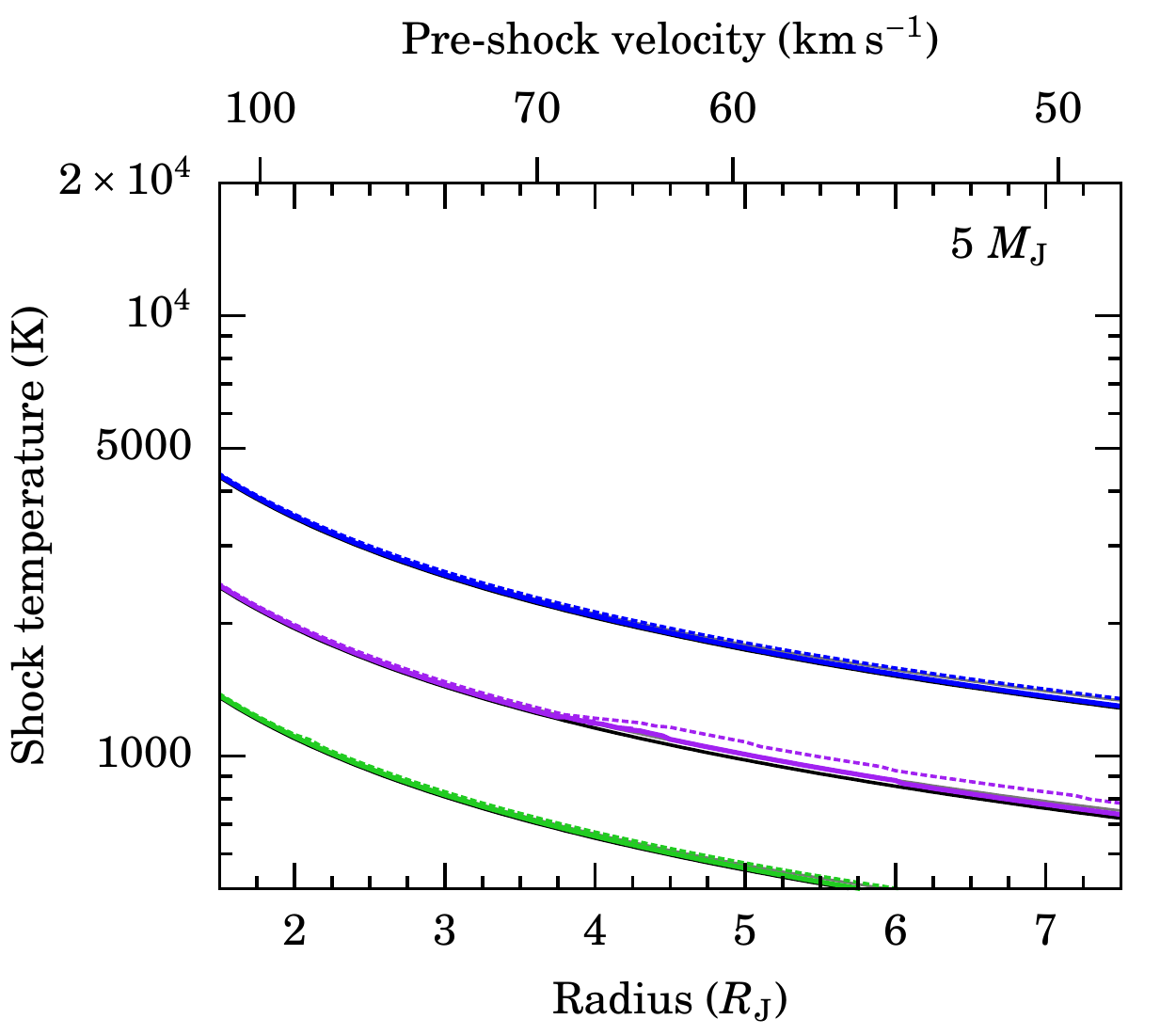}\\
\epsscale{0.55}
 \plotone{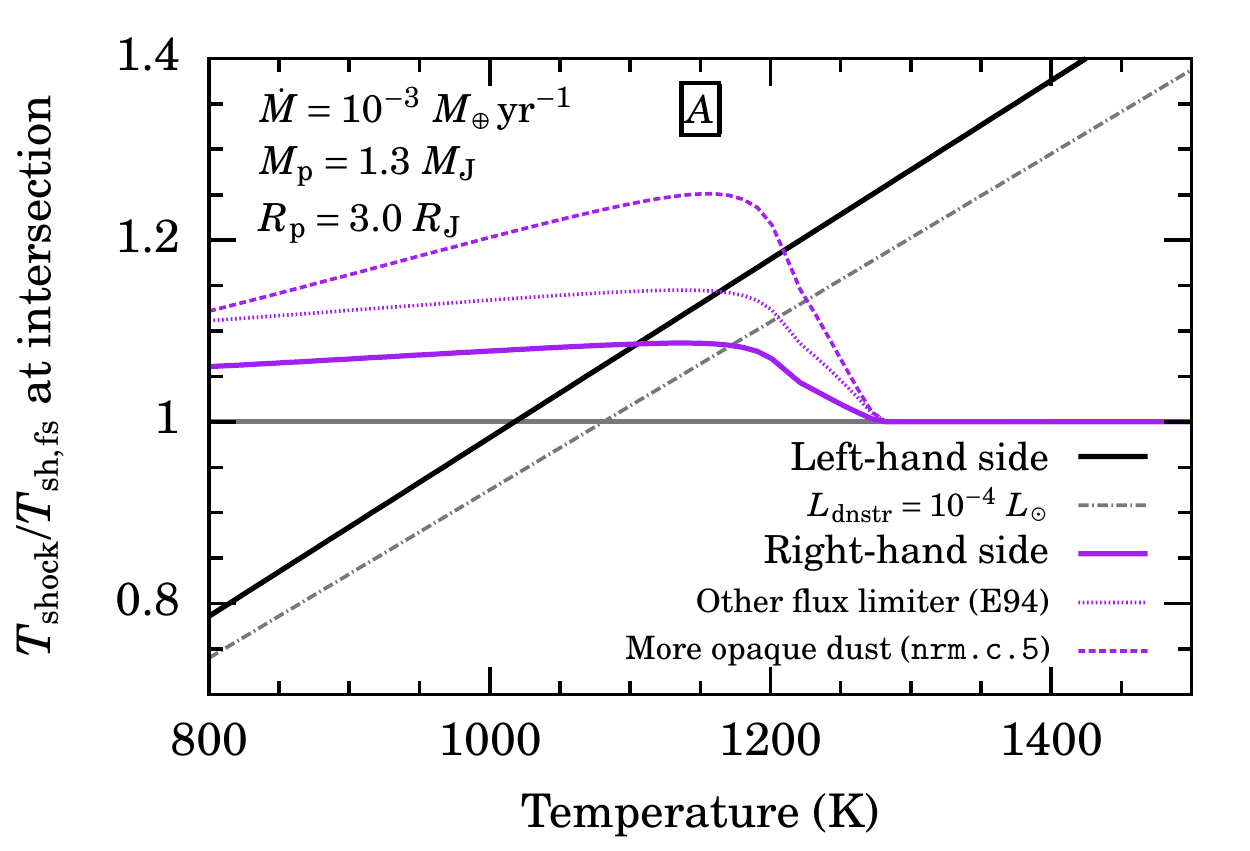}%
\caption{
\textit{Top row:}
Analytical shock temperature $\TSchock$ taking upstream opacity effects into account,
obtained by solving implicitly Equation~(\ref{Gl:TSchock implizit approx, fuer Abb}).
We use accretion rates of $\MPunkt=10^{-4}$ to $10^{-2}~\ME\,\mbox{yr}^{-1}$
(\textit{bottom to top}).
Planet masses of 1.3~$\MJ$ (\textit{left panel}) and 5~$\MJ$ (\textit{right}) are shown.
Both $\Llinks=0$ (\textit{full lines})
and $\log_{10}\left(\Llinks/\LSonne\right)=\log_{10}\left(\MPunkt/\ME\,\mbox{yr}^{-1}\right)-1$ (\textit{dashed lines})
is shown.
The ``free-streaming solution'' (Equation~(\ref{Gl:TSchockff}))
is shown by solid lines (\textit{black and grey, respectively}).
Along the top axes, the corresponding free-fall velocities are indicated.
An example for label~\textit{A} in the left panel is shown in the bottom row.
\textit{Bottom row:}
Graphical solution of Equation~(\ref{Gl:TSchock implizit approx, fuer Abb}) for $\MPunkt=10^{-3}$
when $\MP=1.3~\MJ$ and $\RP=3~\RJ$.
We plot the fourth root of the left- and right-hand sides.
We vary (see legend)
$\Llinks$ (nominal: $\Llinks=0$), the flux limiter (nominal: \citetalias{lever81}),
and the opacity model (nominal: \texttt{nrm.h.s}), which leads to different solutions.
The denominator of the quantity on the $y$ axis, $\TSchockff$, is evaluated at $\Llinks=0$.
The shock temperature is given by the intersection of the left- and right-hand sides.
}
\label{Abb:Timplizit(RP)}
\end{figure*}

In Figure~\ref{Abb:Timplizit(RP)} we display the shock temperature as a function of planet radius
according to Equation~(\ref{Gl:TSchock implizit approx, fuer Abb}).
We consider $\MP=1.3$ and 5~$\MJ$, take $\MPunkt=10^{-4,-3,-2}~\ME\,\mbox{yr}^{-1}$,
and vary $\RP$ from~1 to about 7~$\RJ$, truncating however at lower $\RP$ for the lowest accretion rate.
The downstream luminosity is taken to be zero or smaller than but not entirely negligible compared
to the accretion luminosity, $\log_{10}\left(\Llinks/\LSonne\right)=\log_{10}\left(\MPunkt/\ME\,\mbox{yr}^{-1}\right)-1$
for definiteness.
For reference, the corresponding pre-shock (free-fall) velocity is indicated on the top axis.
The solution is also compared to the free-streaming shock temperature $\TSchockff$ (Equation~(\ref{Gl:TSchockff})).

Figure~\ref{Abb:Timplizit(RP)} reveals that opacity effects can alter the shock temperature
only for a relatively small range of parameters.
Specifically, for $\MPunkt\sim10^{-3}$--$10^{-2}~\ME\,\mbox{yr}^{-1}$, planets of low mass ($\MP\lesssim3~\MJ$)
with moderate to large radii ($\RP\approx3$--10~$\RJ$) could have their $\TSchock$ increased
by up to tens of percent, by a few 100~K.
This might be important especially for a non-zero $\Lint$,
leading to a higher shock temperature than expected naively.

\subsubsection{Regimes of the shock temperature}

To understand these results graphically, we show in the bottom row of Figure~\ref{Abb:Timplizit(RP)}
the fourth root of the left- and right-hand sides of Equation~(\ref{Gl:TSchock implizit approx, fuer Abb}).
We see that
the points deviating in Figure~\ref{Abb:Gitter} can do so for two reasons.
First of all, a high constant (temperature- and density-independent) opacity will
increase the shock temperature by reducing the effective speed of light ahead of the shock.
The threshold for this is $\kapR\rhoFf\RP\sim1$, as mentioned above,
and an example is for $(\MPunkt=10^{-3}~\ME\,\mbox{yr}^{-1}, \MP=1.3~\MJ, \RP=3~\RJ)$,
labelled ``A'',
where the opacity of the most refractory dust component leads to a higher shock temperature
than $\TSchockff=1000$~K:
$\TSchock\approx1080$~K for the nominal dust model
and $\TSchock\approx1200$~K for the curve sticking out most in Figure~\ref{Abb:kappaSchnitt}.
Second of all, the case of \textit{non-constant} high ($\kapR\rhoFf\RP\gtrsim1$) opacities
opens up the possibility of multiple solutions.

If $\kapR\rho\RP\ll1$ at the free-streaming shock temperature $\TSchockff$,
it is necessary and sufficient for the opacity slope
\begin{equation}
 \alpha_\kappa\equiv \left(\frac{\partial\ln\kapR}{\partial\ln T}\right)_\rho
\end{equation}
to satisfy $\alpha_\kappa>\alpha_\kappa^{\rm crit}=1$
at higher temperatures and for a sufficient $T$ range
in order to have one high-$T$ solution. %
A drop of $\alpha_\kappa$ below $\alpha_\kappa^{\rm crit}$ at a higher temperature
will lead to a third solution.
The stronger the slopes, the closer these higher-$T$ solutions will be
to $\TSchockff$.

If on the other hand $\kapR\rho\RP\gtrsim1$ at $\TSchockff$
(i.e., the pre-shock gas would be diffusive already if it were at that temperature),
$\TSchockff$ will not be a solution.
Provided the opacity slope $\alpha_\kappa<1$ at some higher temperature,
there will only be a single\footnote{
  In the case that the opacity shows several non-monotonicities at higher temperatures
  the number of roots is of course different and the discussion would need to be adapted.
  This is presumably the case at the iron opacity ``bump'' at $\log T\approx5.3$ \citep{iglesias92,jiang15},
  and in principle also for a very metal-rich gas.}
solution at high temperature.
Again, the stronger the slopes $|\alpha_\kappa|$ in the increasing and decreasing parts,
the closer the solution will be to $\TSchockff$.

If however $\kapR\rho\RP\gtrsim1$ at $\TSchockff$
and $\alpha_\kappa$ remained $>1$ at higher temperatures there would be formally
no solution: try it as it may by increasing its temperature, the shock would not
be able to radiate away the kinetic energy.
It is not clear though whether in this case we would still find a Mach number
such that $\etaklassisch=1$, or whether the model otherwise breaks down.  %
Fortunately, for gas mixtures as considered here $\alpha_\kappa$ does drop again,
so that the situation does not arise.

Qualitatively, these results do not depend much on different model settings.
Including a small downstream luminosity (coming from compression, an interior luminosity, or both)
does not change significantly the shock temperature(s)
of Figure~\ref{Abb:Timplizit(RP)}.
Changing the opacity (the dust model or, at higher temperature, the abundance of water; \citealp{malygin14})
or the flux limiter also has a clear but limited effect.

\subsubsection{Mach number}
 \label{Theil:Mach analytisch implizit}

Within the restriction of negligible $\Llinks$,
the Mach number $\Mach$ increases with increasing $\MP$ or $1/\RP$
but, perhaps counterintuitively, decreases with increasing accretion rate.
Using Equation~(\ref{Gl:TSchock_rauh}),
\begin{subequations}
 \label{Gl:Mach analytisch}  %
\begin{align}
 \Mach = &~2\left(\frac{G^3\pi\sigSB\Delta\fred}{\etaklassisch}\right)^{1/8}\sqrt{\frac{\mu\mH}{\gamma\kB}} \left(\frac{\MP^3}{\MPunkt\RP}\right)^{1/8}\\
                      =&~30 \left(\frac{\MP}{1.3~\MJ}\right)^{3/8} \left(\frac{\mu}{2.29}\right)^{1/2}\left(\frac{\gamma}{1.44}\right)^{-1/2}\notag \\
                              & \times \left(\frac{\MPunkt}{10^{-2}~\ME\,\mathrm{yr}^{-1}}\right)^{-1/8} \left(\frac{\RP}{1.5~\RJ}\right)^{-1/8},
\end{align}
\end{subequations}
where we took $\etaklassisch=1$ and $\Delta\fred=1$ for the second expression.
Thus the Mach number depends only moderately on the mass ($\Mach\propto\MP^{3/8\approx0.4}$)
and accretion rate (given that the latter ranges over several orders of magnitude)
and barely on the radius. %

\begin{figure}
 \epsscale{1.1}
 \plotone{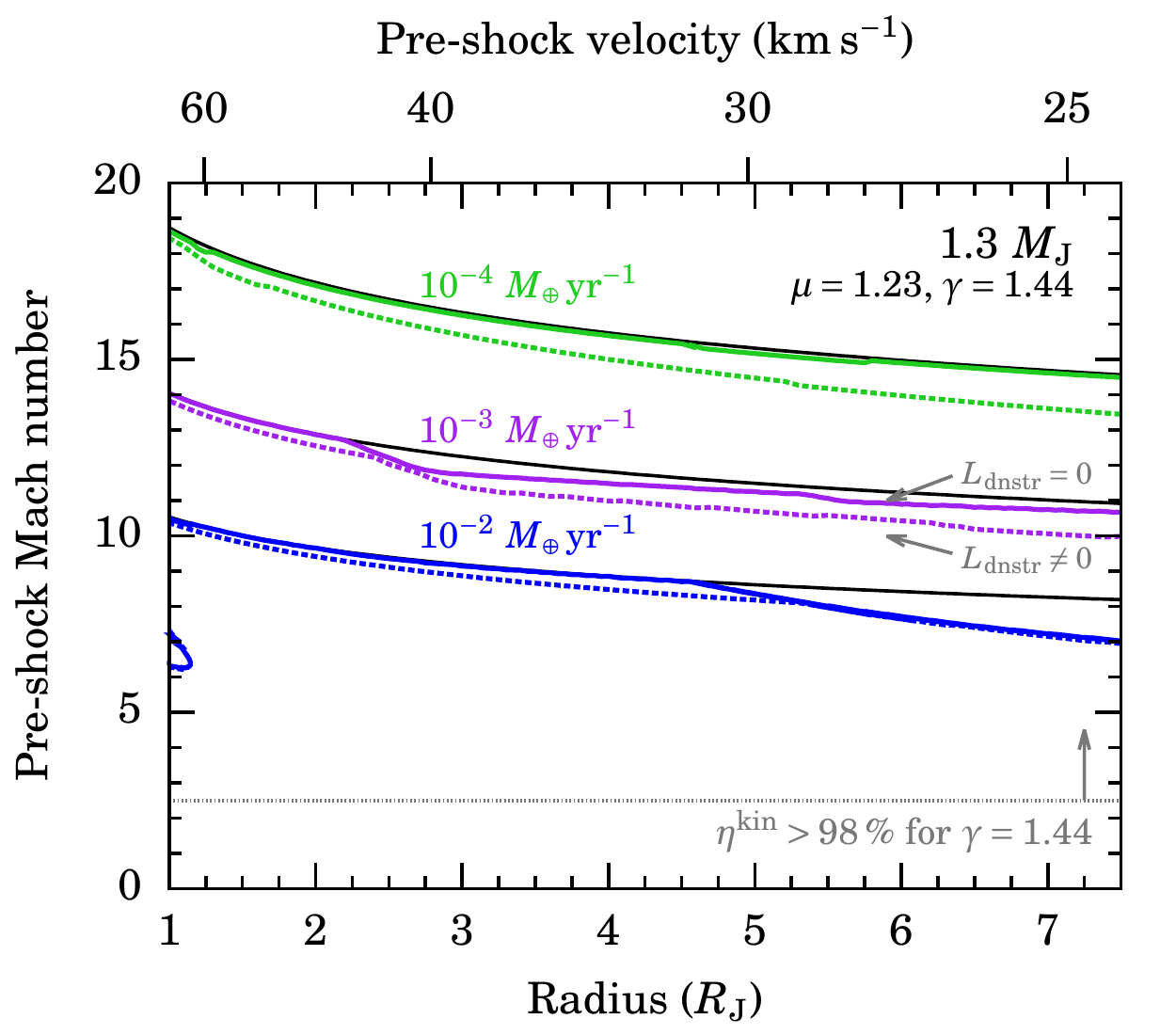}
\caption{
Analytical pre-shock Mach number.
The mass is $\MP=1.3~\MJ$ and the radius and accretion rate are varied.
We use $\mu=1.23$ (different from our nominal case)
and $\gamma=1.44$ to obtain a rough lower bound on the Mach number
and thus also on $\etaklassisch$.
As in Figure~\ref{Abb:Timplizit(RP)},
coloured solid (short-dashed) lines are for
$\Llinks=0$ ($\Llinks\neq0$),
using the implicit shock temperature from Equation~(\ref{Gl:TSchock implizit approx}).
The limiting case of $\Delta\fred=1$ (which implies simultaneously $\Llinks=0$ and the free-streaming limit for the shock temperature;
Equation~(\ref{Gl:Mach analytisch})) is also shown (solid black lines).
The value $\Mach=2.5$ is highlighted
since above this, $\etaklassisch$ is essentially 100\,\%\ (\textit{dotted line}).
}
\label{Abb:Mach implizit}
\end{figure}

We show in Figure~\ref{Abb:Mach implizit} the pre-shock Mach number
obtained from Equation~(\ref{Gl:TSchock implizit approx}) with $\mu=1.23$,
and also plot Equation~(\ref{Gl:Mach analytisch}) for reference.
Mach numbers increase with decreasing accretion rate, reaching $\Mach\sim30$
(in particular at higher masses, not shown) for $\MPunkt=10^{-4}~\ME\,\mathrm{yr}^{-1}$.
The Mach number at $\TSchockff$, given by Equation~(\ref{Gl:Mach analytisch}),
provides an upper bound but
$\Mach$ still remains securely above $\Mach\approx2.5$.
There, $\etaklassisch$ is near 100\,\%, which justifies \textit{a posteriori}
our approximation.
Also, taking $\Llinks$ not large (see Figure~\ref{Abb:Mach implizit}) but non-zero barely changes these results.

\subsubsection{Discussion of the analytical shock temperature}

The preceding analysis has shown that, for an isothermal radiative shock with negligible downstream luminosity,
the shock temperature is determined by the conditions immediately upstream.
There can be small to very important deviations from the free-streaming result $\TSchockff$.
These occur either because of the dust contribution to the opacity (at low temperature)
or because of the high opacity and high pre-shock density at high temperature and accretion rates.
Nevertheless, the $\TSchockff$ solution remains valid for a large part of parameter space.

Interestingly, our derivations do not depend explicitly on the equation of state (EOS),
specifically on the choice of and (non-)constancy of $\mu$ and $\gamma$ across or ahead of the shock.
Thus the expressions~(\ref{Gl:TSchock implizit beide}--\ref{Gl:TSchock fuer kapparhor>>1}) %
for the shock temperature could apply also to the case of a general EOS.
This is should certainly be the case for $(\MPunkt,\MP,\RP)$ combinations for which the pre- and post-shock
points have the same $\gamma$ and $\mu$ values.
It might also hold more generally but this will be investigated in a subsequent article.

Finally, note that
the importance of a shock temperature increased with respect to $\TSchockff$
for lower planet masses and larger radii than shown here, as might be relevant during the early stages
of detachment (runaway accretion), will have to be assessed with separate formation
calculations as in \citet{berardo17}.
\section{Importance of the Equation of State and opacity} %
 \label{Theil:einzelne}

Next we verify the robustness of our results by varying different parts of the microphysics
that go into our simulations.
This also gives us occasion to show global profiles of the accretion flow;
we had restricted ourselves in figure~2 of \citetalias{m16Schock}
to the vicinity of the shock.

\subsection{Dependence of pre-shock-region quantities on the EOS}  %
 \label{Theil:L,eta von ZG}

While the use of an ideal but non-perfect EOS is deferred to a later article,
we study in this section the importance of the constant $\mu$ and $\gamma$.
For conciseness, we will refer to a particular $(\mu,\gamma)$ combination as an EOS.  %

\begin{figure*}
 \epsscale{1.1}
 \plotone{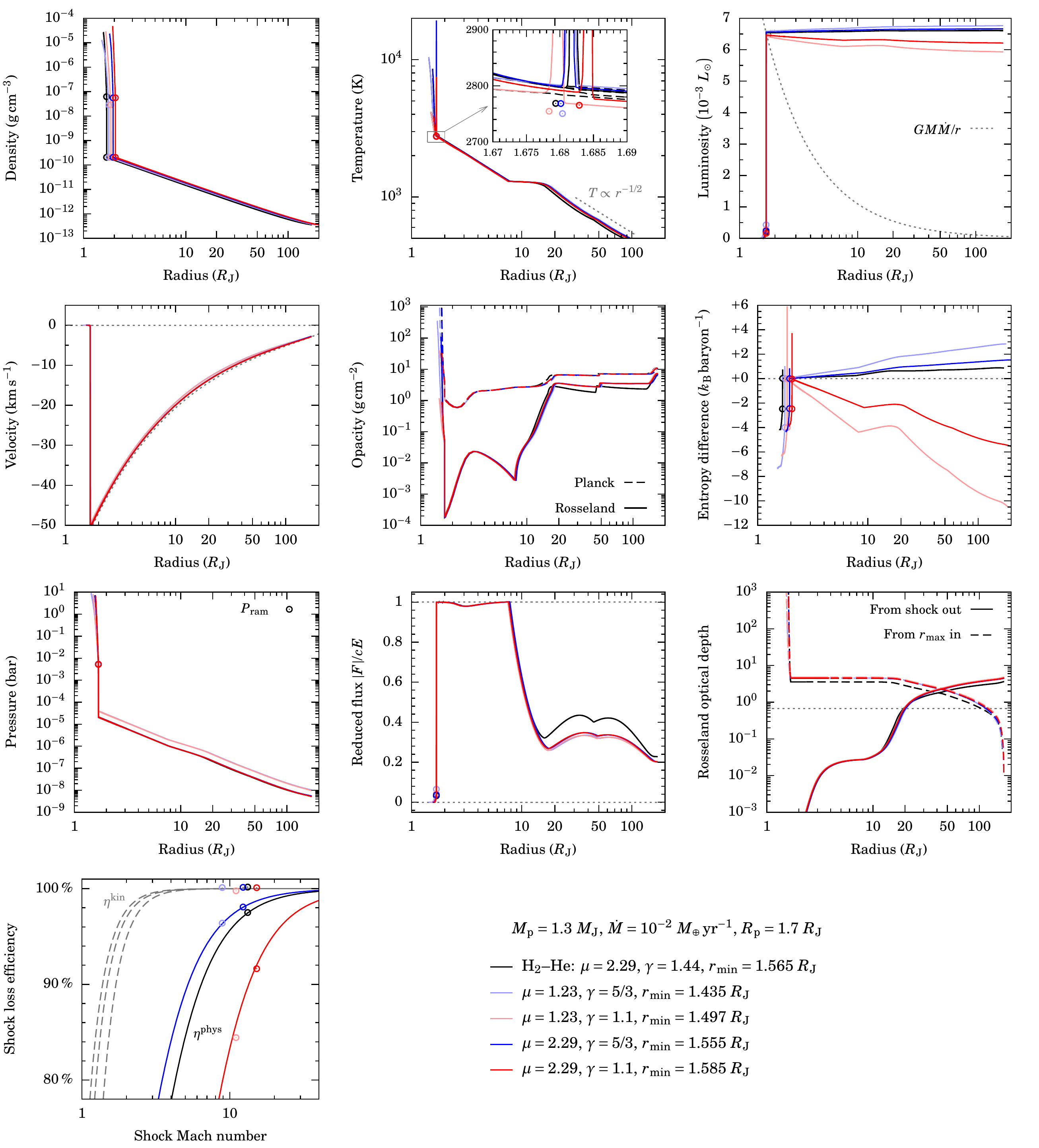}
\caption{
Accretion and shock profiles (see axis labels) using a perfect equation of state with different mean molecular weights $\mu$
and ratios of specific heat $\gamma$ (see legend).
The black line corresponds to the nominal case of an H$_2$--He mixture.
The \citet{malygin14} and \citet{semenov03} opacities are used.
The simulations differ only in $\mu$, $\gamma$, and the inner edge of the computational domain $\rmin$ (see legend),
the latter chosen to have the shocks coincide at a time $t=5\times10^7$~s
after the start of accretion at $t=0$.
All but the first profile in the density and entropy-difference panels
are shifted horizontally for legibility.
Grey dotted lines highlight values of 0, 1, or 2/3 in the $v$, $\fred$, $S$ and $\Delta\tau$ panels,
as appropriate.
Circles in the temperature panel and its inset show the estimate from Equation~(\ref{Gl:TSchock_rauh a}),
whereas in the density, luminosity, pressure, and $\fred$ panels they highlight the post-shock conditions
and in part the pre-shock conditions.
In the $\etaphys(\Mach)$ panel on the bottom left, the solid (dashed) curves show the physical (kinetic) efficiency
for $\gamma=5/3$, 1.44, 1.1 (top to bottom) compared to the simulation results (circles).
}
\label{Abb:Abbgammamu}
\end{figure*}

Figure~\ref{Abb:Abbgammamu} shows the profiles for the four extreme combinations $(\mu,\gamma)$,
where $\mu=1.23$ (atomic) or $\mu=2.29$ (molecular), and $\gamma=1.1$ (occurring for dissociation or ionization)
or $\gamma=5/3$ (monatomic, no dissociation or ionization)
together with the nominal case $(\mu=2.29,\gamma=1.44)$, i.e., for a perfect-gas H$_2$--He mixture.
The density in the accretion flow being set by mass conservation,
it is independent of the EOS,
but the jump across the shock scales as $\Delta\rho\propto\mu$.  %
The pressure in the accretion flow does scale as $P\propto 1/\mu$ but,
being a dynamical quantity, the ram pressure (i.e., the post-shock pressure)
is the same in all cases.
While this might be counterintuitive, also the shock temperature (see inset) %
is independent of $(\mu,\gamma$),
which was to be expected from the analytical estimates of the shock temperature
since the microphysical parameters do not enter anywhere.
The opacities $\kapR$ and $\kapP$ are the same here because we use in all cases the same tables
(computed using a non-perfect EOS)
while the opacity is fundamentally a function of temperature
and density (but not mean molecular weight).

\subsubsection{Entropy}

One of the quantities which changes with the EOS is
the entropy calculated with the chosen $(\mu,\gamma)$. %
Note that it differs from the entropy that would be obtained
with non-perfect EOS
shock simulations.
This $(\mu,\gamma)$-dependent entropy is given in general
(but for constant $(\mu,\gamma)$, i.e., outside of chemical reactions such as conversion of ortho- to parahydrogen,
dissociation, or ionization)
by
\begin{equation}
 \label{Gl:Entro knst ZG}
 s = s_0 + \frac{\gamma}{\gamma-1} \frac{\ln(10)}{\mu}\log_{10}\frac{T}{T_0} - \frac{\ln(10)}{\mu}\log_{10}\frac{P}{P_0},
\end{equation}
where $s_0$ is a constant, %
and $T_0$ and $P_0$ are an arbitrary reference temperature and pressure, respectively.
The form is similar as a function of $(P,\rho)$ (e.g., \citealp{rafikov16}).

The radial profile of the entropy given by Equation~(\ref{Gl:Entro knst ZG})
is shown in Figure~\ref{Abb:Abbgammamu} for various EOS.
We plot in fact the entropy difference with respect to the pre-shock location,
since we are interested in the radial profiles of $s$ for the individual $(\mu,\gamma)$ cases.
It increases outwards
for $\gamma=5/3$ since it is greater than the critical value of~4/3 (section~3.3.2 of \citealp{m16Schock})
but decreases for $\gamma=1.1$ (since then $\gamma<4/3$) outwards
in the accretion flow. Convection is however not expected there
because of the supersonic motion.

\subsubsection{Luminosity}
 \label{Theil:L(r; ZG)}
 
Another difference between the various EOS is in the luminosity profile,
as Figure~\ref{Abb:Abbgammamu} shows.
We now proceed to explain it.
The luminosity $L$ varies throughout
the accretion flow by an amount which depends on the EOS, \textit{not} directly on the optical depth.
This perhaps surprising statement can be derived from considering the
steady-state (time-independent) version of the evolution equation
for the total energy (so that the $\pm\Lambda$ terms cancel
even in non-equilibrium radiation transport; Equation~(\ref{Gl:GG2T})) and writing the gravity field term as a potential:
\begin{subequations}
 \label{Gl:dLdr}
\begin{align}
 \frac{\dd L}{\dd r} &= |\MPunkt| \left(\frac{\dd h}{\dd r} + \frac{\dd}{\dd r} \left[\frac{1}{2}v^2 - \frac{G\MP}{r}\right]\right) \label{Gl:dLdr a}\\
                     &= |\MPunkt| \left(\frac{\dd h}{\dd r} - \frac{1}{\rho} \frac{\dd P}{\dd r}\right) \label{Gl:dLdr b}\\
                     &= |\MPunkt| \left(\frac{\dd\eint}{\dd r} + P\frac{\dd(1/\rho)}{\dd r}\right)\\
                     &= |\MPunkt| T\frac{\dd s}{\dd r}, \label{Gl:dLdr d}
\end{align}
\end{subequations}
where $h=H/\rho=\eint+P/\rho$ is the specific enthalpy per mass and is given by $h=\gamma/(\gamma-1)\kB T/(\mu\mH)$ for a perfect EOS.
Only Equation~(\ref{Gl:dLdr a}) was derived in \citetalias{m16Schock}.  %
It %
did not require identifying the gas and radiation temperatures with each other,
leaving the derivation valid also for 2-$T$ radiation transport.
Equation~(\ref{Gl:dLdr b}) is obtained from $\MPunkt=4\pi r^2 \rho v$ (Equation~(\ref{Gl:GGa}))
and the time-independent momentum equation (Equation~(\ref{Gl:GGb})).
Equation~(\ref{Gl:dLdr d}) follows from the first law of thermodynamics
\begin{equation}
T\dd s = \dd\eint + P\dd\left(\frac{1}{\rho}\right),
\end{equation}
with the entropy $s$ here not normalised by $\kB\,\mH^{-1}$.
Note that, \nnBegut{in the pre-shock region,}
it is the small deviation of the velocity from a free-fall
profile which makes $\dd L/\dd r\propto T\dd s/\dd r$ and not $\propto\dd h/\dd r$.

In \citetalias{m16Schock}, we had argued that the radial non-constancy (over $\sim$~Hill-sphere scales)
of the luminosity can be understood from the enthalpy profile
when the second term in Equation~(\ref{Gl:dLdr a}) is negligible.
While this is formally true, Equation~(\ref{Gl:dLdr}) provides a more complete derivation
which reveals that the relevant quantity is the \textit{entropy},
without requiring restrictions on any term.
Thus the radial luminosity profile is set not directly by the optical depth.
Nevertheless, the analysis in Section~\ref{Gl:T-Profile im Akkfl} has shown that
there is a link between the local opacity and its slope on the one hand
and the temperature (and thus the entropy) profile on the other.

In the case of a perfect EOS, Equation~(\ref{Gl:dLdr d}) becomes
\begin{subequations}
\label{Gl:dLdr mit gamma}
\begin{align}
 \frac{\dd L}{\dd r} &= \frac{|\MPunkt|}{\gamma-1}\frac{\kB T}{\mu\mH}\left[\frac{\dd\ln T}{\dd r}-(\gamma-1)\frac{\dd\ln\rho}{\dd r}\right]\\
                &= \frac{\kB |\MPunkt|}{\mu\mH}\frac{T}{r}\left[\frac{1}{\gamma-1}\frac{\dd\ln T}{\dd \ln r}+\frac{3}{2}\right],\label{Gl:dLdr mit gamma b}
\end{align}
\end{subequations}
where the second line holds for a free-fall density profile.
Since the factor $T/r$ in Equation~(\ref{Gl:dLdr mit gamma b}) decreases outwards, it is the layers closest
to the shock which contribute most to the change in $L$ between $\rSchock$ and $\rmax$,
at least for a constant logarithmic temperature slope.
Equation~(\ref{Gl:dLdr mit gamma}) reveals that, all other things being equal,
the change in luminosity from the shock to the Hill sphere
will be more important as $\gamma$ tends to~1 (i.e., more isothermal), %
or for smaller mean molecular weight $\mu$.
The choice of $\gamma$ can change the sign of $\dd L/\dd r$ whereas $\mu$ cannot.

In fact, an estimate for the change in luminosity between the shock
and a given distance in the flow, in particular the accretion radius, can be derived
by using the constant-$(L/\fred)$ temperature profile (see Equation~(\ref{Gl:dlnT^4dlnr})),
\begin{equation}
 \label{Gl:T(r) konst L/fred}
 T(r)=\TSchock \left(\frac{\rSchock}{r}\right)^{-1/2},
\end{equation}
where $\TSchock$ is the shock temperature,
which by Equation~(\ref{Gl:TSchock implizit approx, fuer Abb}) can be larger
than the free-streaming temperature.
(Note that $T\propto r^{-1/2}$ can hold not only for free-streaming.)
In Figure~\ref{Abb:Jakobus}c or~\ref{Abb:Abbgammamu},
the $T(r)$ profiles are rarely steeper than Equation~(\ref{Gl:T(r) konst L/fred}),
so that it will provide an upper bound.
Inserting Equation~(\ref{Gl:T(r) konst L/fred}) in Equation~(\ref{Gl:dLdr mit gamma}) yields
\begin{equation}
 \frac{\dd \LkonstLfred}{\dd r} = -\frac{|\MPunkt|}{\gamma-1}\frac{\kB \TSchock}{\mu\mH}\frac{4-3\gamma}{2\RP} \left(\frac{r}{\RP}\right)^{3/2},
\end{equation}
using the subscript ``(\ref{Gl:T(r) konst L/fred})'' on $L$ to remind
that a temperature profile given by Equation~(\ref{Gl:T(r) konst L/fred}) was assumed.
Thus, the change (drop or increase) in $\LkonstLfred$ between $\RP$ and $\RAkk$,
neglecting terms of order $\sqrt{\RP/\RAkk}\ll1$,
is given by
\begin{equation}
 \label{Gl:Delta L bis RAkk approx}
 \Delta \LkonstLfred = \int_\RP^\RAkk \frac{\dd L}{\dd r} = \frac{3\gamma-4}{\gamma-1}\frac{\kB \TSchock|\MPunkt|}{\mu\mH}, %
\end{equation}
with the luminosity reaching the local circumstellar disc (the nebula) equal to
\begin{equation}
 \label{Gl:L(RAkk)}
 L(\RAkk) \approx \Lint + \LAkk + \Delta \LkonstLfred. %
\end{equation}
Equation~(\ref{Gl:Delta L bis RAkk approx}) is more general than equation~(32) in \citetalias{m16Schock}.
In the usual limit $\RAkk\gg\RP$, the luminosity change $\Delta \LkonstLfred$ is thus independent of $\RAkk$,
and thus also of $\kLiss=1/3$ (as we assume here)
for the accretion radius $\RAkk\approx\kLiss\RHill$.
For $\gamma>4/3$, the luminosity increases outwards ($\Delta \LkonstLfred>0$),
whereas for $\gamma<4/3$ there is a decrease in luminosity ($\Delta \LkonstLfred<0$).

Equation~(\ref{Gl:Delta L bis RAkk approx}) is an upper bound in the case $\gamma<4/3$
because any flattening of the luminosity profile
as in Figure~\ref{Abb:Abbgammamu} will slow down the decrease of $T(r)$.
For $\gamma>4/3$ the estimate is more like a lower bound but
since the critical $\gamma=4/3\approx1.33$ is not far from $\gamma=1.44$ for molecular hydrogen with helium,
it is also roughly equal to the actual change in that case.
This can be seen graphically in Figure~\ref{Abb:Abbgammamu}.

Thus,
in Figure~\ref{Abb:Abbgammamu},
$L(r)$ increases outwards for the high-$\gamma$ cases ($\gamma=5/3$)
as well as for the nominal case ($\gamma=1.44$),
while it overall decreases slightly for the low-$\gamma$ cases ($\gamma=1.1$).
The mean molecular weight also plays a role,
and, in all, $L$ at roughly the Hill radius varies by roughly 10~percent
for this specific $(\MPunkt,\MP,\RP)$
case over all $(\gamma,\mu)$ combinations.
In Section~\ref{Theil:Disk ZG} we look at this more generally.

However, the size of the luminosity jump $\Delta L$ at the shock is very nearly the same
across all simulations in Figure~\ref{Abb:Abbgammamu}.
Since in all cases the immediate upstream region is in the free-streaming regime,
the small differences in $L(\rSchock^+)$ are related to the slightly different
downstream luminosities or, equivalently, $\fred(\rSchock^-)$.

\subsubsection{Global physical efficiency}

Finally, depending on both $\gamma$ and $\mu$, the global physical efficiency
ranges from $\etaphys=84\,\%$ to 98~\%, increasing with $\mu$ but decreasing
with $\gamma$.
This large difference comes from different $\gamma$ values but also
the changing Mach number (through $\mu$),
as can be seen from Equation~(\ref{Gl:Mach analytisch}).
One can wonder whether a low $\mu$ caused by ionization could lead to much
lower Mach numbers and thus efficiencies.
However, since the Mach numbers we find are all high ($10\lesssim\Mach\propto1/\sqrt{\mu}$),
even a change by a factor of at the very most $\sqrt{0.6/2.3}=0.5$ (from ionized to molecular gas)
would leave $\Mach\gtrsim5>2.5$ and thus the shock supercritical
for a given $\TSchock$ and $\vFf$.
This statement should still be revisited
with simulations using the full EOS since $\Mach$ is of course a function of $\TSchock$.

\subsubsection{Other quantities}

The other panels of Figure~\ref{Abb:Abbgammamu} are shown for completeness.
The profiles are qualitatively similar between
all simulations (see the detailed description in \citetalias{m16Schock})  %
and we can mention here that we find the gas and radiation are always well coupled,
which is presumably related to the sufficiently high opacity.
Also, as in \citetalias{m16Schock}, the radiative precursor to the shock
is larger than the simulation box, as even a cursory comparison
to standard supercritical and subcritical shock structures (e.g., \citealp{ensman94})
reveals.

\subsubsection{Summary of the effect of the EOS}

In summary, depending on the EOS, a different luminosity reaches the accretion radius
but the variation is moderate (tens of percent) across the relevant input parameter range.
However, both the shock temperature and the ram (post-shock) pressure are independent of the perfect EOS,
at least in the limit of an isothermal shock.
This is an important result.
We conjecture that the post-shock temperature and pressure
will not be different when using instead an ideal but non-perfect EOS.
This will need to be verified but if if holds, and if the shock is still isothermal, it implies that
the post-shock entropy $\Snach$, which depends only on $T$ and $P$, will be the same as found here.
As for the shock efficiency, it clearly depends on the EOS (see Figure~\ref{Abb:Abbgammamu}).
\subsection{Influence of the dust opacity}

\begin{figure*}
 \epsscale{1.1}   %
 \plotone{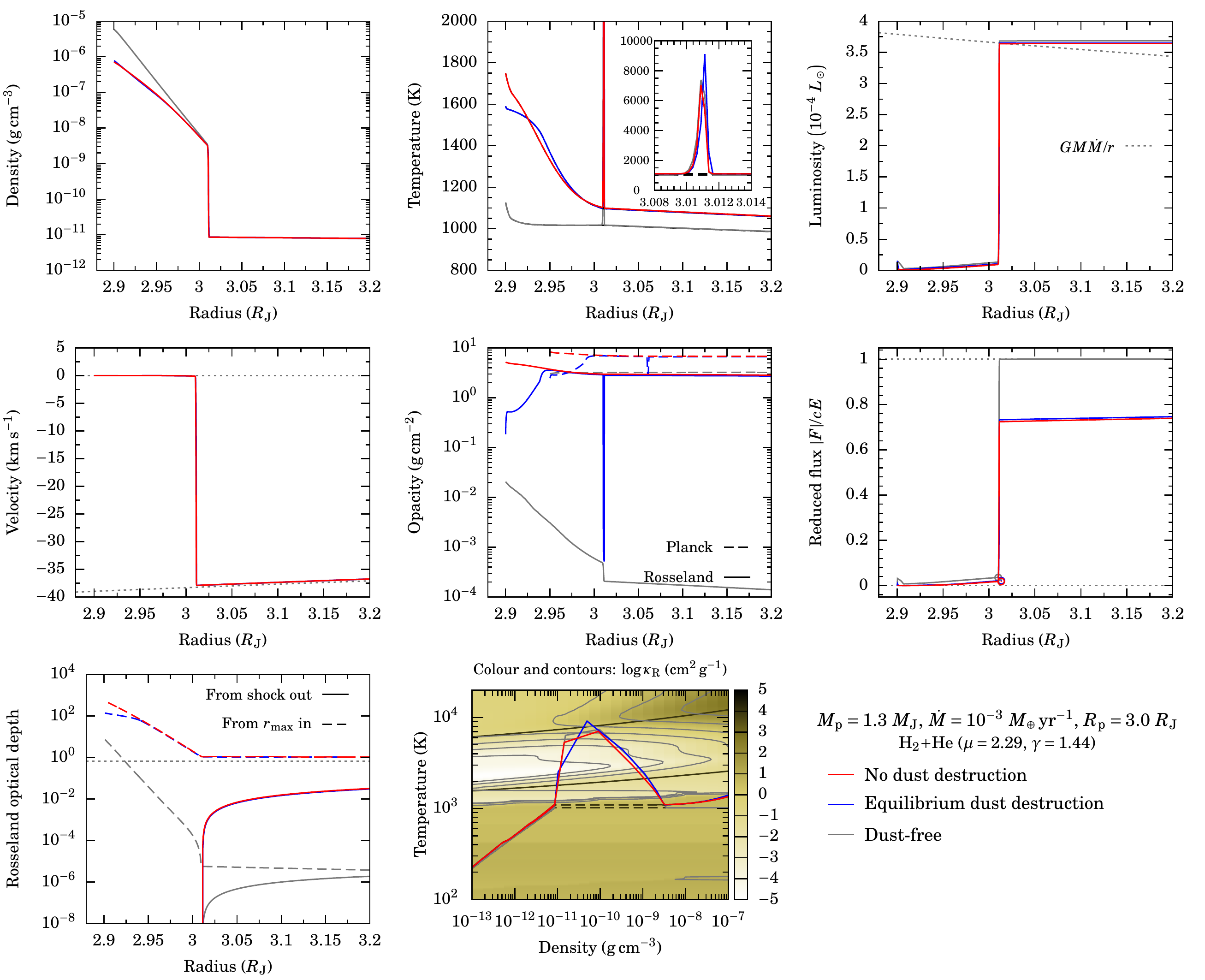}
 \caption{
Structure of the shock for different behaviours of the dust in the Zel'dovich spike:
no destruction (\textit{red curves})
or equilibrium destruction (nominal case; \textit{blue curves}).
For comparison, a case with no dust opacity at all is also shown (\textit{grey}).
In the top and bottom middle panels, the radiation temperature is shown as dashed black lines.
In the opacity panel, the Planck curves are shifted by 0.05~$\RJ$ to the right
for clarity.
Note that the Planck opacity is high ($\kapP\gtrsim3$~cm$^2$\,g$^{-1}$) even in the absence of dust
and that the gas and radiation are tightly coupled, with $\Trad\neq\TGas$ only
in the Zel'dovich spike (see the middle panels in the top and bottom rows).
The solid black lines in the bottom middle panel display for reference
where hydrogen is dissociated or ionized to 50~\%\ \protect\citepalias{scvh}.
}
\label{Abb:mitohneStaub}
\end{figure*}

Simulations such as the ones presented here do not spatially resolve the physical
Zel'dovich spike (see appendix~B of \citealp{vaytetgonz13}),
which is typically much thicker than the mean free path of the gas particles
yet still orders of magnitude smaller than a photon mean free path \citep{zeldovich67,drake07}.
The peak temperatures should be very high but the cooling behind the peak
also very quick, so that the fate of dust grains passing through this shock
is not obvious \textit{a priori}.
Calculating their time-dependent sublimation and recondensation including non-equilibrium effects
is beyond the scope of this work.
Since our standard assumption is to use time-independent (equilibrium) abundances
from the simple model of \citet{isella05},
in which the destruction temperature is simply
$\TZerst=1220\times{\rho_{-11}}^{0.0195}$~K, where $\rho_{-11}\equiv\rho\times10^{11}~\mathrm{cm}^3\,\mathrm{g}^{-1}$,
we consider in this section the other extreme, namely that the grains are not destroyed.
Note that the question of the dust \mbox{(non-)}destruction in the shock poses itself
only for shocks at low enough temperature that solid grains are still present the incoming material
(e.g., the $\approx1~\MJ$ cases at $\MPunkt=10^{-3}~\ME\,\mbox{yr}^{-1}$ in Figure~\ref{Abb:Gitter}).

Figure~\ref{Abb:mitohneStaub} shows the resulting shock structure for constant dust abundance in red and
equilibrium abundances in blue. In all relevant quantities ($\rho$, $v$,
$T$, $L$, $\fred$), with the obvious exception of $\kapR$ and $\kapP$, the profiles are essentially identical
for these two cases.
This holds also throughout the simulation domain (not shown).
In particular, the shock temperature is the same, and the Zel'dovich spike stays very
optically thin (as it should), increasing from, very roughly, $\Delta\tauR\sim10^{-5}$ to $10^{-2}$.
Note that the opacity is too high in the always-dust case (see blue solid line in opacity panel)
but that this does not affect the temperature structure of higher post-shock regions,
i.e., directly below the shock.

As a more extreme case, we also switched off entirely the dust contribution
shown by grey profiles in Figure~\ref{Abb:mitohneStaub}.
This represents the limiting case of the reduction of \citet{szul17} by a factor of ten relative
to the ISM, on the grounds that the growth of dust grains into larger aggregates
diminishes the opacity.
Also, as pointed out by \citet{uyama17}, dust tends to settle to the midplane while accretion comes
from higher up in the circumstellar disc, so that the accretion flow onto a planet is
actually possibly devoid of dust.

Leaving out the dust opacity lowered the Rosseland mean by four orders of magnitude near the shock
but the Planck mean only by a factor of two due to the high Planck opacity of the gas.
The consequence was a decrease of the shock temperature by only 100~K,
associated with a jump in the reduced flux increasing from $\Delta\fred\approx0.7$
to $\Delta\fred\approx1$.

Our conclusion from this test is that the upstream $\fred$ (given $\fred\approx0$ downstream),
and hence ultimately the opacity there,
is important in setting the shock temperature, but that the dust destruction
in the Zel'dovich spike is not.
This seems to imply that one would need to follow the time-dependent evaporation
of the dust \textit{approaching the shock} in each simulation, %
since the outer parts are always cold enough that dust should be present\footnote{It
   is easy to estimate, assuming $T\propto r^{-1/2}$, that only unrealistic
   parameter combinations could lead to evaporated dust (i.e., $T\gtrsim1000$~K, taking the density dependence into account)
   at $r\sim\kLiss\RHill$ or $r\sim\RBondi$.}.
However, this will affect only a very small fraction of cases.
Indeed, it requires
(i)~pre-shock temperatures above the (density-dependent) dust destruction temperature
but (ii)~not too high to not already have the dust destroyed a large distance ahead from shock,
in which case the dust would certainly be evaporated.
The transition region is very narrow in temperature ($\Delta T\sim100$~K; \citealp{semenov03})
compared to the range of shock temperatures we can expect (see, e.g., Figure~\ref{Abb:Gitter}).
For these cases where it is relevant, a zeroth-order approach would be to compare the flow
timescale $\tFliess=r/v$ to the evaporation time as given by the Polanyi-Wagner formula
(see, e.g., equation~(20) of \citealp{grassi17}).
However, this is an only marginally important consideration
and it will not be investigated further.

\section{Discussion}
 \label{Theil:Disk}

We discuss some of the results presented here.

\subsection{Hot starts or cold starts?}
 \label{Theil:H oder K?}

The main question driving this work is whether shock processing of the accreting gas
leads to hot starts or to cold starts \citep{marl07}.
While a detailed coupling of the shock results to formation calculations
is needed to resolve this,
there are several ways of estimating the outcome beforehand.
We now consider them in turn.

\subsubsection{Luminosity-based argument:\\ shock heating vs.\ internal luminosity}
 \label{Theil:H oder K? Leuchtkraft}

The classical understanding of the extreme outcomes
considers that the accreting gas carries only kinetic energy
and that for cold starts,
\textit{all} of this incoming energy is radiated away at the shock,
with no energy brought into the planet,
while for hot starts none of the incoming energy should leave the system
but instead be added to the planet.
However, this view neglects the thermal energy of the gas that is brought into the planet,
which comes from pre-heating by the radiation from the shock. %
In \citetalias{m16Schock} we have shown that this is in fact not negligible,
leading to the definition of $\etaphys$ instead of $\etaklassisch$ (see Section~\ref{Theil:eta(Mach)}).
Therefore, it is possible for most of the accretion energy to be radiated away at the shock
but to still have an important heating of the planet by the shock
if the inward heating is large compared to the internal luminosity.

To quantify this,
at least when following only the global energetics,
one can look at
the effective heating by the shock $\QSchock$ (to be defined shortly) relative to the internal luminosity $\Lint$,
\begin{equation}
 \qSchock\equiv\frac{\QSchock}{\Lint}.
\end{equation}
A large $\qSchock$ would suggest that the shock can heat up the planet appreciably.

The effective heating of the planet by the shock, in turn,
is the net rate of total energy which is not lost from the system
and therefore added to the planet.
By the definition of $\etaphys$ (Equation~(\ref{Gl:etaphys_DeltaLE})),
this is
\begin{equation}
 \label{Gl:QSchock-Def}
 \QSchock = \left(1-\etaphys\right)|\EPkt(\rmax)| = |\EPkt(\rSchock^-)|,
\end{equation}
where $\rSchock^-$ is immediately downstream of the shock.
From the definition of $\EPkt$ (Equation~(\ref{Gl:EPkt(r)})), we have
in the isothermal limit applicable to our results
\begin{subequations}
\label{Gl:QSchock(Mach)}
\begin{align}
 \QSchock &=\MPunkt\left[h(\rSchock^-) + \ekin(\rSchock^-)\right]\\
          &=\left(1+\frac{\gamma-1}{2\gamma^2\Mach^2}\right) \MPunkt h(\TSchock) \label{Gl:QSchock(Mach) mit h}\\
          &=\left(1+\frac{\gamma-1}{2\gamma^2\Mach^2}\right)\frac{\gamma}{\gamma-1}\frac{\kB\TSchock}{\mu\mH}\MPunkt\\
          &=\left(\frac{2}{(\gamma-1)\Mach^2} + \frac{1}{\gamma^2\Mach^4}\right)\LAkkmax,
\end{align}
\end{subequations}
using also that for an isothermal shock the post- and pre-shock densities
are related by $\rho_2 = \rho_1 \gamma\Mach^2$ and that $\rho_2v_2=\rho_1v_1$.
We recall that $\LAkkmax=G\MP\MPunkt/\RP$.
Equation~(\ref{Gl:QSchock(Mach) mit h}) in particular shows that what is added to the planet
is mostly the enthalpy of the gas, with a small additional term corresponding
to the leftover kinetic energy (from the post-shock settling velocity of the gas).
The latter is vanishingly small in the limit of a large pre-shock Mach number.

\begin{figure*}
 \epsscale{1.1}
 \plottwo{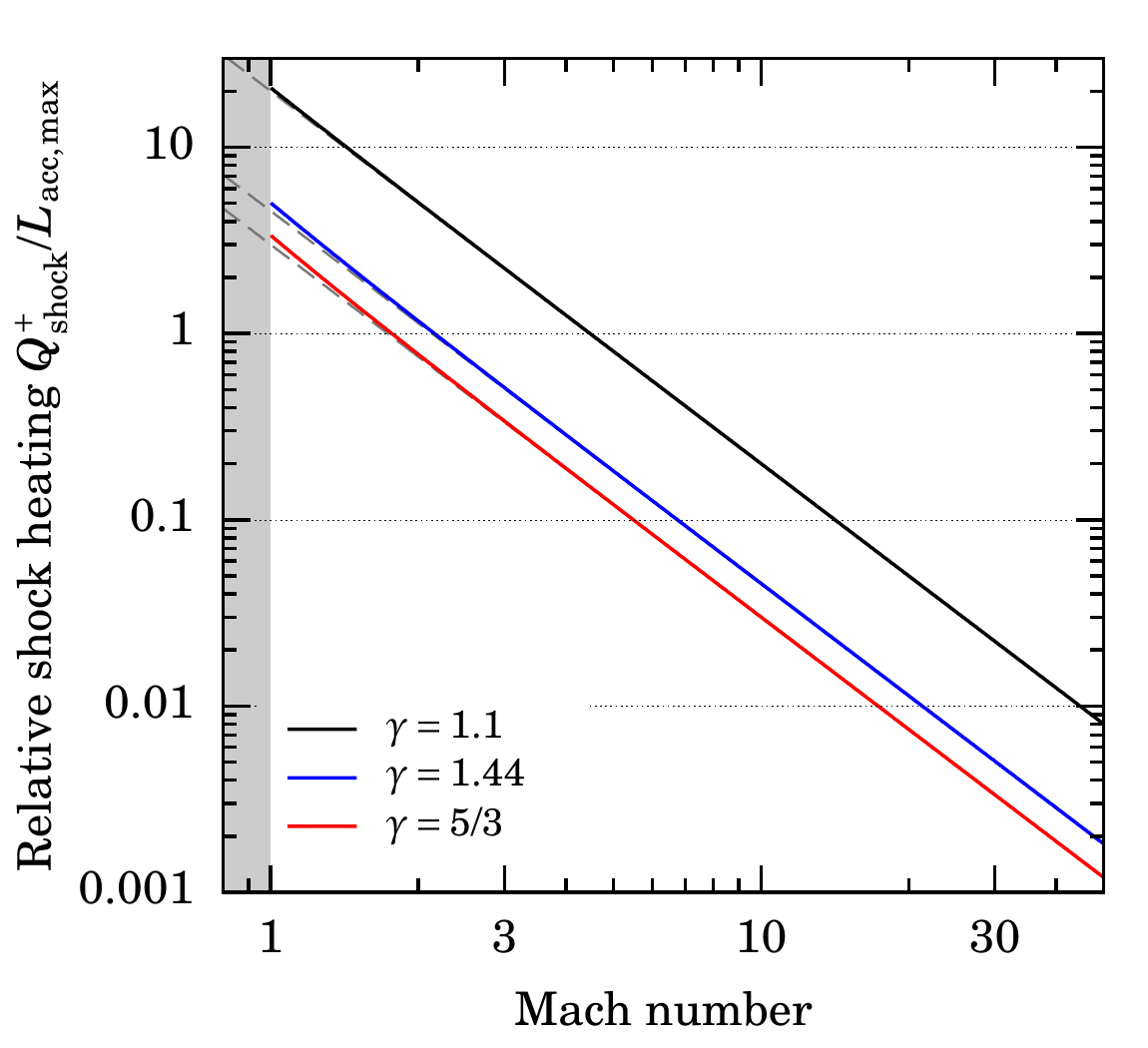}%
{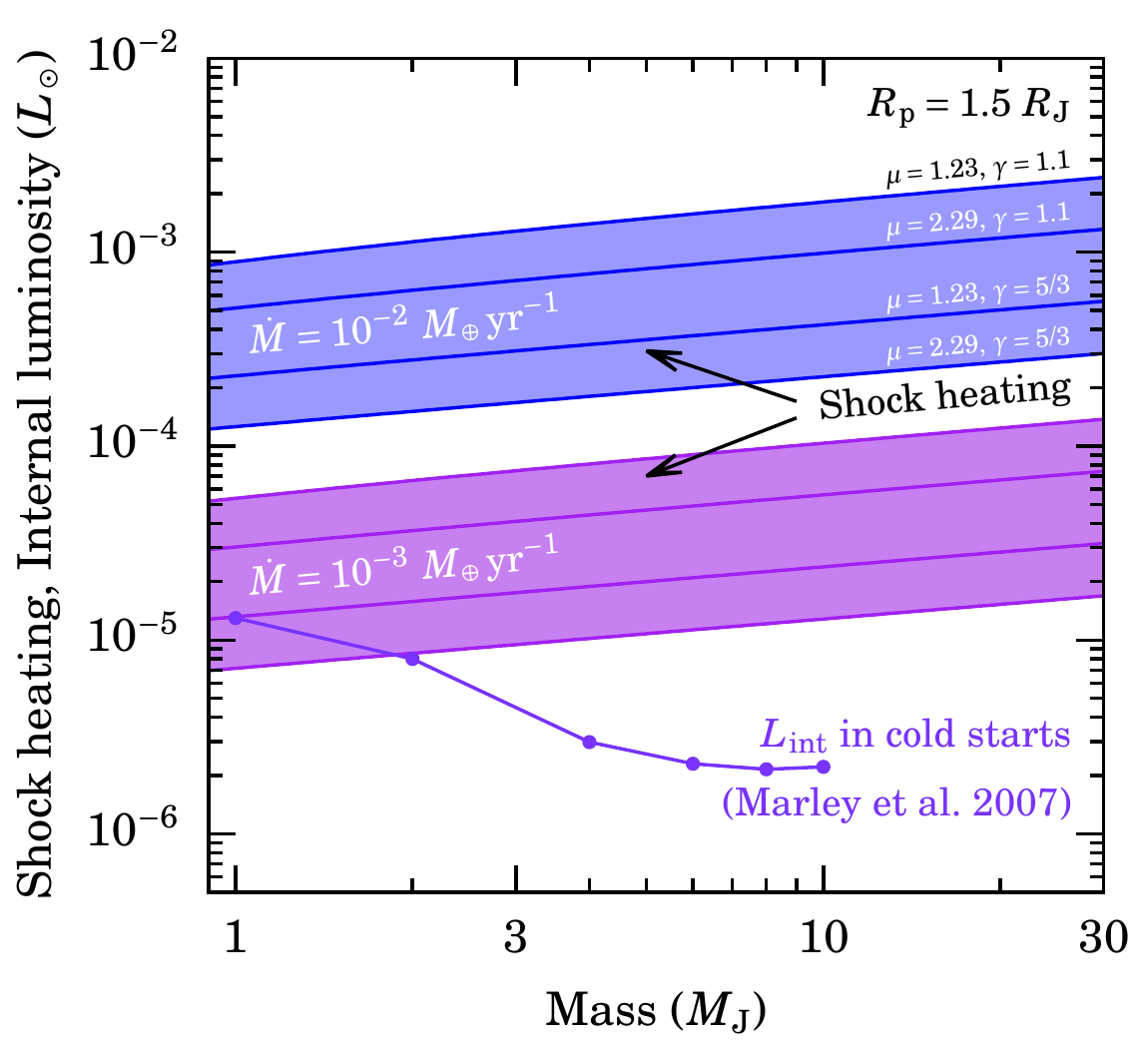}
\caption{
Estimated heating of the planet by the shock.
\textit{Left panel:} Heating $\QSchock$ relative to the maximal accretion luminosity $\LAkkmax=G\MP\MPunkt/\RP$
for an isothermal shock (Equation~(\ref{Gl:QSchock(Mach)}))
as a function of the pre-shock Mach number for different $\gamma$.
Below $\Mach=1$ there is no shock (greyed-out region).
The large-$\Mach$ limit $\QSchock/\LAkk=2/[(\gamma-1)\Mach^2]$ is also shown (dashed grey lines).
\textit{Right panel:} Absolute $\QSchock$ 
(Equation~(\ref{Gl:Q+Formpar-Werte})) %
for $\MPunkt=10^{-2}$ (blue shaded region) and $10^{-3}~\ME\,\mathrm{yr}^{-1}$ (purple).
We neglect $\Llinks$ in $\QSchock$ and take $\RP=1.5~\RJ$ for all masses.
We also show the approximate internal luminosity $\Lint$ from the \protect\citet{marl07} cold starts
during formation, which accrete mostly at $\MPunkt\approx10^{-2}~\ME\,\mathrm{yr}^{-1}$.
The shock heating clearly dominates over $\Lint$. %
}
\label{Abb:GemmaGalgani MutterSperanza Q+}
\end{figure*}

The shock heating relative to $\LAkkmax$ is plotted against Mach number
in Figure~\ref{Abb:GemmaGalgani MutterSperanza Q+}a.
It may seem surprising that for low Mach numbers $\Mach<2$--4 (depending on $\gamma$),
we have that $\QSchock>\LAkkmax$.
However, this simply comes from the fact that the gas at $\rmax$ does not bring
only kinetic energy but also enthalpy with it.

For the range $\Mach\approx7$--35 (looking at Figures~\ref{Abb:Gitter-etaphys} and~\ref{Abb:Mach implizit})
and with %
$\gamma=1.44$ as we used here, we see from Figure~\ref{Abb:GemmaGalgani MutterSperanza Q+}a that
$\QSchock\approx(10$\,\%--0.4\,\%$)\LAkk$ respectively.  %
In this high-Mach number limit (valid actually already for $\Mach\gtrsim1$),
the effective heating rate can be simplified to
\begin{subequations}
\begin{align}
 \QSchock =\,& \frac{2}{\gamma-1}\frac{\LAkk}{\Mach^2} \\
          =\,& \frac{\gamma}{\gamma-1}\frac{\kB}{\mu\mH}\left(\frac{G\MP\MPunkt^5}{16\pi\sigSB\RP^3}\right)^{1/4} \ell^{1/4}\label{Gl:Q+Formpar}\\
          = \,&3\times10^{-4}~\LSonne~\frac{\gamma_{1.44}}{\gamma_{1.44}-1} \left(\frac{\mu}{2.29}\right)^{-1} \ell^{1/4}\notag\\
                  & \times  \left(\frac{\MPunkt}{10^{-2}~\ME\,\mathrm{yr}^{-1}}\right)^{5/4} \left(\frac{\RP}{1.5~\RJ}\right)^{-3/4}\notag\\
                     & \times \left(\frac{\MP}{10~\MJ}\right)^{1/4}, \label{Gl:Q+Formpar-Werte}
\end{align}
\end{subequations}
where again $\ell=1+\Llinks/\LAkkmax$, the
downstream luminosity $\Llinks=\Lint+\LKomp$ being the sum of the luminosity coming from the deep interior
and the compression luminosity,
and writing $\gamma_{1.44}\equiv\gamma/1.44$.
For Equations~(\ref{Gl:Q+Formpar}) and~(\ref{Gl:Q+Formpar-Werte}), we took the free-streaming limit
for the shock temperature (Equation~(\ref{Gl:TSchockff})), which was seen to hold over most of parameter space.
As in Equation~(\ref{Gl:TSchockff}), $\ell$ formally depends on $(\MPunkt,\MP,\RP)$
but this dependence is negligible in the limit of $\Llinks\ll\LAkkmax$.

\citet{mordasini17} have demonstrated that not only a ``hot accretion''
but even a ``cold nominal'' formation scenario leads to warm starts.
Therefore, we estimate $\qSchock$ with the cold starts of \citet{marl07},  %
which represent a conservative lower limit
to the entropies and luminosities of planets during their formation.
Since \citet{marl07} do not report the internal luminosities of their planets during formation,
we need to estimate them.
Their figure~4 shows that right after formation, the cold starts stay
within 90\,\%\ of the initial luminosity for a timescale
$\tnitti\lesssim1$~Myr for 1~$\MJ$,
1--2~Myr for 2~$\MJ$,
and $\sim6$~Myr and increasing for 4~$\MJ$ and up\footnote{
  Alternatively, one could look at the $e$-folding times of the luminosity,
  which are 4~Myr for $\texp=1~\MJ$, 10~Myr for 2~$\MJ$, 60~Myr for 4~$\MJ$, and increasing.
  Surprisingly,
  the Kelvin--Helmholtz times $\tKH=G\MP^2/\RP\LP=1~\mathrm{Gyr}~(\MP/4~\MJ)^2/[(\RP/1.3~\RJ)(\LP/3\times10^{-6}~\LSonne)]$
  are longer than this $e$-folding time
  by roughly a large factor of 30 (i.e., ca.\ 1.5~dex).
  The reason for this difference is not clear.
  Note that $\tKH$ is within ca.~$\pm0.5$~dex of the cooling time
  $t_S=\MP\overbar{T}S/\LP$ defined in \protect\citet{mc14},
  where $S$ is the entropy of the planet and $\overbar{T}$ its mass-averaged temperature.
  Thus both $t_S$ and $\tKH$ are hardly adequate proxies for $\tnitti$ or $\texp$,
  and one must exercise caution when using $\tKH$ in timescale-based arguments.
}.
The time spent in runaway accretion in their models is around
$\tAkk=0.1$~Myr to 0.5~Myr for 1~to 10~$\MJ$, respectively.
Thus even for $\MP=1~\MJ$ the cooling timescale $\tnitti$ is much larger than $\tAkk$,
and it should be a reasonable approximation to assume that the internal luminosity
and the radius
do not change considerably between the last stages of the runaway accretion phase---when,
for example, more than half the planet mass has been assembled---and
right after.

Figure~\ref{Abb:GemmaGalgani MutterSperanza Q+}b displays the approximate heating by the shock
(Equation~(\ref{Gl:Q+Formpar-Werte})).
Considering the extreme combinations of $\mu$ and $\gamma$
as in Section~\ref{Theil:einzelne},
we find $\QSchock\approx10^{-4}$--$10^{-3}~\LSonne$
for $\MPunkt\approx10^{-2}~\ME\,\mathrm{yr}^{-1}$
and $\QSchock\approx10^{-5}$--$10^{-4}~\LSonne$
for $10^{-3}~\ME\,\mathrm{yr}^{-1}$.
The radius is fixed at $\RP=1.5~\RJ$, a typical value for forming planets \citep{morda12_II},
but varying it in a reasonable range $\RP\approx1$--3~$\RJ$ does not change the outcome qualitatively.

We compare in Figure~\ref{Abb:GemmaGalgani MutterSperanza Q+}b the shock heating to
the internal luminosity of the \citet{marl07} cold starts.
In their model, most of the mass is accreted with $\MPunkt\approx10^{-2}~\ME\,\mathrm{yr}^{-1}$
with a linear decrease of $\MPunkt$ towards the end \citep{hubickyj05,marl07}.
Thus, the relevant heating is around $\QSchock>10^{-4}~\LSonne$.
This is one to two orders of magnitude higher~(!) than the (post-)formation luminosities
$\Lint\approx10^{-6}~\LSonne$ of \citet{marl07}.
At $1~\MJ$ the heating could be only moderate but for $\MP\gtrsim2~\MJ$
the conclusion becomes more secure.
Also, taking $\Llinks\neq0$ into account for the estimate would only lead to
a lower Mach number and thus a higher $\QSchock$.

Thus, based on this \textit{a posteriori} comparison of the internal luminosity
and of the energy input rate, the shock is expected to heat up
planets in the ``cold classical'' approach \citep{marl07,mordasini17}.
The importance of the shock should increase with planet mass.

\subsubsection{Shock-temperature-based argument}
\citet{berardocumming17}
and \citet{cumming18} followed the time-dependent internal structure of accreting planets
with constant accretion rates. They specified their outer boundary conditions for the planet structure
as a temperature $T_0$ at a pressure equal to the ram pressure, $P_0=\Pram$.
\citet{berardocumming17} report that setting $T_0$ as a fraction\footnote{These authors
  write ${T_0}^4 = \chi \TAkk^4 + \Teff^4$
  but their $\TAkk$ differs from our equivalent quantity, $\TSchockff$,
  with $T_{\mathrm{acc}}=4^{1/4}\TSchockff\approx1.4\TSchockff$ in the case of
  $\ell=0$ in Equation~(\ref{Gl:TSchockff}).
  We therefore use here $f$ and not $\chi$.
  Note also that $\chi$ in \citet{cumming18} was written as $\eta$
  in \citet{berardocumming17} but that it should not be confused with the efficiency.
  Our $f$ here can be larger than~1, when the pre-shock $\kapR\rho\RP$ is large
  (see Section~\ref{Theil:TSchock analytisch}).
}
$f$ of the free-streaming temperature $\TSchockff$
(plus a relative contribution from the internal luminosity)
led to fully radiative interiors
at the end of formation for $f$ above a certain $\fmin$.
Smaller values of $f$ resulted in convective interiors.
The
minimum fraction $\fmin$ was lower for larger accretion rates,
with $\fmin\approx1$ for $\MPunkt=10^{-3}$ and
$\fmin\approx0.1^{1/4}\approx0.6$ for $\MPunkt=10^{-2}$~$\ME\,\mathrm{yr}^{-1}$.
Since we find $f\geqslant1$---i.e., the temperature at the ram pressure matches
the temperature in the $\etaklassisch=100\,\%$ limit (Equation~(\ref{Gl:TSchockff}))---we
expect formation calculations using our results to lead to radiative planets.
Note that even though we considered only $\Lint=0$ here,
we expect the same result ($\etaklassisch=1$ and thus $f=1$) to hold for non-zero interior luminosity.
It should however still be explored systematically, especially since the result
of \citet{berardocumming17} was for a specific choice of pre-runaway entropy of the planet $S_i$,
and it is the contrast between this entropy $S_i$ and the immediate post-shock entropy $S_0=S(T_0,P_0)$
that matters \citep{berardo17}.

\subsubsection{Entropy-based argument}
Since our post-shock entropies are larger or much larger than post-formation entropies
(and thus, neglecting cooling, entropies during formation) of \citet{mordasini17},
we certainly do not expect
the shock to be able to cool the planet as it accretes.
\citet{berardo17} found that they needed extremely low shock entropies
(with temperatures of order $\TSchock\approx150$~K)
to reproduce the cold starts of \citet{marl07}.
We however find $\TSchock>1000$~K, down to $\MPunkt=10^{-3}~\ME\,\mathrm{yr}^{-1}$.
Thus the high post-shock entropy will at least slow down the cooling of the planet during its formation
(the ``stalling'' regime of \citealp{berardo17}),
if not heat the planet (``heating regime''), but should not allow for any decrease of the entropy
(``cooling regime'').
From this argument too, then, cold starts seem unlikely.

\subsection{Opacities and gas--radiation coupling}
 \label{Disk:kappa}

Next we discuss the opacities. %
We have seen in detail in this work that the Rosseland mean controls the extent
to which radiation is diffusing as opposed to freely streaming.
As for the Planck opacity, it is an important factor in determining
the extent to which the opacity carrier (gas or dust)
and the radiation are in equilibrium.
Indeed, the inflowing matter will equilibrate with the outgoing radiation,
leading to $\TGas\approx\Trad$,
if the energy exchange time (controlled by the absorption coefficient $\rho\kapP$;
cf.\ \citealp{malygin17})
is shorter than the time needed for the gas to reach the shock (controlled by $\vFf$).
Therefore we take a critical look at uncertainties concerning the opacities.

\begin{itemize}

\item
We find that everywhere except in the shock (in the Zel'dovich spike),
the pre-shock temperatures of the gas and radiation are equal.
Simulations which use non-equilibrium radiation transport
with tables with unrealistically low Planck mean values (cf.\ Figure~\ref{Abb:kappaSchnitt}b)
might not find that the shock is able to pre-heat the gas.
Thus it is important result that the Planck values are high enough
for the radiation and gas to be coupled.

\item
A quite
uncertain aspect of dust opacity computations is the distribution
of dust grains properties (size, porosity, composition, etc.)
and also their sublimation. However,
we found that the exact opacity in the accretion flow
does not matter for the shock temperature.
This effectively removes a source of uncertainty
and makes the derived shock temperatures more robust.

\item
Nevertheless, the presence of dust in the accretion flow
was seen to affect the temperature at and beyond the dust destruction front
(cf.\ \citealp{stahlerI,vaytet13}) and thus also the luminosity at the Hill sphere.
With a dust destruction front, the temperature there and beyond remains higher
by up to a factor of several
compared to the expression for a constant-$(L/\fred)$ profile,
Equation~(\ref{Gl:T(r) konst L/fred})
(albeit with the same powerlaw dependence).
The decrease or increase in \textit{luminosity} between the shock and the Hill sphere
is also different from the case
without a dust destruction front (see Section~\ref{Theil:L,eta von ZG}).
However,
if the dust in the incoming flow is strongly depleted relative to the interstellar medium
abundance assumed in \citet{semenov03},
the flow will tend more to be in the free-streaming regime
and its temperature thus given by Equation~(\ref{Gl:T(r) konst L/fred}).

\item
We conducted tests as in \citetalias{m16Schock}
with constant low opacity. Typically, as we verified separately (not show),
at lower values $\kapR=\kapP\sim0.01$~cm$^2$\,g$^{-1}$
the radiation and gas temperatures stay decoupled even in the high-density post-shock region.
This is however entirely unrealistic given that at those densities ($\rho\approx10^{-10}$--$10^{-8}~\textrm{g\,cm}^{-3}$)
and temperatures ($\TGas\approx500$--5000~K; cf.\ figure~4 of \citetalias{m16Schock})
the Planck opacity
is rather of order $\kapP\sim10$~cm$^2$\,g$^{-1}$ (Figure~\ref{Abb:kappaSchnitt}b).

\item
Finally, we related the behaviour of the opacity to that of the temperature
and thus also of the luminosity (Section~\ref{Gl:T-Profile im Akkfl}).
This makes it now possible to understand the ``bursts'' in $L$
seen, e.g., at 0.3~au in figure~8 of \citet{vaytet13}, who also use FLD,
however keeping the frequency dependency.
These bursts are associated with sharp opacity transitions
in the respective wavelength band (in particular at the dust destruction front)
and with slight changes of slope in the temperature.

\end{itemize}

We note that
it is in the Zel'dovich spike that non-equilibrium (2-$T$) effects
lead to the formation of observable spectral tracers of accretion onto protostars
and brown dwarfs
\citep{hartmann16,santamar19}.
This emission is discussed for the shock onto the circumplanetary disc in \citet{aoyama18}
and for the accretion shock on the planet surface in a forthcoming publication (Aoyama et al., in prep.).

\subsection{Equation of State}
 \label{Theil:Disk ZG}
 
In these first two papers (\citealt{m16Schock}, this work) we have restricted ourselves
to a perfect equation of state (EOS; constant $\mu$ and $\gamma$).
To first order, this should not affect our main results.
However, (i)~the luminosity in the accretion flow (and thus at the Hill sphere),
(ii)~the post-shock compression luminosity, and thus also
(iii)~the more precise value of $\TSchock$ (through the $\Delta\fred$ factor)
should all be affected to some extent by the EOS,
at least for some combinations of $(\MPunkt,\MP,\RP)$.

Concerning item~(i),
we studied in Section~\ref{Theil:L(r; ZG)} an estimate $\Delta \LkonstLfred$ of the drop or increase in the luminosity
between the shock and the Hill radius.
In the limit of a perfect EOS and of temperature profile $T\propto r^{-1/2}$,
Equation~(\ref{Gl:Delta L bis RAkk approx})
shows that the ratio $|\Delta \LkonstLfred|/\LAkk$ is highest for high accretion rates,
low masses, and high radii.
Over $\MPunkt=10^{-3}$--$10^{-2}~\ME\,\mathrm{yr}^{-1}$, $\MP=1$--10~$\MJ$, and $\RP=1$--5~$\RJ$
(an even wider parameter space than what we consider for our simulations),
the relative drop or increase is never larger than $|\Delta \LkonstLfred|/\LAkk\approx10$--15\,\%,
taking the extreme case of $(\mu=1.23,\gamma=1.1)$.
Taking the actual temperature profile into account (as opposed to assuming $T\propto r^{-1/2}$ everywhere)
will change this somewhat but usually only to make it smaller.
In any case, the variation in $L$ across the Hill sphere is unimportant
compared to the effect of other simplifications of our model.
We note that for molecular hydrogen and neutral helium (our default case),
the actual change in $L$ (i.e., as measured from the simulation and not using a simple $T\propto r^{-1/2}$ profile)
is less than 2\,\%\ for any $(\MPunkt,\MP,\RP)$ considered here.

We report already here that preliminary estimates suggest that for simulations with a full EOS,
the change in luminosity across the Hill radius is at most approximately 20\,\%,
which is thus noticeable but also not large.
Details will be presented in a forthcoming publication. %

Finally, the relative smallness of the luminosity change $\Delta\LkonstLfred$
justifies \textit{a posteriori} the assumption of constant $(L/\fred)$ made to derive it.
Indeed, a relative change $|\Delta \LkonstLfred|/\LAkk\approx10$--15\,\%\ over 2--3~dex in radius
(from $\RP\approx1$--3~$\RJ$ to $\RAkk\approx\kLiss\RHill(\MP)\approx250$--500~$\RJ$ for $\MP\approx1$--10~$\MJ$
at 5.2~au)
corresponds to an approximate average slope (see Equation~(\ref{Gl:beta-Def}))
of at most $|\beta|\approx\log_{10}(1.15)/2=0.04$ if $\fred$ is constant
or $|\beta|\approx0.4$ at most if $\fred$ also changes by a factor of ca.~3 as in Figure~\ref{Abb:Abbgammamu}.
Thus even for these conservative estimates we find $|\beta|\ll2$, justifying \textit{a posteriori}
the assumption of a constant $L/\fred$ used to derive the change in $L$.
\section{Summary and conclusions}
 \label{Theil:Zus}

In this series of papers (\citealp{m16Schock}; this work; Marleau et al., in prep.)
we take a detailed look at the physics of the accretion shock
in planet formation. In this second paper, %
we have updated to
disequilibrium (2-$T$) radiation transport (i.e., following the gas and radiation
energy densities separately)
and modern opacities, especially for the gas:
we use the dust opacities of \citet{semenov03}
but the gas opacities of \citet{malygin14},
avoiding the too-low Planck mean opacities %
normally included in \citet{semenov03}.%
We have also now surveyed a range
of values for the formation parameters $(\MPunkt,\MP,\RP)$,
assuming negligible $\Lint$ (see Figure~\ref{Abb:Lint/LSchock}),
namely
$\MPunkt=10^{-3}$--$10^{-2}~\ME\,\mathrm{yr}^{-1}$,
$\MP\approx1$--10~$\MJ$,
$\RP\approx1.6$--3~$\RJ$.
This has motivated us to several semi-analytical derivations
along with comparisons to simulation outputs.
We have kept the simplification of a perfect equation of state and
focused on the case of molecular hydrogen
with a cosmic admixture of helium ($\mu=2.29$, $\gamma=1.44$).  %

We now summarize our primary findings on different aspects.
Concerning the thermal and radiative properties of the accretion flow:
\begin{enumerate}
 \item Both our simulations and analytical theory show
       that the behaviour of the luminosity in the accretion flow
       is not the direct result of radiative transfer effects
       but rather depends on the equation of state (Section~\ref{Theil:L(r; ZG)}).
       The luminosity turns out to be radially constant to $\approx2$\,\%\ for values
       of $\mu$ and $\gamma$ appropriate for H$_2$\,+\,He.
       Taking other values of $\mu$ and $\gamma$
       increases or decreases the change in $L$
       between the shock and roughly the Hill sphere.  %
       However, the maximum change is relatively small
       with $|\Delta \LkonstLfred|/\LAkk\lesssim15$\,\%\ across
       the relevant parameter space and for any ($\gamma\geqslant1.1,\mu\geqslant1.23$) combination
       (Section~\ref{Theil:Disk ZG}).
       We highlight that the $\Delta\tauR\sim1$ surface is not of any particular significance
       (Section~\ref{Theil:Delta tau bis rmax}).

 \item
      Thanks to the sufficiently high Planck mean opacities (for which a contribution from the dust is not needed),
      the matter and radiation are very well coupled both ahead of and behind the shock,
      i.e., everywhere except in the Zel'dovich spike (Figures~\ref{Abb:kappaSchnitt} and~\ref{Abb:Abbgammamu}).
       In fact, non-equilibrium (2-$T$) radiation transport could be neglected when studying only
       the post-shock temperature and pressure or the global energetics.
 
 \item
 As found in \citetalias{m16Schock} and confirmed here,
       the radiative precursor to the shock \citep{mihalas84,commer11}
       is larger than the simulation domain, which is roughly the Hill sphere, even in the case
       of somewhat high Rosseland optical depth ($\Delta\tauR\sim10$).
 \item
 The pre-shock region close to the planet, out to some Rosseland optical depth,
       is usually in the free-streaming regime
       and not, as one would expect for a supercritical shock, in the diffusion limit (e.g., figure~8 of \citealt{vaytetgonz13}).
       Thus the shock is a thick--thin shock in the classification of \citet{drake06}.
       At low shock temperatures ($T\lesssim1500$~K) the dust is still present,  %
       making the pre-shock region somewhat diffusive and raising the shock temperature.

\end{enumerate}
The shock properties were a focus of this study and we found the following:
\begin{enumerate}
 \item
 As in \citetalias{m16Schock}, all shocks are isothermal and supercritical,
       and the Mach numbers are high enough for $\etaklassisch\approx100\,\%$ of the incoming kinetic energy flux
       to be converted to radiation \textit{locally at the shock} (see grey lines in Figure~\ref{Abb:Gitter-etaphys}
       and Figure~\ref{Abb:Mach implizit}).
       The post-shock pressure is equal to the ram pressure $\Pram$.
 \item
 The free-streaming analytical estimates
       of the shock temperature (Equation~\ref{Gl:TSchock_rauh b}) and of the upstream luminosity
       (Equation~\ref{Gl:LAkk ab unendlich}) were seen to hold very well over a large portion
       of parameter space.
       Importantly, we found out that this holds also for high optical depths between the shock
       and the nebula.
       Deviations of $\sim5\,\%$ in $T$ occur at low shock temperatures (Figure~\ref{Abb:Gitter}a).  %
 \item
       An important analytical development was the derivation of
       an implicit equation (Equation~(\ref{Gl:TSchock implizit approx, fuer Abb})) for the shock temperature $\TSchock$
       given a Rosseland mean opacity function $\kapR(\rho,\TSchock)$.
       We solved this numerically (Figure~\ref{Abb:Timplizit(RP)}).

 \item
       Based on our analysis,
       $\TSchock$ should not be affected to first order
       by the use of a non-perfect ideal EOS (i.e., considering dissociation and ionization),
       since $\gamma$ and $\mu$ do not enter in the derivation of $\TSchock$ (see Equation~(\ref{Gl:TSchock implizit beide})).
       However, (i)~the luminosity in the accretion flow (and thus at the Hill sphere),
       (ii)~the post-shock compression luminosity, and thus also
       (iii)~the more precise value of $\TSchock$ (through the $\Delta\fred$ factor; Equation~(\ref{Gl:TSchock_rauh}))
       should all be depend somewhat on the EOS.
       This will be assessed in a subsequent paper (Marleau et al., in prep.).
 \item
       We calculated the post-shock entropies immediately below the shock
       using an EOS taking dissociation and ionisation into account (appendix~A of \citealp{berardo17}).
       While this is formally not consistent with our (perfect-EOS) simulations,
       we argued this is likely accurate since $\TSchock$ and $\Ppost=\Pram$
       are probably independent of the EOS.
       The immediate post-shock entropies were found
       to be between approximately 13 and 20~$\kB\,\mH^{-1}$ for our range of parameters
       (Section~\ref{Theil:Spost}, Figure~\ref{Abb:Gitter Joseph}).
       These values are high compared to the post-formation entropy of planets,
which is at most around 10--14~$\kB\,\mH^{-1}$ according to current, though not definitive,
predictions \citep{berardocumming17,berardo17,morda13,mordasini17}.
However, we caution and emphasize that
the actual entropy of the gas added to the planet, below the post-shock settling layer,
is different from this immediate post-shock entropy.
This is explored in \citet{berardo17} and \citet{berardocumming17}.
\end{enumerate}
Finally, a key output of our simulations was the efficiency of the shock:
\begin{enumerate}
 \item 
       We have measured
       the physical efficiency $\etaphys$
       as a function of accretion rate $\MPunkt$, planet mass $\MP$, and planet radius $\RP$
       (Figure~\ref{Abb:Gitter Joseph}) and derived it semi-analytically
       (Equations~(\ref{Gl:etaisoth_phys}) and~(\ref{Gl:Mach analytisch})).
       This efficiency captures the global energy recycling occurring in the pre-shock region \citepalias{m16Schock}.
       We saw (Figure~\ref{Abb:Gitter Joseph}) that the efficiencies are always greater than roughly 97\,\%\
       for the range of parameters considered here.

 \item
       Naively, a high $\etaphys$ could suggest that the gas is added ``cold''
       but the part not escaping (i.e., the heating of the planet heating by the shock)
       turns out to be much larger than the internal luminosity in the \citet{marl07}
       extreme cold starts (Section~\ref{Theil:H oder K? Leuchtkraft}).

\end{enumerate}

The semi-analytical work presented here revealed
the reduced flux $\fred=\Frad/(c\Erad)$ (Equation~(\ref{Gl:fred-Def}))
to be a powerful quantity for understanding the behaviour of the radiation field
(free streaming or diffusing, often termed approximately ``optically thin'' and ``optically thick'').
This holds at least for the grey treatment of radiation transport used in this work
in a spherically symmetric geometry.
When $L/\fred$ is sufficiently constant radially, we showed in Equation~(\ref{Gl:fred og kapparhor})
that there is a simple relation between the reduced flux $\fred$ and the opacity,
which provides an intuitive understanding of $\fred$.

The main results of our simulations are post-shock $(P,T)$ values
and global efficiencies $\etaphys$.
This is useful respectively for detailed modeling of the structure of accreting planets as in \citet{berardo17},
\citet{berardocumming17}, and \citet{cumming18},
as well as for the one-zone, global approach of, e.g., \citet{hartmann97}.
The Bern model \citep{alibert05,morda12_I,mordasini17} is currently in between,
calculating detailed planet structures but with the assumption of a radially constant luminosity.
Note that the modelling of the energy transfer at the accretion shock
is also relevant in the context of star formation (e.g., \citealp{baraffe12,geroux16,baraffe17,jensen18}).
Researchers interested in using our simulation results can take the semi-analytical formul\ae\
presented above, including the opacity effects for the temperature,  %
under the assumption of a perfect EOS.

The other main outcome is
the amount of radiation reaching the Hill sphere.
This
should be useful input for studies of the thermo-chemical
feedback of planets on the local protoplanetary disc,
for instance as in a number of recent papers \citep{cleeves15,cridland17,stamatellos18,rab19}.
Within the simplification of a spherical accretion geometry,
our results show that essentially all of the accretion shock
luminosity is expected to reach the local nebula,
and that a high Rosseland optical depth, at least up to $\Delta\tauR\sim10$,
does not lead to significant extinction of the bolometric shock luminosity in the accretion flow.

Finally, we have explored by different means whether our results point
towards hot starts or cold starts (Section~\ref{Theil:H oder K?}).
As discussed above,   %
the heating of the planet by the shock $\QSchock$ (i.e., the flow rate of inward-going energy; Equation~\ref{Gl:QSchock-Def})
was estimated to be much larger than the internal luminosity for the \citet{marl07} classical cold starts.
This suggests that they are not entirely realistic.
As for the ``nominal cold start'' or the ``hot start'' assumption during accretion,
\citet{mordasini17} showed that both lead to warm or even hot starts.
Taken together, all of this might explain why direct imaging observations,
with the sole exception of 51~Eri~b \citep{macintosh15,nielsen19},
have not been finding evidence for planets even consistent with cold starts.

\acknowledgments

\small
We acknowledge the valuable support of Th.~Henning and W.~Benz for this project.
It is a pleasure to thank
A.~Mignone and collaborators for their excellent open-source code \PLUTO.
We also thank
W.~Kley and
H.~Klahr for several helpful discussions;
N.~Malygin for alacritous and competent help
as well as for providing routines to read and use his opacity tables in \texttt{Makemake};
D.~Semenov for publicly-available \texttt{fortran} code for his opacity data;
N.~Turner,
A.~Cumming,
C.~P.~Dullemond, %
R.~P.~Nelson,    %
T.~Tanigawa,     %
M.~Ikoma,
D.~N.~C.~Lin,    %
J.~Bouwman,
and
V.~Elbakyan  %
for interesting questions, comments, and discussions;
and A.~Emsenhuber (Bern), V.~Lutz and J.~Kr\"uger (T\"ubingen), and U.~Hiller (MPIA) for rapid and patient help
with the respective computing clusters.
\nnBegut{We thank the referee for a useful report which lead to several clarifications
and motivated a deeper analysis of some aspects.}
G-DM and RK acknowledge the support of the DFG priority program SPP 1992 ``Exploring the Diversity of Extrasolar Planets'' (KU 2849/7-1).
G-DM and CM acknowledge support from the Swiss National Science Foundation under grant BSSGI0\_155816 ``PlanetsInTime''.
Parts of this work have been carried out within the framework of the NCCR PlanetS supported by the Swiss National Science Foundation.
RK acknowledges financial support within the Emmy Noether research group on
``Accretion Flows and Feedback in Realistic Models of Massive Star Formation''
funded by the German Research Foundation (DFG) under grants no.~KU 2849/3-1 and KU 2849/3-2.
Part of the simulations presented here were performed on the \texttt{ba(t)chelor} cluster at the MPIA.
The authors acknowledge support by the High Performance and Cloud Computing Group at the Zentrum f\"ur Datenverarbeitung
of the Universit\"at T\"ubingen, the state of Baden-W\"urttemberg through bwHPC, and the German Research Foundation (DFG) through grant no.~INST 37/935-1 FUGG.
All figures were produced using \texttt{gnuplot}'s terminal \texttt{epslatex} with the font package \texttt{fouriernc}.

\software{\PLUTO\ \citep{mignone07,mignone12},
\mkmk\ (\citealp{kuiper10}; Kuiper, Yorke, \&\ Mignone, in prep.)%
}

\bibliographystyle{yahapj_nodash}
\bibliography{std.bib}{}

\end{document}